\documentclass[graybox]{svmult}
\usepackage{mathptmx}       % selects Times Roman as basic font
\usepackage{helvet}         % selects Helvetica as sans-serif font
\usepackage{courier}        % selects Courier as typewriter font
\usepackage{makeidx}         % allows index generation
\usepackage{graphicx}        % standard LaTeX graphics tool
\usepackage{multicol}        % used for the two-column index
\usepackage[bottom]{footmisc}% places footnotes at page bottom
\usepackage{amsmath,amssymb,amsfonts}        % AMS math packages
\usepackage{hyperref}
\usepackage{tikz}
\usepackage{wasysym}
\makeindex             % used for the subject index
\usepackage{dsfont}
\newcommand{\de}{\text{d}}
\newcommand{\extder}{\text{d}}
\newcommand{\Ad}{\text{Ad}}
\newcommand{\Tr}{\text{Tr}}
\newcommand{\STr}{\text{STr}}
\newcommand{\alg}{\mathfrak}
\newcommand{\diag}{\text{diag}}
\newcommand{\nln}{\nonumber\\}

\renewcommand{\L}{\mbox{\tiny L}}
\newcommand{\R}{\mbox{\tiny R}}

\begin{document}
\title*{AdS3 Integrability, Tensionless Limits, and Deformations: A Review}
\author{Fiona K.~Seibold and Alessandro Sfondrini}
\institute{Fiona K.~Seibold \at Blackett Laboratory, Imperial College London, SW7 2AZ London, United Kingdom, and DESY, Notkestrasse 85, 22607 Hamburg, Germany.
\email{fiona.seibold@desy.de} (MATRIX Simons Young Scholar)
\and Alessandro Sfondrini \at Dipartimento di Fisica e Astronomia, Università degli Studi di Padova, via Marzolo 8, 35131 Padova, Italy, and
Istituto Nazionale di Fisica Nucleare, Sezione di Padova, via Marzolo 8, 35131 Padova, Italy. \email{alessandro.sfondrini@unipd.it}
(MATRIX Simons Fellow)}

\maketitle

\abstract{Motivated by the recent advances in the understanding of integrability for $AdS_3$ backgrounds, we present a lightning review of this approach, with particular attention to the ``tensionless'' limits (with zero and one unit of NSNS flux), and to the many integrable deformations of the supergravity backgrounds. Our aim is to concisely but comprehensively take stock of the state of the art in the field, in a way accessible to non-experts, and to highlight outstanding challenges. Along the way we reference where the various derivation of these results, which we mostly omit, can be found in full detail.
}

\section{Introduction}
\label{sec:introduction} 
It is in general very difficult to quantise any superstring model and compute non-protected observables, even in the planar limit, in the presence of Ramond-Ramond background fluxes, see e.g.~\cite{Gopakumar:2022kof}. 
We know from the correspondence between type IIB $AdS_5\times S^5$ superstrings that computing the spectrum of free strings as function of the string tension is ``as difficult'' as computing the spectrum of planar $\mathcal{N}=4$ supersymmetric Yang-Mills theory as a function of the 't~Hooft coupling; remarkably both tasks can be accomplished by means of \textit{integrability}, see~\cite{Arutyunov:2009ga,Beisert:2010jr} for reviews.
By this we mean that one can find a system of equations which may be solved numerically at high precision for any state and any value of the 't Hooft coupling (or tension); in special limits, the equations can also be solved order-by-order analytically or semi-analytically.

Even if historically integrability was first discovered~\cite{Minahan:2002ve} (and arguably, understood~\cite{Beisert:2005fw,Beisert:2005tm}) on the gauge theory side of the duality, it is very helpful to think of it as arising from the worldsheet of the string in lightcone gauge (as it is presented in~\cite{Arutyunov:2009ga}). This is because there is a plethora of integrable string backgrounds that can be studied classically, semiclassically or even (under some assumptions) at the quantum level, for which we do not even know the form of the dual theory.

A particularly interesting class of superstring backgrounds is that of the $AdS_3$ type.%
\footnote{%
A rather extensive review of integrability for that setup was recently written~\cite{Demulder:2023bux}; this paper can be though of as a more agile overview of the topics explained in detail there, which also includes a discussion about the weak-tension limits and dual CFTs, as well as of the deformations.
}
They can be supported by NSNS flux in which case the worldsheet theory is a supersymmetric WZW model, and can be solved exactly in the RNS formalism. From the point of view of the planar spectrum, these are almost free theories --- the spectrum is freely generated by acting with the Ka\v{c}-Moody generators, up to imposing the Virasoro cosntraint, which can be done via the Sugawara construction, see~\cite{Maldacena:2000hw}. The spectrum of the pure-NSNS theory is quite atypical; it is very degenerate, and it features both a discrete part (corresponding to short-string states) and a continuum (long strings).

There are then several ways to deform this setup. The simplest modification is to do a T-duality--shift--T-duality transformation (TsT) of the background~\cite{Lunin:2005jy}. Such a transformation can be reabsorbed in a modification of the boundary conditions of the fields~\cite{Frolov:2005dj,Alday:2005ww}.
%, which makes it straightforward to find the deformed spectrum. 
It is then straightforward to obtain the spectrum, as long as the TsT transformation is performed along directions that correspond to Cartan generators of the symmetry algebra. The other cases are more challenging to study~\cite{Guica:2017mtd}.  %v2
In terms of the WZW model, %they
TsT transformations %v2
can be understood as current-current deformations which do not spoil the holomorphic structure of the model~\cite{Forste:2003km,Borsato:2018spz,Borsato:2023dis}.
In the modern understanding of integrable deformations, these are called ``homogeneous'' Yang-Baxter deformations, as we shall review. 

There are however many other types of deformations, which generically turn on moduli corresponding to RR fluxes.
The simplest of these are those which preserve all the (super)isometries of the model --- in fact, they do not affect the metric at all, but only the Kalb-Ramond field. From the string non-linear sigma model point of view, this amounts to playing with the coefficients if front of the sigma model and Wess-Zumino piece of the bosonic action, as we shall see; in string theory, they arise by turning on an axion in the geometry, see~\cite{OhlssonSax:2018hgc}. The whole collection of superstring actions thus obtained is classically integrable~\cite{Cagnazzo:2012se}. These are usually called ``mixed (NSNS/RR) flux  backgrounds'', where the parameter~$k$ associated to the NSNS flux is discrete, and the one associated to the axion~$h$ is continuous. The string tension, in dimensionless units, is
\begin{equation}
    T= \sqrt{\frac{k^2}{4\pi^2}+h^2}\,.
\end{equation}
It is possible also to have backgrounds with $k=0$ (and hence, no Kalb-Ramond field); this background arises as the near-horizon limit of D1-D5 branes~\cite{Maldacena:1997re}.

For the $AdS_3\times S^3\times T^4$ backgrounds with arbitrary NSNS/RR flux (as well as, in principle, for their TsT deformations) the $T\gg1$ limit is relatively well understood, both in terms of semiclassical string NLSM computations and in the near-pp-wave expansions~\cite{Berenstein:2002jq}.
For some of these backgrounds, owing to integrability, we have recently understood how to compute the spectrum at arbitrary tension~$T$; indeed, we expect (or hope) this to be possible soon for \textit{all} these backgrounds.

One particularly interesting limiting case for this setup is the ``tensionless'' limit, where $T$ is small and the dual theory should become simpler. In fact there is more than one such setup. One is $k=0$ and $h\ll1$, which can be studied by integrability (by means of the ``mirror Thermodynamic Bethe Ansatz''~\cite{Brollo:2023pkl}), and whose dual should be related to weakly-coupled strings in the D1-D5 system~\cite{OhlssonSax:2014jtq}. Another notable setup is $k=1$ and $h\ll1$. Here the dual is (a marginal deformation of) the symmetric-product orbifold CFT of four bosons and fermions~\cite{Giribet:2018ada,Gaberdiel:2018rqv,Eberhardt:2018ouy}. Also in this case integrability can be identified, both at $h=0$~\cite{Frolov:2023pjw} and in principle at the leading order in $h\ll1$~\cite{Gaberdiel:2023lco,Frolov:2023pjw}, as we will review.%
\footnote{Note that it was argued in~\cite{Fabri:2025rok} that the derivation in~\cite{Gaberdiel:2023lco} presents technical issues that put into question the derivation of the integrability structure.}

There are also more general deformations that one may consider. They introduce RR flux when starting from pure-NSNS backgrounds \textit{and} break (or rather, deform) some of the superisometries of the background. They correspond to inhomogeneous Yang-Baxter deformation~\cite{Klimcik:2002zj,Delduc:2013qra}.%
\footnote{Of course one may consider the inhomogeneous Yang-Baxter deformation of a pure-RR background; for $AdS_3$ backgrounds this generates a Kalb-Ramond field $B$ such that $H=\de B$ vanishes.}
In the simplest setup 
(say, a sigma model on the three-sphere~$S^3$) they would correspond to a \textit{squashing} of the sphere, so that the isometry algebra is deformed from $\mathfrak{su}(2)$ to the \textit{quantum group} $\mathfrak{su}(2)_q$ where $q$ is the deformation parameter. %v2
In the particular case of $AdS_3$ superstrings, the zoo of possible deformations is quite rich, even when preserving part of the supersymmetry~\cite{Hoare:2022asa}. The dual CFTs of these models are thus far mysterious, and recently even more general multiparametric deformations have been constructed too --- namely, of the ``elliptic'' type~\cite{Hoare:2023zti}.

Below we will try to concisely review these developments. We will begin by summarising several basic facts about $AdS_3$ in the lightcone gauge, in Section~\ref{sec:lcgauge}; this part is explained in complete detail in other reviews, see in particular~\cite{Demulder:2023bux}. We will then discuss the S~matrix and spectrum of the mixed flux (but otherwise undeformed) model in Section~\ref{sec:symmetries-smatrix}; this material is mostly standard  but the presentation is amended to make it easier to match with the symmetric product orbifold CFT at $k=1$ (following~\cite{Frolov:2023pjw}). In Section~\ref{sec:tensionless} we discuss the latest insights on the $k=0$ and $k=1$ models at $h\ll1$; this presentation, while based on existing results, is rather new.
Then, in Section~\ref{sec:deformations} we discuss the landscape of integrable deformations of string sigma models. While some of this material can also be found in the review~\cite{Hoare:2021dix}, we mainly focus on the string theory interpretation of the deformations. Many of these deformations can also be applied to for instance strings on $AdS_5 \times S^5$, some of them are particular to $AdS_3$ and present interesting new features. In particular, in Section~\ref{sec:biYB-WZ} we focus on the bi-Yang-Baxter deformation, and the bi-Yang-Baxter-plus-Wess-Zumino (bi-YB+WZ) deformation; for the former, we also present the fluxes in a particularly simple form which has not appeared elsewhere. In~\ref{sec:elliptic} we review the elliptic deformation, which further generalises the (trigonometric) ``quantum'' deformations.
We conclude in Section~\ref{sec:outlook} by recapping some intriguing open problems in the field.

\section{AdS3 strings in uniform lightcone gauge}
\label{sec:lcgauge}
We are interested in studying strings backgrounds supported by a mixture of RR and NSNS flux. There are three families of $AdS_3$ backgrounds with maximal supersymmetry (i.e., 16 Killing spinors). They are $AdS_3\times S^3\times T^4$, $AdS_3\times S^3\times K3$ and $AdS_3\times S^3\times S^3\times S^1$. Among these, the first background is by far the most studied, and we will focus on it for now.

\subsection{Superstring action and integrability}
As discussed in the introduction, in the presence of RR fluxes it is difficult to use the RNS formalism to compute the string spectrum. Our starting point is instead the Green-Schwarz action, 
% which for this background is of the form
% \begin{equation}
%     S_{\text{GS}} = S_{\text{bos}}+ S_{(2)}+S_{(4)}\,,
% \end{equation}
% where the first term on the right-hand side does not include Fermions, the second is quadratic in Fermions, and the last is quartic. The form of the last two terms can be found in~\cite{Wulff:2013kga}.
which takes the form
\begin{equation}
     S_{\text{GS}} = S_{\text{bos}}+ S_{(2)}+S_{(4)} + \dots + S_{(32)}\,.
 \end{equation}
The first term on the right-hand side does not include Fermions, the second is quadratic in Fermions, the third is quartic in the fermions, and so on. While for generic backgrounds the action may go up to order 32 in Fermions, the explicit expressions are only known up to quartic order~\cite{Wulff:2013kga}. For several models including the one at hand, it is possible to make a specific choice of (lightcone) $\kappa$-gauge, so that the action truncates at quartic order in the Fermions~\cite{Wulff:2013kga}.

It is worth saying that this action can also be related to a supercoset \`a la Metsaev-Tseytlin~\cite{Metsaev:1998it} for the $AdS_3\times S^3$ part, plus free Bosons on~$T^4$. 
More specifically, Cagnazzo and Zarembo~\cite{Cagnazzo:2012se} introduced a generalisation of the Metsaev-Tseytlin action. Let us first recall that $G/H$ is a symmetric space if $H$, $G$ are Lie groups with Lie algebras $\mathfrak{h},\mathfrak{g}$ so that $\mathfrak{h}$ is a subalgebra of $\mathfrak{g}$, $\mathfrak{g}\cong\mathfrak{h}\oplus \mathfrak{h}_{\perp}$ and there exists a $\mathbb{Z}_2$ algebra automorphism $\Omega$, $\Omega^2=\mathbf{1}$, so that for $h\in\mathfrak{h}$, $\Omega h=h$.
For a (Bosonic) sigma model whose target space is a symmetric space~$G/H$, one can decompose the Maurer-Cartan form as
\begin{equation}
    J = g^{-1}\text{d} g = J^{(0)}+J^{(2)}\,,
\end{equation}
simply by demanding that $J^{(0)}$ and $J^{(2)}$ have eigenvalue $+1$ and $-1$, respectively, under~$\Omega$. For a supercoset, one instead needs a $\mathbb{Z}_4$ superalgebra automorphism so that the decomposition reads 
\begin{equation}
    J = g^{-1}\text{d} g = J^{(0)}+J^{(1)}+J^{(2)}+J^{(3)}\,,
\end{equation}
where $J^{(1)}$, $J^{(3)}$ are in the odd (Fermionic) part of the superalgebra, and the $J^{(n)}$ have eigenvalues $i^n$ under~$\Omega$. Then on can write the Metsaev-Tseytlin action schematically as
\begin{equation}
\label{eq:MT}
S_{\text{MT}}=\frac{1}{2}\int\limits_{\Sigma}\,
    \text{Str}\left[J^{(2)}\wedge *J^{(2)}+J^{(1)}\wedge J^{(3)}\right]\,,
\end{equation}
where $\Sigma$ is the two-dimensional string worldsheet.
This expression can be used to represent the GS action on ``semi-symmetric'' target spaces, such as famously
\begin{equation}
    G/H,\qquad G=PSU(2,2|4)\,,\qquad H=SO(1,4)\times SO(5)\,
\end{equation}
which gives the $AdS_5\times S^5$ action. For the case of $AdS_3$ backgrounds one needs a more general action, due to Cagnazzo and Zarembo,
\begin{align}
S_{\text{CZ}}=&\frac{1}{2}\int\limits_{\Sigma=\partial B}\,
    \text{Str}\left[J^{(2)}\wedge *J^{(2)}+\hat{q}\;J^{(1)}\wedge J^{(3)}\right]\\
    &+
    q\int\limits_{B}\,
    \text{Str}\left[J^{(2)}\wedge J^{(2)}\wedge J^{(2)}+J^{(1)}\wedge J^{(3)}\wedge J^{(2)} +J^{(3)}\wedge J^{(1)}\wedge J^{(2)}\right],
    \nonumber
\end{align}
where $\hat{q}^2=\sqrt{q^2 - 1}$ and $0\leq q\leq  1$ is a new parameter of the model. One then takes
\begin{equation}
     G=PSU(1,1|2)\times PSU(1,1|2)\,,\qquad H=SO(1,2)\times SO(3)\,,
\end{equation}
which gives the $AdS_3\times S^3$ supergeometry (the $T^4$ part can be added ``by hand''). 
The existence of this formulation was very important to argue for the classical integrability of this model~\cite{Cagnazzo:2012se} and it will be very useful in discussing its integrable deformations. It however should be noted that matching this action with the GS one for the whole 10-dimensional background requires  a choice of the $\kappa$~gauge fixing (which, moreover, is not the one compatible with lightcone gauge~\cite{Sundin:2012gc}). This is discussed also in~\cite{Borsato:2014hja}. In any case, classical integrability holds for the whole superstring model. At this stage let us bypass this discussion; moreover, for simplicity (and to illustrate the role of the parameter $q$) let us focus on the Bosonic action.

Let us indicate the bosonic fields as
\begin{equation}
    X^\mu,\quad \mu=0,\dots 9,\qquad
    X^\mu=(t,z_1,z_2,\ \varphi,y_1,y_2,\ x_1,x_2,x_3,x_4).
\end{equation}
The first three fields are from $AdS_3$, the next three from $S^3$, and the remaining four from $T^4$.

The geometry $G_{\mu\nu}(X)$ is specified by the metric, whose line element is (in the notation of~\cite{Lloyd:2014bsa})
\begin{align} 
    \de s^2=\,& -\left(\frac{4+z_1^2+z_2^2}{4-z_1^2-z_2^2}\right)^2\de t^2+\left(\frac{4}{4-z_1^2-z_2^2}\right)^2\left(\de z_1^2+\de z_2^2\right)
    \nonumber\\
    &+\left(\frac{4-y_1^2-y_2^2}{4+y_1^2+y_2^2}\right)^2\de \varphi^2+\left(\frac{4}{4+y_1^2+y_2^2}\right)^2\left(\de y_1^2+\de y_2^2\right)
    \nonumber\\
    &+\sum_{j=1}^4 \de x_j^2\,,
\end{align}
and by the Kalb-Ramond field $B_{\mu\nu}(X)$, given by the two-form%
\footnote{This background has a constant dilaton which we set to zero.}
\begin{align}
    B=&\frac{16q}{(4-z_1^2-z_2^2)^2}\left(z_1\de z_2-z_2\de z_1\right)\wedge \de t
    \nonumber\\
    &+\frac{16q}{(4+y_1^2+y_2^2)^2}\left(y_1\de y_2-y_2\de y_1\right)\wedge \de \varphi\,.
\end{align}
This highlights that~$q$, and indeed the second line of the CZ action, is related to the Kalb-Ramond field.%
\footnote{%
This particular choice for the parametrisation of the metric  be obtained by an appropriate parametrisation of the group elements in the CZ action --- though of course at the Bosonic level it is easy to just work with the Polyakov action by picking a given metric.
}
As for NSNS and RR fluxes, which will appear in $S_{(2)}$ and $S_{(4)}$ are
\begin{equation}
H=\de B\,,\qquad F_3=-\sqrt{q^{-2}-1}\,H\,,    
\end{equation}
respectively. Both are proportional to the sum of the volume forms on $AdS_3$ and $S^3$. The proportionality involves the parameter~$q$, $0\leq q\leq 1$, which interpolates between the RR case ($q=0$) and the NSNS case ($q=1$). Notice that the radii of $AdS_3$ and $S^3$ are the same (as dictated by the supergravity equations) and we have normalised them by the string length (or tension) so that the bosonic action is
\begin{equation}
    S_{\text{bos}}=-\frac{T}{2}
    \int\limits_{-\infty}^{+\infty}\de\tau
    \int\limits_{0}^{r}\de\sigma
    \left[\left(\gamma^{ab}G_{\mu\nu}(X)+\varepsilon^{ab}B_{\mu\nu}(X)\right)\partial_aX^\mu\partial_bX^\nu\right],
\end{equation}
where $\gamma^{ab}$ is the unit-determinant inverse metric on the worldsheet and $\varepsilon^{ab}$ is the Levi-Civita tensor yielding the Wess-Zumino term. Because the sphere is compact, the overall coefficient of the Wess-Zumino must be quantised. Specifically, it must be
\begin{equation}
    T q =\frac{k}{2\pi}\,,\qquad k\in\mathbb{Z}\,.
\end{equation}
In what follows, without loss of generality, we will assume $k$ to be a non-negative integer. It is sometimes useful to treat $k$ as the ``amount of NSNS flux'' and introduce a new coupling $h\geq0$ representing the ``amount of RR flux''. It is important to note that the latter is not quantised.%
\footnote{%
A better way to understand $h$ when $k\neq0$ is to start with a background with NSNS flux and no RR flux, which can be obtained by a system of $N_1\gg1$ fundamental strings and $k$ NS5 branes, and turn on a constant RR zero-form~$c_0$; then, $F_3=-c_0\,H$ without modifying the D1-D5 charge of the background. In the special case where $k=0$, then $h$ is a combination of the RR flux times the dilaton. See e.g.~\cite{OhlssonSax:2018hgc} for a discussion of the brane realisation and moduli of these backgrounds.
}
In these terms, the dimensionless string tension~$T$ is sourced by both $h$ and $k$ as
\begin{equation}
\label{eq:tension}
    T=\sqrt{h^2+\frac{k^2}{4\pi^2}}\,.
\end{equation}
Note finally that in the special case of $h=0$ the action is that of a Wess-Zumino-Witten model.

\subsection{Gauge fixing and lightcone Hamiltonian}
We now want to study this model in the uniform lightcone gauge~\cite{Arutyunov:2005hd}, using the isometries $t$ and $\varphi$ and constructing the combinations
\begin{equation}
    X^-=\varphi-t,\qquad
    X^+=a\,\varphi+(1-a)t\,,\quad 0\leq a\leq1\,.
\end{equation}
The parameter $a$, which may look a little unusual, is introduced for later convenience. To introduce our desired gauge fixing we might use the first order formalism or, equivalently, go to a T-dual frame for $X^-$; here we will do the former, see e.g.~\cite{Sfondrini:2019smd} for a discussion of the latter approach. The conjugate momenta are
\begin{equation}
    P_\mu = \frac{\delta S_{\text{GS}}}{\delta (\partial_\tau X^\mu)}=\frac{\delta S_{\text{bos}}}{\delta (\partial_\tau X^\mu)}+\text{Fermions}\,,
\end{equation}
and the gauge-fixing condition is%
\footnote{%
In the full model, this is to be supplemented by the lightcone $\kappa$-gauge fixing condition $\Gamma^+\Theta^I=0$, where $\Theta^I$ with $I=1,2$ are the type IIB spinors.}
\begin{equation}
\label{eq:lcgauge}
    X^+=\tau\,,\qquad P_-=2\,.
\end{equation}
Using this condition is possible to eliminate the worldsheet metric~$\gamma^{ab}$ as well as the longitudinal modes of the string, leaving only the 8 transverse fields
\begin{equation}
    X^{j},\quad j=1,\dots 8,\qquad
    X^j=(z_1,z_2,y_1,y_2,x_1,x_2,x_3,x_4)\,.
\end{equation}
The action reads
\begin{equation}
    S_{\text{bos}}=
    \int\limits_{-\infty}^{+\infty}\de\tau
    \int\limits_{0}^{r}\de\sigma
    \left(P_j\,\partial_\tau{X}^j-\mathcal{H}\right),
\end{equation}
where $\mathcal{H}=-P_{+}$ is the worldsheet Hamiltonian density. Due to the lightcone gauge condition~\eqref{eq:lcgauge}, the worlsdheet Hamiltonian is related to a combination of Noether charges in the target-space:
\begin{equation}
    \mathbf{H}=\int\limits_{0}^{r}\de\sigma\,\mathcal{H}=
    \int\limits_{0}^{r}\de\,\sigma(-P_+)=
    \int\limits_{0}^{r}\de\sigma\left(-P_{t}-P_{\varphi}\right)=\mathbf{E}-\mathbf{J}\,,
\end{equation}
where $\mathbf{E}$ is the target-space energy and $\mathbf{J}$ is the angular momentum on $S^3$.
This is desirable because the combination $\mathbf{E}-\mathbf{J}$ is precisely what appears in the BPS bound of the super-isometry algebra of $AdS_3\times S^3\times T^4$.%
\footnote{Other choices for the gauge fixing are in principle possible, and they have been recently explored in~\cite{Borsato:2023oru}.}
Moreover, note that
\begin{equation}
    r= \int\limits_{0}^{r}\de\sigma\,2=
    \int\limits_{0}^{r}\de\sigma\, P_-=\mathbf{J}+a\,\mathbf{H}\,,
\end{equation}
which is unsurprising as the reparametrisation invariance is lost. This also suggests that $a=0$ is a particularly simple choice, whereby the size of the worldsheet is the (quantised) angular momentum of a reference geodesic on the sphere (in fact, of a reference half-BPS state).
Readers familiar with $T\bar{T}$ deformations~\cite{Smirnov:2016lqw,Cavaglia:2016oda} will have noticed that the gauge parameter $a$ is somewhat reminiscent of the parameter of such deformations; this relation was discussed in~\cite{Baggio:2018gct,Frolov:2019nrr,Frolov:2019xzi}.  

It is easy to compute the explicit form of $\mathcal{H}$ as a function of the fields and of their conjugate momenta (see~\cite{Lloyd:2014bsa} for the result including the Fermion contributions). The expression is rather unwieldy (of the Nambu-Goto form) and we do not report here. It is interesting however to look at the expression order by order in the fields. Such an expansion can be obtained as a large-tension expansion ($T\gg1$, by rescaling the fields by $T^{-1/2}$) and interpreted as a near-pp-wave expansion~\cite{Berenstein:2002jq}.%
\footnote{
Namely, the pp-wave geometry is the one obtained from the metric and B-field described above by expanding around the lightcone geodesic $t(\tau,\sigma)\sim\tau$ and $\varphi(\tau,\sigma)\sim\tau$.
}
At leading order we get
\begin{align}
    \mathcal{H} =&\, P_zP_{\bar{z}}+P_yP_{\bar{y}}+P_{u}P_{\bar{u}}+P_{v}P_{\bar{v}}+|\acute{z}|^2+|\acute{y}|^2+|\acute{u}|^2+|\acute{v}|^2
    \nonumber\\
    &
    -iq\,(\bar{z}\acute{z}-z\acute{\bar{z}}+\bar{y}\acute{y}-y\acute{\bar{y}})+|z|^2+|y|^2+\mathcal{O}(T^{-1})\,,
\label{eq:nearBMN}
\end{align}
where we introduced complex fields ($u,v$ are two complex fields related to $x_j$) and denote by primes the~$\partial_\sigma$ derivatives.
In principle this theory needs to be quantised in finite volume~$r=J/2$ (where $J$ is the eigenvalue of~$\mathbf{J}$) but this is made very difficult by the presence of the $\mathcal{O}(T^{-1})$ interaction terms.%
\footnote{%
Note that if the interaction terms are dropped altogether, this is indeed the lightcone Hamiltonian of the pp-wave string background, which can indeed be quantised without issue as it is a free theory.
}

A limit whereby we retain the interaction terms but significantly simplify the model is the \textit{decompactification limit}, whereby $r\to\infty$. In that case, the $\mathcal{O}(T^{-1})$ correspond to quartic and higher vertices of a tree-level S-matrix. In principle, one can use them to compute loop corrections to the S~matrix on the $(1+1)$-dimensional worldsheet of the string.%
\footnote{%
This computation was performed in~\cite{Sundin:2016gqe}, see also references therein, but it should be taken with a grain of salt as it is fraught with divergences both in the UV and in the IR.
}
Leaving aside the discussion of interactions, it is interesting to look at the momentum-space expression for the free Hamiltonian which provides some information on the particle spectrum of the model
\begin{equation}
\label{eq:nearbmnH}
    \mathbf{H}=\int\limits_{-\infty}^{+\infty}\de p\, \sum_{j=1}^8\omega(p,\mu_j)\,a^{\dagger}_j(p)\,a_j(p)\,,\qquad
    \omega(p,\mu)=\sqrt{p^2+2q\mu\,p+\mu^2}\,.
\end{equation}
The values of the quantum number $\mu$ (which is a combination of $AdS_3$ and $S^3$ spin) are
\begin{equation}
    \mu_j=(+1,-1,\,+1,-1,\,0,0,0,0)\,.
\end{equation}
where the first pair of numbers refers to $z,\bar{z}$, the second pair to $y,\bar{y}$, and the last four to the $T^4$ modes.
It is worth mentioning that, had we included Fermions, we would have found a similar quadratic Hamiltonian, with the same dispersion relation and mass spectrum~$\mu_j$. This is due to supersymmetry, in particular due to the fact that our choice of the gauge fixing preserves half of the supersymmetry. 

Rather than discussing in more details this perturbative expansion (or other similar expansions, such as the semiclassical one), we will see how symmetries and integrability allow to fix the S~matrix almost uniquely.

\section{Symmetries, S matrix, and spectrum}
\label{sec:symmetries-smatrix}

The fact that the classical theory underlying $AdS_3\times S^3\times T^4$ strings is integrable gives us hope to quantise the model exactly. The idea is to treat the lightcone gauge-fixed model as one would treat an integrable model like the Sinh-Gordon model. In that case, a powerful approach~\cite{Zamolodchikov:1978xm} is to assume that integrability carries over in the quantum theory and derive consequences of this assumptions. The most dramatic consequences appear for the scattering matrix: the scattering is elastic, without any macroscopic particle production, and any $N$-to-$N$ scattering process can be broken down into a sequence of $N(N-1)/2$ 2-to-2 scattering processes.%
\footnote{
Note that when we talk about scattering we refer to scattering on the two dimensional worldsheet, where the coordinate $\sigma$ is decompactified, $r\to\infty$.
}
If that is the case, the 2-to-2 scattering matrix is all is needed to completely construct the scattering, and therefore solve the model in the infinite volume limit ($r=\infty$). Along with the usual unitarity requirements, the 2-to-2 S~matrix $\mathbf{S}(p_1,p_2)$ has to satisfy the celebrated Yang-Baxter equation,
\begin{align}
\nonumber
&\mathbf{S}(p_2,p_3)\otimes\mathbf{1} \cdot
\mathbf{1}\otimes\mathbf{S}(p_1,p_3) 
\cdot
\mathbf{S}(p_1,p_2)\otimes\mathbf{1}\\
&\qquad\qquad\qquad\qquad\qquad
=
\mathbf{1}\otimes\mathbf{S}(p_1,p_2) 
\cdot
\mathbf{S}(p_1,p_3)\otimes\mathbf{1}
\cdot
\mathbf{1}\otimes\mathbf{S}(p_2,p_3)\,.
\label{eq:qYBE}
\end{align}
The various requirement on the S~matrix can be formalised in terms of the Faddeev-Zamolodchikov algebra, see for instance~\cite{Arutyunov:2009ga} for a review.
All in all, we have rephrased the problem of quantising the model to the one of finding an S~matrix for any value of the parameters $h,k$. This can (essentially) be done by symmetry.

\subsection{Symmetries}

The original superisometry algebra of the $AdS_3\times S^3\times T^4$ is $\alg{psu}(1,1|2)_{\L}\oplus\alg{psu}(1,1|2)_{\R}$ (where L, R, stand for ``left'' and ``right'') is generated by the sixteen supercharges
\begin{equation}
    Q^{\alpha\,m\,a}_{\L}\,,\qquad
    Q^{\dot{\alpha}\,\dot{m}\,a}_{\R}\,,
\end{equation}
where $\alpha$ is a fundamental index of $\alg{su}(2)_{\L}$, $m$ is a spin-$1/2$ index of $\alg{su}(1,1)_{\L}\cong\alg{sl}(2,\mathbb{R})_{\L}$ and the dotted indices correspond to ``right'' generators. Notice that the index~$a$ is common to left and right supercharges, and it corresponds to a $\alg{su}(2)$ outer automorphism. Geometrically, these rotations can be identified with a subalgebra of the $\alg{so}(4)$ rotations which (up to boundary conditions) ``rotate'' the four flat directions.%
\footnote{In fact, for all intents and purposes we are considering a the model in a limit where $T^4$ is replaced by $\mathbb{R}^4$. This is sufficient to study the sector of the model without any momentum or winding modes on the torus.}
Conventionally, one writes $\alg{so}(4)\cong\alg{su}(2)_{\bullet}\oplus \alg{su}(2)_{\circ}$, where $\alg{su}(2)_{\bullet}$ is the aforementioned automorphism and $\alg{su}(2)_{\circ}$ commutes with all generators in $\alg{psu}(1,1|2)_{\L}\oplus\alg{psu}(1,1|2)_{\R}$. There are also four $\alg{u}(1)$ isometries, but they will not be important for us here.

In the lightcone gauge-fixed theory we will have fewer symmetries. 
In terms of the Cartans of $\alg{psu}(1,1|2)_{\L}\oplus\alg{psu}(1,1|2)_{\R}$ we have that the light-cone Hamiltonian%
\footnote{
To suitably confuse the reader, we indicate the generators of the symmetries in the lightcone gauge with bold letters, and the other generators with regular letters.
}
\begin{equation}
    \mathbf{H} = L^0_{\L}+L^0_{\R}-J^{12}_{\L}-J^{12}_{\R}\,.
\end{equation}
Remark that the BPS bound of $\alg{psu}(1,1|2)$ is precisely
\begin{equation}
    L^0-J^{12}\geq0\,,
\end{equation}
and that it is saturated precisely on the highest-weight states of 1/2-BPS representations. This guarantees that the ground state of the lightcone Hamiltonian is 1/2-BPS (tough more precisely there is a whole Clifford module of such ground states, as we will see).
Half of the supercharges survive, namely
\begin{equation}
    \mathbf{Q}_{\L}^a = Q_{\L}^{+-a},\quad
    \mathbf{S}_{\L}^a = Q_{\L}^{-+a},\qquad
    \mathbf{Q}_{\R}^a = Q_{\R}^{\dot{+}\dot{-}a},\quad
    \mathbf{S}_{\R}^a = Q_{\R}^{\dot{-}\dot{+}a},
\end{equation}
along with the Cartan generators which we bundle as it follows
\begin{align}
    \mathbf{M}&=L^0_{\L}-L^0_{\R}-J^{12}_{\L}+J^{12}_{\R}\,,\nonumber\\
    \mathbf{B}&=L^0_{\L}-L^0_{\R}+J^{12}_{\L}-J^{12}_{\R}\,,\nonumber\\
    \mathbf{R}&=L^0_{\L}+L^0_{\R}+J^{12}_{\L}+J^{12}_{\R}\,.
\end{align}
The first two are combinations of the $AdS_3$ and $S^3$ spin; the last one is related to $\mathbf{H}$ and to the total R-charge $\mathbf{J}$ as $\mathbf{R}=\mathbf{H}+2\mathbf{J}$.

The 1/2-BPS vacuum of the lightcone gauge, which we denote by~$|0\rangle_r$, satisfies
\begin{equation}
    \mathbf{H}|0\rangle_r=\mathbf{M}|0\rangle_r=\mathbf{B}|0\rangle_r=0\,,\qquad
    \mathbf{R}|0\rangle_r=2r\,|0\rangle_r\,,
\end{equation}
i.e., it has R-charge~$r$.

The surviving subalgebra of $\alg{psu}(1,1|2)_{\L}\oplus\alg{psu}(1,1|2)_{\R}$ is
\begin{equation}
    \{\mathbf{Q}^a_{\L},\,\mathbf{S}^b_{\L}\}=\frac{1}{2}
   \varepsilon^{ab}(\mathbf{H}+\mathbf{M}),\qquad 
    \{\mathbf{Q}^a_{\R},\,\mathbf{S}^b_{\R}\}=\frac{1}{2}
   \varepsilon^{ba}(\mathbf{H}-\mathbf{M}).
\end{equation}
Remarkably, there is another set of nonvanishing commutation relations which \textit{are not part of the original symmetry algebra}, namely
\begin{equation}
\label{eq:centralext}
        \{\mathbf{Q}^a_{\L},\,\mathbf{Q}^b_{\R}\}=
   \varepsilon^{ab}\,\mathbf{C},\qquad 
    \{\mathbf{S}^a_{\L},\,\mathbf{S}^b_{\R}\}=
   \varepsilon^{ba}\,\mathbf{C}^\dagger.
\end{equation}
The central extensions~$\mathbf{C}$, $\mathbf{C}^\dagger$ are similar to Beisert's~\cite{Beisert:2005tm}  and their existence and form can be determined through a semiclassical analysis~\cite{Lloyd:2014bsa}, similar to~\cite{Arutyunov:2006ak}. It is worth noting that 
the central extension do not commute with $\mathbf{R}$ (which measures the overall R-charge of the vacuum) and therefore generate \textit{length-changing effects} of the type
\begin{equation}
\label{eq:lenghtchanging}
    \mathbf{C}\,|0\rangle_{r}=|0\rangle_{r+1}\,,\qquad
    \mathbf{C}^\dagger|0\rangle_{r}=|0\rangle_{r-1}\,,
\end{equation}
much like in Beisert's case. In other words, we can only study the centrally extended algebra in the decompactification limit, $r\to\infty$.
In the case of $AdS_3$ this central extension was found in~\cite{Borsato:2014exa} (see also~\cite{Hoare:2013pma,Hoare:2013lja} for mixed-flux backgrounds) following~\cite{Arutyunov:2006ak}.
We will denote the centrally extended algebra that survives (and thrives!) in the lightcone gauge-fixed theory as $\mathcal{A}$. It is isomorphic to
\begin{equation}
\label{eq:Aalgebra}
    \mathcal{A}\cong \big[\mathfrak{u}(1)_{\mathbf{B}}\oplus\mathfrak{su}(2)_\bullet\big] \ltimes \Big\{\big[\mathfrak{su}(1|1)^{\oplus2}/\mathfrak{u}(1)\big]^{\oplus2}\rtimes\big[\mathfrak{u}(1)_{\mathbf{C}}\oplus\mathfrak{u}(1)_{\mathbf{C}^\dagger}\big]\Big\} \,,
\end{equation}
where the  symbol $\ltimes$ indicates the central extension by $\mathbf{C},\mathbf{C}^\dagger$, while $\mathfrak{u}(1)_{\mathbf{B}}$ generated by~$\mathbf{B}$ and $\mathfrak{su}(2)_{\bullet}$ act as  outer automorphisms; we omitted the commuting algebra generated by $\mathfrak{su}(2)_{\circ}$. Notice that, importantly, $\mathbf{R}$ decouples in the $r\to\infty$ limit.

Again through a large-tension (near-BMN and semiclassical) analysis  it is possible to determine that lightcone-gauge excitations (i.e., particles) transform in short representations of the above algebra~$\mathcal{A}$. Such representations are four-dimensional and obey the constraint~\cite{Borsato:2012ud,Borsato:2013qpa}
\begin{equation}
\label{eq:shortening}
    \mathbf{H}^2 = \mathbf{M}^2+4\mathbf{C}^\dagger\mathbf{C}\,.
\end{equation}
Moreover it is possible to determine that, on a single-particle state of worldsheet momentum~$p$, we have
\begin{equation}
    \mathbf{M}\,|p,\mu\rangle=\left(\frac{k}{2\pi}p+\mu\right)|p,\mu\rangle\,,\qquad
    \mathbf{C}\,|p,\mu\rangle=\frac{ih}{2}\left(e^{ip}-1\right)|p,\mu\rangle\,,
\end{equation}
and similarly for $\mathbf{C}^\dagger$. Here $\mu$ labels the various representations. Namely, there is a representation with $\mu=+1$ containing the excitations related to $Y,Z$; one with $\mu=-1$ containing the excitations related to $\bar{Y},\bar{Z}$; two with $\mu=0$ containing the torus fields.
These formulae imply that the dispersion relation is~\cite{Hoare:2013lja}
\begin{equation}
\label{eq:dispersion}
    H(p,\mu)=\sqrt{\left(\frac{k}{2\pi}p+\mu\right)^2+4h^2\left(\sin\frac{p}{2}\right)^2}\,,
\end{equation}
which is related to the BMN result by a large-tension, small-momentum limit (recall that in the large tension limit $h,k\to\infty$ with $q$ fixed): 
\begin{equation}
    p=\frac{p'}{T},\qquad H(p,m)= \omega(p',m)+\mathcal{O}(1/T)\,.
\end{equation}

Notice that the central extension must vanish on physical states, which obey the level-matching condition $p=0\ \text{mod}2\pi$. This is clearly the case for single-particle states, and on multi-particle states we should get~\cite{Arutyunov:2006ak,Borsato:2013qpa}
\begin{equation}
    \mathbf{C}\,|p_1,\dots p_n,\mu_1,\dots \mu_n\rangle=\frac{ih}{2}\left(e^{i(p_1+\cdots+p_n)}-1\right)|p_1,\dots p_n,\mu_1,\dots \mu_n\rangle\,,
\end{equation}
which requires a non-trivial coproduct~$\Delta$, which n the supercharges takes the form~\cite{Borsato:2013qpa}
\begin{equation}
    \Delta[\mathbf{Q}] = \mathbf{Q}\otimes\mathbf{1}+e^{\tfrac{i}{2}\mathbf{p}}\otimes\mathbf{Q}\,,\qquad
    \Delta[\mathbf{S}] = \mathbf{S}\otimes\mathbf{1}+e^{-\tfrac{i}{2}\mathbf{p}}\otimes\mathbf{S}\,,
\end{equation}
where we kept the Fermion signs implicit and omitted the L,R, $a$ indices as the coproduct is blind to them. This is similar to the ``string-frame'' coproduct of $AdS_5\times S^5$~\cite{Arutyunov:2006ak}.

\begin{table}
\caption{Charges of the excitations corresponding to the eight transverse modes of $AdS_3\times S^3\times T^4$ strings and their superpartners. Each block is a representation, labeled by $\mu$. Bound state representations would have higher values of $|\mu|$. In this table we present the case $k\neq0$ because then one needs to consider separately whether $p> - 2\pi \mu/k$ or $p< - 2\pi \mu/k$, for which we introduce the chirality $c=\pm1$ for the condition $cp> - 2\pi\,c\, \mu/k$.
Note finally that the representations are only defined in the limit of infinte worldsheet size ($r$, $r'$, or $r''$), i.e.~up to length-changing effects. Hence, the length might shift across the representation depending on the supercharges used, so that it may be $r'=r\pm1/2$, and so on.}
\label{tab:mixedcharges}       % Give a unique label
%
% Follow this input for your own table layout
%
\renewcommand{\arraystretch}{1.4}
\begin{center}
\begin{tabular}{|l|c|c|c|c|c|c|c|c|}
\hline
Excitation &$c$& $\mu$ & $L^0_{\L}$ &$L^0_{\R}$ &$J^{12}_{\L}$ &$J^{12}_{\R}$ & $\mathfrak{su}(2)_\bullet$ &$\mathfrak{su}(2)_\circ$\\
\hline
$|Y(p)\rangle_r$ & $+1$& $+1$ & $\tfrac{r}{2}+\tfrac{k}{2\pi}p+\tfrac{1}{2}\delta H$ &$\tfrac{r}{2}+\tfrac{1}{2}\delta H$ & $\tfrac{r-2}{2}$ & $\tfrac{r}{2}$ & $\mathbf{1}$& $\mathbf{1}$\\
$|\eta^a(p)\rangle_{r'}$& $+1$& $+1$ &  $\tfrac{r'+1}{2}+\tfrac{k}{2\pi}p+\tfrac{1}{2}\delta H$ &$\tfrac{r'}{2}+\tfrac{1}{2}\delta H$& $\tfrac{r'-1}{2}$ & $\tfrac{r'}{2}$ & $\mathbf{2}$& $\mathbf{1}$\\
$|Z(p)\rangle_{r''}$ & $+1$& $+1$ &$\tfrac{r''+2}{2}+\tfrac{k}{2\pi}p+\tfrac{1}{2}\delta H$ &$\tfrac{r''}{2}+\tfrac{1}{2}\delta H$ & $\tfrac{r''}{2}$ & $\tfrac{r''}{2}$ & $\mathbf{1}$& $\mathbf{1}$\\
\hline
$|Y(p)\rangle_r$ & $-1$& $+1$ & $\tfrac{r}{2}+\tfrac{1}{2}\delta H$ &$\tfrac{r}{2}-\tfrac{k}{2\pi}p+\tfrac{1}{2}\delta H$ & $\tfrac{r}{2}$ & $\tfrac{r+2}{2}$ & $\mathbf{1}$& $\mathbf{1}$\\
$|\eta^a(p)\rangle_{r'}$& $-1$& $+1$ &  $\tfrac{r'}{2}+\tfrac{1}{2}\delta H$ &$\tfrac{r'-1}{2}-\tfrac{k}{2\pi}p+\tfrac{1}{2}\delta H$& $\tfrac{r'}{2}$ & $\tfrac{r'+1}{2}$ & $\mathbf{2}$& $\mathbf{1}$\\
$|Z(p)\rangle_{r''}$ & $-1$& $+1$ &$\tfrac{r''}{2}+\tfrac{1}{2}\delta H$ &$\tfrac{r''-2}{2}-\tfrac{k}{2\pi}p+\tfrac{1}{2}\delta H$ & $\tfrac{r''}{2}$ & $\tfrac{r''}{2}$ & $\mathbf{1}$& $\mathbf{1}$\\
\hline
$|\bar{Z}(p)\rangle_r$ & $+1$& $-1$ & $\tfrac{r-2}{2}+\tfrac{k}{2\pi}p+\tfrac{1}{2}\delta H$ &$\tfrac{r}{2}+\tfrac{1}{2}\delta H$ & $\tfrac{r}{2}$ & $\tfrac{r}{2}$ & $\mathbf{1}$& $\mathbf{1}$\\
$|\bar{\eta}^a(p)\rangle_{r'}$& $+1$& $-1$ &  $\tfrac{r'-1}{2}+\tfrac{k}{2\pi}p+\tfrac{1}{2}\delta H$ &$\tfrac{r'}{2}+\tfrac{1}{2}\delta H$& $\tfrac{r'+1}{2}$ & $\tfrac{r'}{2}$ & $\mathbf{2}$& $\mathbf{1}$\\
$|\bar{Y}(p)\rangle_{r''}$ & $+1$& $-1$ &$\tfrac{r''}{2}+\tfrac{k}{2\pi}p+\tfrac{1}{2}\delta H$ &$\tfrac{r''}{2}+\tfrac{1}{2}\delta H$ & $\tfrac{r''+2}{2}$ & $\tfrac{r''}{2}$ & $\mathbf{1}$& $\mathbf{1}$\\
\hline
$|\bar{Z}(p)\rangle_r$ & $-1$& $-1$ & $\tfrac{r}{2}+\tfrac{1}{2}\delta H$ &$\tfrac{r+2}{2}-\tfrac{k}{2\pi}p+\tfrac{1}{2}\delta H$ & $\tfrac{r}{2}$ & $\tfrac{r}{2}$ & $\mathbf{1}$& $\mathbf{1}$\\
$|\bar{\eta}^a(p)\rangle_{r'}$& $-1$& $-1$ &  $\tfrac{r'}{2}+\tfrac{1}{2}\delta H$ &$\tfrac{r'+1}{2}-\tfrac{k}{2\pi}p+\tfrac{1}{2}\delta H$& $\tfrac{r'}{2}$ & $\tfrac{r'-1}{2}$ & $\mathbf{2}$& $\mathbf{1}$\\
$|\bar{Y}(p)\rangle_{r''}$ & $-1$& $-1$ &$\tfrac{r''}{2}+\tfrac{1}{2}\delta H$ &$\tfrac{r''}{2}-\tfrac{k}{2\pi}p+\tfrac{1}{2}\delta H$ & $\tfrac{r''}{2}$ & $\tfrac{r''-2}{2}$ & $\mathbf{1}$& $\mathbf{1}$\\
\hline%
$|\chi^{\dot{a}}(p)\rangle_r$ & $+1$& $0$ & $\tfrac{r-1}{2}+\tfrac{k}{2\pi}p+\tfrac{1}{2}\delta H$ &$\tfrac{r}{2}+\tfrac{1}{2}\delta H$ & $\tfrac{r-1}{2}$ & $\tfrac{r}{2}$ & $\mathbf{1}$& $\mathbf{2}$\\
$|T^{a\dot{a}}(p)\rangle_{r'}$& $+1$& $0$ &  $\tfrac{r'}{2}+\tfrac{k}{2\pi}p+\tfrac{1}{2}\delta H$ &$\tfrac{r'}{2}+\tfrac{1}{2}\delta H$& $\tfrac{r'}{2}$ & $\tfrac{r'}{2}$ & $\mathbf{2}$& $\mathbf{2}$\\
$|\tilde{\chi}^{\dot{a}}(p)\rangle_{r''}$ & $+1$& $0$ &$\tfrac{r''+1}{2}+\tfrac{k}{2\pi}p+\tfrac{1}{2}\delta H$ &$\tfrac{r''}{2}+\tfrac{1}{2}\delta H$ & $\tfrac{r''+1}{2}$ & $\tfrac{r''}{2}$ & $\mathbf{1}$& $\mathbf{2}$\\
\hline%
$|\chi^{\dot{a}}(p)\rangle_r$ & $-1$& $0$ & $\tfrac{r}{2}+\tfrac{1}{2}\delta H$ &$\tfrac{r+1}{2}-\tfrac{k}{2\pi}p+\tfrac{1}{2}\delta H$ & $\tfrac{r}{2}$ & $\tfrac{r+1}{2}$ & $\mathbf{1}$& $\mathbf{2}$\\
$|T^{a\dot{a}}(p)\rangle_{r'}$& $-1$& $0$ &  $\tfrac{r'}{2}+\tfrac{1}{2}\delta H$ &$\tfrac{r'}{2}-\tfrac{k}{2\pi}p+\tfrac{1}{2}\delta H$& $\tfrac{r'}{2}$ & $\tfrac{r'}{2}$ & $\mathbf{2}$& $\mathbf{2}$\\
$|\tilde{\chi}^{\dot{a}}(p)\rangle_{r''}$ & $-1$& $0$ &$\tfrac{r''}{2}+\tfrac{1}{2}\delta H$ &$\tfrac{r''-1}{2}-\tfrac{k}{2\pi}p+\tfrac{1}{2}\delta H$ & $\tfrac{r''}{2}$ & $\tfrac{r''-1}{2}$ & $\mathbf{1}$& $\mathbf{2}$\\
\hline
\end{tabular}
\end{center}
\end{table}

\begin{table}
\caption{Charges of the excitations corresponding to the eight transverse modes of $AdS_3\times S^3\times T^4$ strings and their superpartners for the case of $k=0$. Each block is a representation, labeled by $\mu$. Bound state representations would have higher values of $|\mu|$. With respect to the case $k\neq0$, the particles with $|\mu|>0$ are not chiral at $h=0$. This could also be seen by observing that when formally taking $k\to0^+$ only the cases  $p> - 2\pi \mu/k$ for $\mu>0$ and $p< - 2\pi \mu/k$ for $\mu<0$ are possible.
As above, the representations are only in the limit of infinte worldsheet size.}
\label{tab:RRcharges}       % Give a unique label
%
% Follow this input for your own table layout
%
\renewcommand{\arraystretch}{1.4}
\begin{center}
\begin{tabular}{|l|c|c|c|c|c|c|c|c|}
\hline
Excitation &$c$& $\mu$ & $L^0_{\L}$ &$L^0_{\R}$ &$J^{12}_{\L}$ &$J^{12}_{\R}$ & $\mathfrak{su}(2)_\bullet$ &$\mathfrak{su}(2)_\circ$\\
\hline
$|Y(p)\rangle_r$ & -& $+1$ & $\tfrac{r}{2}+\tfrac{1}{2}\delta H$ &$\tfrac{r}{2}+\tfrac{1}{2}\delta H$ & $\tfrac{r-2}{2}$ & $\tfrac{r}{2}$ & $\mathbf{1}$& $\mathbf{1}$\\
$|\eta^a(p)\rangle_{r'}$& -& $+1$ &  $\tfrac{r'+1}{2}+\tfrac{1}{2}\delta H$ &$\tfrac{r'}{2}+\tfrac{1}{2}\delta H$& $\tfrac{r'-1}{2}$ & $\tfrac{r'}{2}$ & $\mathbf{2}$& $\mathbf{1}$\\
$|Z(p)\rangle_{r''}$ & -& $+1$ &$\tfrac{r''+2}{2}+\tfrac{1}{2}\delta H$ &$\tfrac{r''}{2}+\tfrac{1}{2}\delta H$ & $\tfrac{r''}{2}$ & $\tfrac{r''}{2}$ & $\mathbf{1}$& $\mathbf{1}$\\
\hline
$|\bar{Z}(p)\rangle_r$ & -& $-1$ & $\tfrac{r}{2}+\tfrac{1}{2}\delta H$ &$\tfrac{r+2}{2}+\tfrac{1}{2}\delta H$ & $\tfrac{r}{2}$ & $\tfrac{r}{2}$ & $\mathbf{1}$& $\mathbf{1}$\\
$|\bar{\eta}^a(p)\rangle_{r'}$& -& $-1$ &  $\tfrac{r'}{2}+\tfrac{1}{2}\delta H$ &$\tfrac{r'+1}{2}+\tfrac{1}{2}\delta H$& $\tfrac{r'}{2}$ & $\tfrac{r'-1}{2}$ & $\mathbf{2}$& $\mathbf{1}$\\
$|\bar{Y}(p)\rangle_{r''}$ & -& $-1$ &$\tfrac{r''}{2}+\tfrac{1}{2}\delta H$ &$\tfrac{r''}{2}+\tfrac{1}{2}\delta H$ & $\tfrac{r''}{2}$ & $\tfrac{r''-2}{2}$ & $\mathbf{1}$& $\mathbf{1}$\\
\hline%
$|\chi^{\dot{a}}(p)\rangle_r$ & $+1$& $0$ & $\tfrac{r-1}{2}+\tfrac{1}{2}\delta H$ &$\tfrac{r}{2}+\tfrac{1}{2}\delta H$ & $\tfrac{r-1}{2}$ & $\tfrac{r}{2}$ & $\mathbf{1}$& $\mathbf{2}$\\
$|T^{a\dot{a}}(p)\rangle_{r'}$& $+1$& $0$ &  $\tfrac{r'}{2}+\tfrac{1}{2}\delta H$ &$\tfrac{r'}{2}+\tfrac{1}{2}\delta H$& $\tfrac{r'}{2}$ & $\tfrac{r'}{2}$ & $\mathbf{2}$& $\mathbf{2}$\\
$|\tilde{\chi}^{\dot{a}}(p)\rangle_{r''}$ & $+1$& $0$ &$\tfrac{r''+1}{2}+\tfrac{1}{2}\delta H$ &$\tfrac{r''}{2}+\tfrac{1}{2}\delta H$ & $\tfrac{r''+1}{2}$ & $\tfrac{r''}{2}$ & $\mathbf{1}$& $\mathbf{2}$\\
\hline%
$|\chi^{\dot{a}}(p)\rangle_r$ & $-1$& $0$ & $\tfrac{r}{2}+\tfrac{1}{2}\delta H$ &$\tfrac{r+1}{2}+\tfrac{1}{2}\delta H$ & $\tfrac{r}{2}$ & $\tfrac{r+1}{2}$ & $\mathbf{1}$& $\mathbf{2}$\\
$|T^{a\dot{a}}(p)\rangle_{r'}$& $-1$& $0$ &  $\tfrac{r'}{2}+\tfrac{1}{2}\delta H$ &$\tfrac{r'}{2}+\tfrac{1}{2}\delta H$& $\tfrac{r'}{2}$ & $\tfrac{r'}{2}$ & $\mathbf{2}$& $\mathbf{2}$\\
$|\tilde{\chi}^{\dot{a}}(p)\rangle_{r''}$ & $-1$& $0$ &$\tfrac{r''}{2}+\tfrac{1}{2}\delta H$ &$\tfrac{r''-1}{2}+\tfrac{1}{2}\delta H$ & $\tfrac{r''}{2}$ & $\tfrac{r''-1}{2}$ & $\mathbf{1}$& $\mathbf{2}$\\
\hline
\end{tabular}
\end{center}
\end{table}

It is worth to briefly mention how the excitations related to transverse coordinates on  $AdS_3\times S^3\times T^4$ --- which perturbatively are given by~\eqref{eq:nearBMN} --- transform under~$\mathcal{A}$ along with their superpartners. To this end, let us note that the dispersion relation~\eqref{eq:dispersion} implies that the energy depends on the continuous parameter~$h$ non-trivially. Because $h$ can be interpreted as a coupling (similar to the 't~Hooft coupling in $\mathcal{N}=4$ SYM), we  introduce the notion of ``anomalous'' part of the dimension (in analogy with $\mathcal{N}=4$ SYM) as
\begin{equation}
    \delta H (p) = H(p)- \big(H(p)\big)\big|_{h=0}\,. 
\end{equation}
Note that at $h=0$ the short representations (i.e., the particles) become ``chiral'': they can be charged under $(\mathbf{Q}^a_{\L},\;\mathbf{S}^a_{\L})$ or under $(\mathbf{Q}^a_{\R},\;\mathbf{S}^a_{\R})$, but not under both sets; this is an immediate consequence of~\eqref{eq:shortening}.

We can now express the structure of the representations in terms of the eigenvalues under the various Cartan elements. Notice because $\mu$ appears in the central charges, different values of $\mu$ will label different irreducible representations. There are four short representations, each containing two bosons and two fermions.
\begin{itemize}
 \item A ``left'' representation has the chiral part of the complex bosons $z,y$ from the transverse modes of the string on $AdS_3\times S^3$. We denote the excitation created by the related creation operators by $|Y(p)\rangle$, $|Z(p)\rangle$; there are also two fermion modes $|\eta^{a}(p)\rangle$. They all have $\mu=+1$.%
\footnote{%
Perturbatively, $|{Y}(p)\rangle$ would go to the excitation $a^{\dagger}_{y}(p)|0\rangle$ created by the free oscillator appearing in~\eqref{eq:nearbmnH} related to $y$, with $\mu=+1$.}
\item
In another representation, which we call ``right''', we have $|\bar{Y}(p)\rangle$, $|\bar{Z}(p)\rangle$; there are also two fermions modes $|\bar\eta^{a}(p)\rangle$. This representation has~$\mu=-1$.
\item 
The remaining two representations contain the torus bosons $T^{a\dot{a}}$, and the fermions $\chi^{\dot{a}}$ and $\tilde{\chi}^{\dot{a}}$; the index $\dot{a}=1,2$ distinguishes the two representations and transforms under~$\mathfrak{su}(2)_\circ$. These two representations are called ``massless'' and have $\mu=0$.
\end{itemize}
If one makes a choice of the lowering operators of~$\mathcal{A}$ so that they are $\mathbf{Q}^a_{\L}$ and $\mathbf{S}_{\R}^a$, the highest-weight state of each four dimensional representation is given by $|Y(p)\rangle$, $|\bar{Z}(p)\rangle$, $|\chi^{\dot{a}}(p)\rangle$, $\dot{a}=1,2$. 

From the point of view of the gauge-fixed GS string, the presence of excitations with no mass gap is perhaps the most striking difference between the $AdS_3$ integrable string models%
\footnote{It is a common feature of $AdS_3\times S^3\times T^4$, $AdS_3\times S^3\times K3$ and $AdS_3\times S^3\times S^3\times S^1$.}
and the better-understood cases of $AdS_5\times S^5$ and $AdS_4\times CP^3$. Massless modes, and in particular massless GS fermions, are actually \textit{necessary} to reproduce the spectrum of BPS states of the dual CFT$_2$, see~\cite{Baggio:2017kza}, so that they cannot be easily argued away by e.g.~picking different gauges. At the same time, they substantially complicate the small-tension analysis of the model (for instance by Bethe ansatz~\cite{Abbott:2015pps,Brollo:2023pkl}, see also~\cite{Abbott:2020jaa}) and introduce infrared divergences in perturbative computations, see e.g.~\cite{Sundin:2016gqe} and references therein.

The charges of various states are contained in Table~\ref{tab:mixedcharges} for $k\neq0$ and in Table~\ref{tab:RRcharges} for $k=0$. It is also worth mentioning that in general (by semi-classical arguments as well as by an analysis of the physical poles of the S matrix, see below), we expect new representations to emerge as \textit{bound states} of the above representations. Specifically, left and left representations would make bound states, as would right with right. In total, for reasons which we will describe in the subsection below, we expect to have two massless representations and $k-1$ massive representation ($k-2$ of which are bound states)~\cite{Frolov:2023lwd,Frolov:2024pkz}.%
\footnote{%
In particular, the right representation with $\mu=-1$ can be related to that with~$\mu=k-1$, which is a bound state of $(k-1)$ left excitations.}
If $k=0$, we expect infinitely many bound states~\cite{Borsato:2013hoa}.

\paragraph{Brief comments on the $AdS_3\times S^3\times S^3\times S^1$ background.}
The family of backgrounds of the type $AdS_3\times S^3\times S^3\times S^1$ also preserve sixteen Killing spinors, which organise themselves in the superalgebra
\begin{equation}
    \mathfrak{d}(2,1;\alpha)_{\L}\oplus\mathfrak{d}(2,1;\alpha)_{\R}\,.
\end{equation}
The parameter $\alpha$, with $0<\alpha<1$, encodes the relative radii of the two spheres. More specifically, if $R_{AdS}$ is the $AdS_3$ radius and $R_1$, $R_2$ are the radii of the two spheres we have
\begin{equation}
    \alpha = \frac{(R_{AdS})^2}{(R_{1})^2}\,,\qquad
    1-\alpha =  \frac{(R_{AdS})^2}{(R_{2})^2}\,,
\end{equation}
so that $\alpha=\tfrac{1}{2}$ is the special case where the two spheres are identical, and $\mathfrak{d}(2,1;\tfrac{1}{2})\cong\mathfrak{osp}(4|2)$. The other notable limit is $\alpha\to0$ (or equivalently $\alpha\to1$) where one of the two spheres becomes flat, $S^3\to\mathbb{R}^3$, and the superalgebra contracts.
This family of  backgrounds, for any $0<\alpha<1$, is integrable~\cite{Babichenko:2009dk}. This remains true even if we introduce a Kalb-Ramond field in the NLSM action~\cite{Cagnazzo:2012se}.
One major difference with the case at hand is that the most supersymmetric BMN solution is 1/4-BPS for this background, rather than 1/2-BPS; this is a geodesic that run time~$t$ in $AdS_3$ and along great circles parametrised by~$\varphi_1$ and $\varphi_2$ along either sphere (with a specific velocity for each space).
As a result the algebra~$\mathcal{A}$ of eq.~\eqref{eq:Aalgebra} contains at most one copy of~$\mathfrak{su}(1|1)^{\oplus2}/\mathfrak{u}(1)$, centrally extended by $\mathbf{C},\mathbf{C}^\dagger$. 
As a result, short representations of~$\mathcal{A}$ are \textit{two}-dimensional (one boson and one fermion) rather than \textit{four}-dimensional, and there are eight, rather than four, distinct representations.
The near-BMN bosonic Hamiltonian still takes the form~\eqref{eq:nearbmnH} but now the allowed values of $\mu$ are
\begin{equation}
    \mu_j=\Big(+1,-1,+\alpha, -\alpha, +(1-\alpha), -(1-\alpha), 0,0\Big)\,.
\end{equation}
The modes with $|\mu|=1$ are transverse modes on $AdS_3$, those with $|\mu|=\alpha$ are transverse modes of the first sphere,   those with $|\mu|=1-\alpha$ are transverse modes of the second sphere, and one of the modes with $\mu=0$ is the $S^1$ boson. The last $\mu=0$ mode comes for a combination of $t$, $\varphi_1$ and $\varphi_2$ fields which is orthogonal to the BMN geodesic.%
\footnote{See also~\cite{Dei:2018yth} for detailed discussion of the near-BMN action around this and more general (non-supersymmetric) geodesics.}
Clearly, as $\alpha\to1$ or $\alpha\to0$ we recover the $AdS_3\times S^3\times T^4$ near-BMN spectrum.

\subsection{S matrix}

The two-to-two S~matrix can be constructed by demanding commutation with the supercharges in the two-particle representation. For each irreducible representation, the S~matrix can be represented by a $16\times16$ matrix. Each such block, which we may label by the values $(\mu_1,\mu_2)$ of the irreducible representation,%
\footnote{%
To be precise, the case of $\mu=0$ is a bit special because there are two representations with that value of~$\mu$, which may be distinguished by they charge under $\mathfrak{su}(2)_\circ$.
}
is fixed by symmetry up to an overall normalisation, the so-called dressing factor. The S~matrix so determined satisfies the Yang-Baxter equation, regardless of the dressing factors, without the need of imposing any further constraint.
The various dressing factors must obey unitarity, crossing and analyticity constraints. However, unlike what happens in relativistic models, it is far from obvious how to introduce sufficiently many constraints (e.g., absence of poles, branch points, \dots) to guarantee that the dressing factors are uniquely determined.

The S~matrix constructed in this way has poles for complex value of the momenta. A semiclassical analysis indicates that some of those poles are physical, and hence should be interpreted as bound states. The bound-state representations can be constructed explicitly, and one finds that, like the representations for fundamental particles, they are four-dimensional --- they just have different values of~$\mu$. This in contrast with e.g.~$AdS_5\times S^5$, where the dimension of the bound-state representations grows linearly in the bound-state number.
Here, one finds that $\mu=+1$ particles lead to bound states with $\mu=+2,+3,\dots$, while  $\mu=-1$ particles lead to bound states with $\mu=-2,-3,\dots$; there are no bound states involving $\mu=0$ particles, there are no bound states from scattering particles with opposite-sign~$\mu$ (i.e., $\mu_1$ and $\mu_2$ yield a bound state with $\mu=\mu_1+\mu_2$ iff $\mu_1 \mu_2>0$).

It is interesting to note that, for $k\neq0$, the energy as well as the other central elements satisfy
\begin{equation}
\label{eq:periodicity}
    H(p+2\pi,\mu) = H(p,\mu-k)\,.
\end{equation}
Hence, from the point of view of the representation structure, if one restricts the momentum to a fundamental region of size $2\pi$ one must consider all values of~$\mu$. Viceversa, if we allow for the momentum to take any real value, one may restricts to $\mu=0,1,\dots,k$.%
\footnote{The reason why we are counting $k+1$, rather than~$k$, particles is that we expect two ``massless'' representations, one of which can be taken to be $\mu=0$, with the other being $\mu=k$.}
Strictly speaking, this is a property of the representation (and hence of the matrix part of the S~matrix), and it is non-trivial to demonstrate that  it is possible to construct dressing factors compatible with the requirement~\eqref{eq:periodicity}.
Notice that instead for pure-RR backgrounds ($k=0$) the dispersion relation is $2\pi$-periodic, and has no (quasi)-periodicity in $\mu$. Indeed, the bootstrap procedure allows to construct representations with any~$\mu\in\mathbb{Z}$, while $p$ can be restricted to a fundamental region.

Finally, it is worth noting that this integrable bootstrap is based on the unproven assumption that the (semi)classical symmetries persist at the quantum level. To validate the construction it is necessary to compare it with the result of perturbative computations, obtained \textit{e.g.}\ semiclassically or perturbatively around the plane-wave background~\cite{Sundin:2013ypa,Hoare:2013lja,Sundin:2016gqe}.

The current state of the art is as it follows
\begin{itemize}
    \item The case of pure-RR was the first to be studied. The matrix part of the all-loop S~matrix was determined in~\cite{Borsato:2014exa} up to the dressing factors. Those were proposed in~\cite{Borsato:2013hoa,Borsato:2016xns}; however, only in 2021 it was realised that such a proposal was flawed and an alternative set of dressing factors was proposed~\cite{Frolov:2021fmj}.
    \item In 2018 the S~matrix for the $h=0$ case was proposed, and the resulting spectrum successfully compared~\cite{Dei:2018mfl} with the one that can be found from the (by then well-known) RNS description~\cite{Maldacena:2000hw}.
    \item In the general mixed-flux case, the all-loop S~matrix up to the dressing factors was proposed in 2014~\cite{Lloyd:2014bsa}, building on~\cite{Hoare:2013pma}. However, only very recently a proposal for the dressing factors was put forward~\cite{Frolov:2024pkz}, see also~\cite{Frolov:2023lwd,OhlssonSax:2023qrk} for related work.
\end{itemize}

\subsection{Spectrum}

The S~matrix describes the theory in the infinite-volume limit, $r\to\infty$. To understand the string spectrum we need however to compute the spectrum for a finite value of $r$, which corresponds to the (quantised) R~charge of a reference 1/2-BPS geodesic. To put the theory back in finite volume one may just impose the Asymptotic Bethe Ansatz (or Bethe-Yang) equations~\cite{Arutyunov:2004vx}, i.e.\ the quantisation conditions for an $M$-particle state, which are schematically%
\footnote{%
Strictly speaking there are several sets of coupled equations, arising from the nested Bethe Ansatz which here takes the form first discussed in~\cite{Borsato:2012ss}, see also~\cite{Seibold:2022mgg} for a more comprehensive discussion.
}
\begin{equation}
    e^{i\, p_j\,r}\prod_{l\neq j}^M S(p_j,\mu_j;p_l,\mu_l)=1\,,\qquad
    j=1,\dots,n\,,
\end{equation}
where $S(p_j,\mu_j;p_l,\mu_l)$ is the two-particle S~matrix, or in logarithmic form
\begin{equation}
\label{eq:logarithmicBYE}
    p_j\,r-i\sum_{l\neq j}^M \log S(p_j,\mu_j;p_l,\mu_l)=2\pi \nu_j\,,\qquad
    j=1,\dots,M\,,
\end{equation}
where different choices of the modes numbers $\nu_j\in\mathbb{Z}$ give in principle different states.%
\footnote{%
The allowed $\nu_j$'s depend on the periodicity of the model (whether or not the momentum is constrained to a fundamental region) and by the statistic of the excitation (no repeated mode numbers for identical Fermions). Moreover, these equations are supplemented by the level-matching constraint stating the sum of all $p_j$'s vanishes modulo~$2\pi$.
}
Due to the string-theory level-matching constraint, the sum of all momenta is also constrained~\cite{Arutyunov:2009ga},
\begin{equation}
\label{eq:levelmatching}
    \sum_{j=1}^M p_j=0\ \text{mod}\,2\pi\,.
\end{equation}

The lightcone energy of such a state is then
\begin{equation}
    H=\sum_{j=1}^M H(p_j,\mu_j)\,.
\end{equation}
Unfortunately, this argument is a bit too simplistic to capture the whole spectrum. In fact, even if in an integrable model no particle production should occur macroscopically, virtual particles exist and they can give rise to ``wrapping effects'' of the type studied by L\"uscher~\cite{Luscher:1985dn,Luscher:1986pf}, see also~\cite{Ambjorn:2005wa}.
In a non-relativistic model such as this one, the kinematics that dictates these effects is given by exchanging time and space and making a double Wick rotation,
\begin{equation}
    \tilde{p}=i\omega\,,\qquad \tilde{\omega}=ip\,,
\end{equation}
resulting in a ``mirror'' model. The dispersion relation of the mirror model, for the case at hand, cannot be written in terms of elementary functions unless $k=0$ or $h=0$. Let us consider the $k=0$ case:
\begin{equation}
    \tilde{\omega}(\tilde{p},\mu)=
    2\,\text{arsinh}\left(\frac{\sqrt{\mu^2+\tilde{p}^2}}{2h}\right)\,.
\end{equation}
Finite-size effects are then suppressed as
\begin{equation}
    e^{-\omega(\tilde{p},\mu)\,r}\lesssim
    \mathcal{O}(h^{2r})\,,\qquad\mu\neq0\,.
\end{equation}
This means that, for models where all particles have $\mu\neq0$ such as $AdS_5\times S^5$, wrapping effects can be overlooked at least in some regime (at small tension) and for states with R-charge $r$ sufficiently large. This is crucial to describe the small-tension $AdS_5\times S^5$ spectrum in terms of a nearest-neighbour spin chain~\cite{Ambjorn:2005wa}.
Our model, however, has gapless $\mu=0$ particles, whose soft modes $|\tilde{p}|\ll1$ contribute to wrapping at any value of the tension. This is one of the reasons why it is believed that, even at small tension, if the model has a quantum-mechanical description that should be in terms of a long-range model.
When turning on~$k>0$ things get even more complicated. To illustrate the point, let us consider $h=0$, $k>0$; we find~\cite{Dei:2018mfl}
\begin{equation}
    \tilde{\omega}(\tilde{p},\mu)=\frac{2\pi}{k}|\tilde{p}|+i\,\mu\,.
\end{equation}
This is a little troubling because \textit{the energy is not real}. However, the imaginary contribution can be interpreted as a chemical potential in the partition function of the model. A similar argument can be done for the general case where $h>0$ and $k>0$, where the mirror dispersion relation can be found from~\eqref{eq:dispersion} as an implicit function. In~\cite{Baglioni:2023zsf} it was argued that even in that case, despite $\tilde{\omega}$ being complex, the energy shifts due to wrapping corrections are real.

Eventually, the way to obtain \textit{exact} equations for the spectrum, i.e.\ equations where both $k$ and $h$ enter as \textit{finite} parameters, is to account for all the finite-volume effects from the get-go by writing Thermodynamic Bethe Ansatz equations~\cite{Yang:1968rm} for the mirror model~\cite{Arutyunov:2007tc}. Strictly speaking these equations would yield the ground-state energy of the model but they can be modified to give the energy level of excited states too, for instance by analytic continuation~\cite{Dorey:1996re}.
The ``mirror TBA'' equations are a set of coupled integral equations for the ``Y~functions'', where each Y~function is related to the density of one type of particles. These equations can be simplified and written in terms of ``T~functions'' and ``Q~functions'', which provides a more compact (and computationally more efficient) set of equations, the so called ``quantum spectral curve''~\cite{Gromov:2013pga}. This simplified set of equations is also called the ``quantum spectral curve'' of the model.

For the case of $AdS_3\times S^3\times T^4$ the status is as it follows:
\begin{itemize}
    \item For the case of pure-NSNS backgrounds ($h=0$), the mirror TBA has been worked out in~\cite{Dei:2018mfl} and it has been shown that the resulting spectrum matches with the one that may be computed from the RNS (i.e., WZW) description.
    \item For the case of pure-RR flux ($k=0$) there are two proposals for the spectrum of the model. The first set of equations to be derived were the quantum spectral curve equations; this was done independently (and simultaneously) in~\cite{Ekhammar:2021pys} and~\cite{Cavaglia:2021eqr}. Unlike what happened for other models, those equations were not obtained by simplifying the mirror TBA, but rather directly conjectured based on symmetry and analyticity considerations. Possibly for this reason it is currently unclear how those equations may describe states which involve gapless ($\mu=0$) excitations.
    Shortly afterwards, a set of mirror TBA equations was derived~\cite{Frolov:2021bwp} which also involves masselss excitations. In fact, using those equations it was argued that massless excitations have the largest energy at small tension, $h\ll1$~\cite{Brollo:2023pkl}.
    \item For mixed-flux backgrounds neither the mirror TBA nor the QSC has been proposed, though the recent progress in understanding the dressing factors~\cite{Frolov:2023lwd,OhlssonSax:2023qrk,Frolov:2024pkz} makes us hopeful that we may do so soon.
\end{itemize}

For the case of $AdS_3\times S^3\times S^3\times S^1$, only the matrix part of the S~matrix is known~\cite{Borsato:2012ud}, along with the asymptotic Bethe equations~\cite{Borsato:2012ss}. The dressing factors are still unknown in general. The only case in which the full S~matrix has been proposed and used to compute the TBA was the pure NSNS model, for which the spectrum was showed to match with the WZW one~\cite{Dei:2018jyj}.

\section{Small-tension spectra and dual CFT}
\label{sec:tensionless}

Here we would like to briefly comment on the small-tension spectrum and on the dual CFT description.
It is generally believed that, when the string tension is small, the degrees of freedom of the worldsheet model (and in particular, of its integrability description) may be interpreted in terms of the ones of a ``weakly-coupled'' dual CFT. This is indeed the case for $AdS_5\times S^5$ and its dual $\mathcal{N}=4$ super-Yang-Mills, where at small tension the 't~Hooft coupling is $\lambda\ll1$ and the magnons on the worldsheet can be related to those of an integrable spin chain representing single-trace operators in the gauge theory~\cite{Minahan:2002ve,Beisert:2005tm}.

In this case things are more complicated because the tension~\eqref{eq:tension} receives contributions from the discrete parameter~$k=0,1,2,\dots$, and we should consider each case separately. The two most interesting cases are those of $k=0$ and $k=1$.

\subsection{The case of \texorpdfstring{$k=0$}{k=0} (pure-RR background)}
\begin{figure}[b]
\sidecaption
% For example, with the graphicx style use
\includegraphics[width=.5\linewidth]{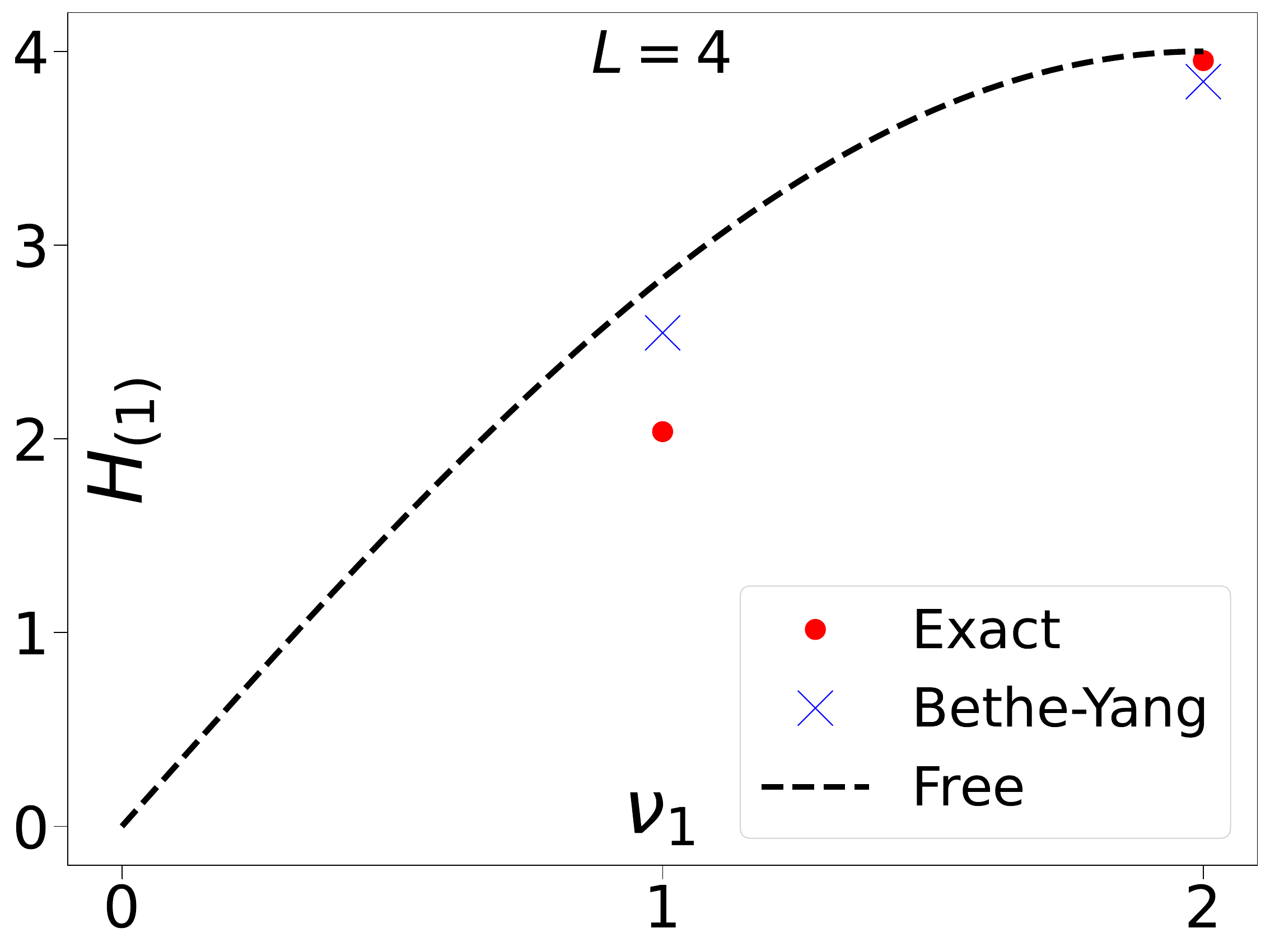}%
\includegraphics[width=.5\linewidth]{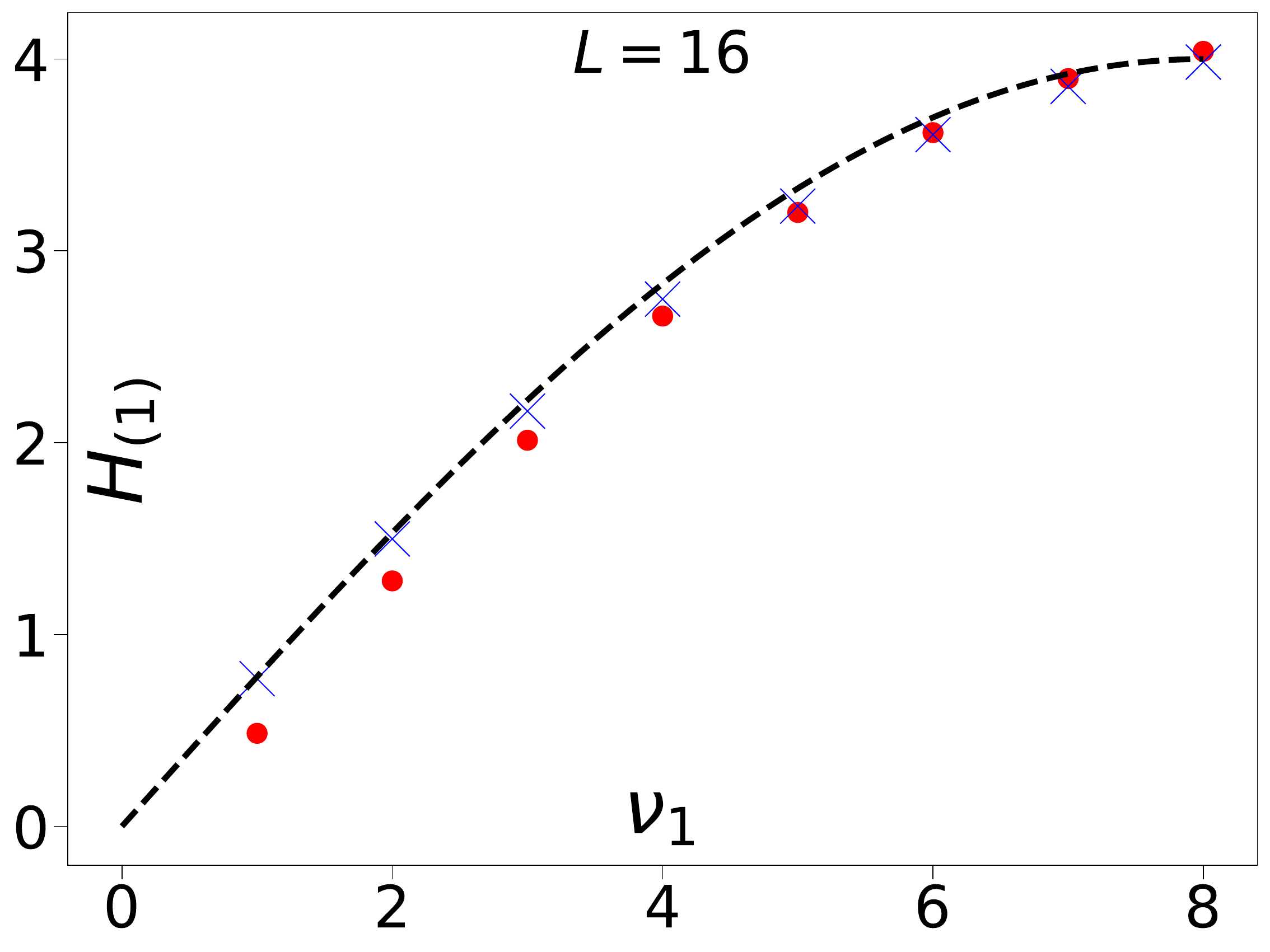}\\
\includegraphics[width=.5\linewidth]{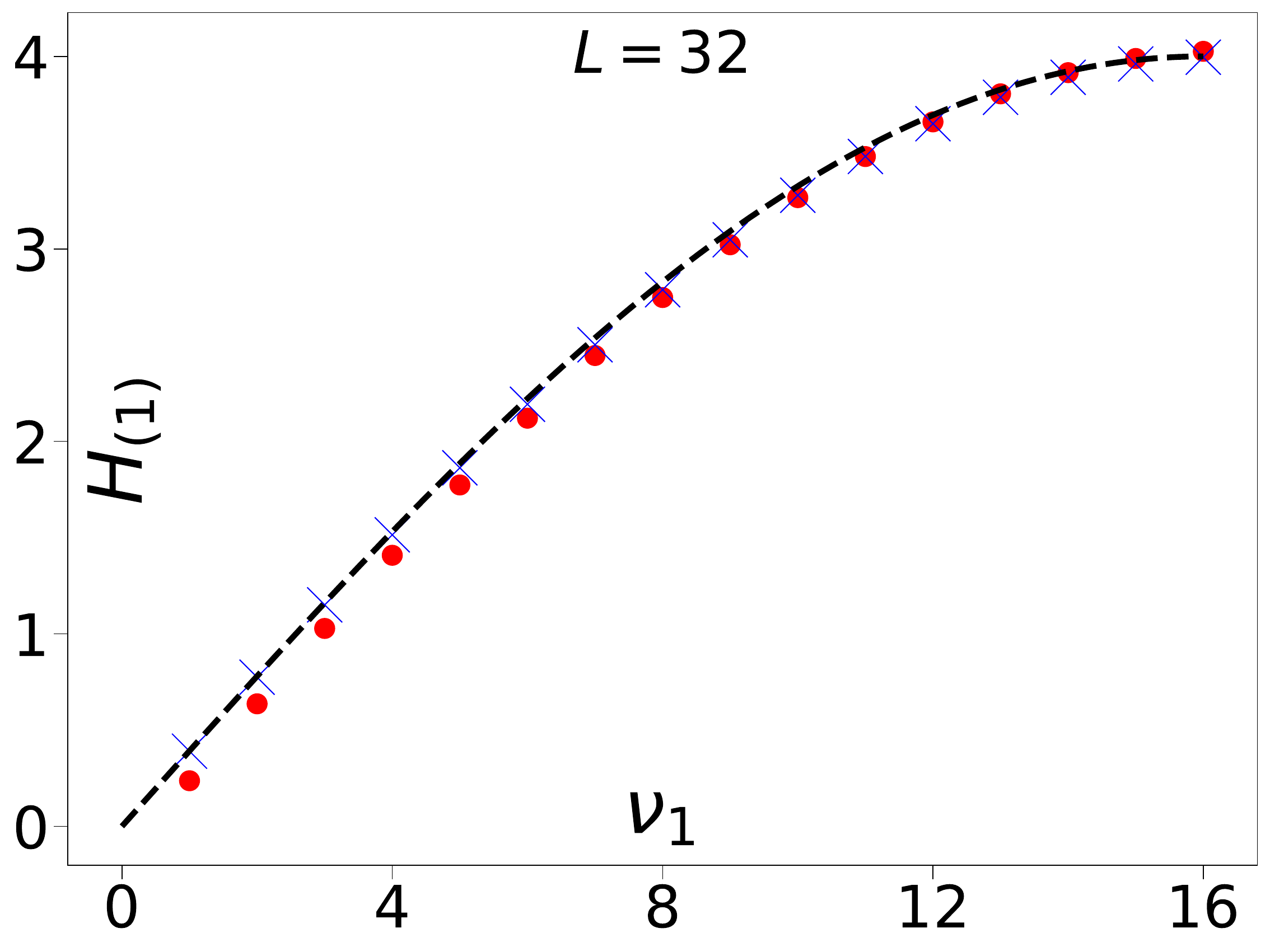}%
\includegraphics[width=.5\linewidth]{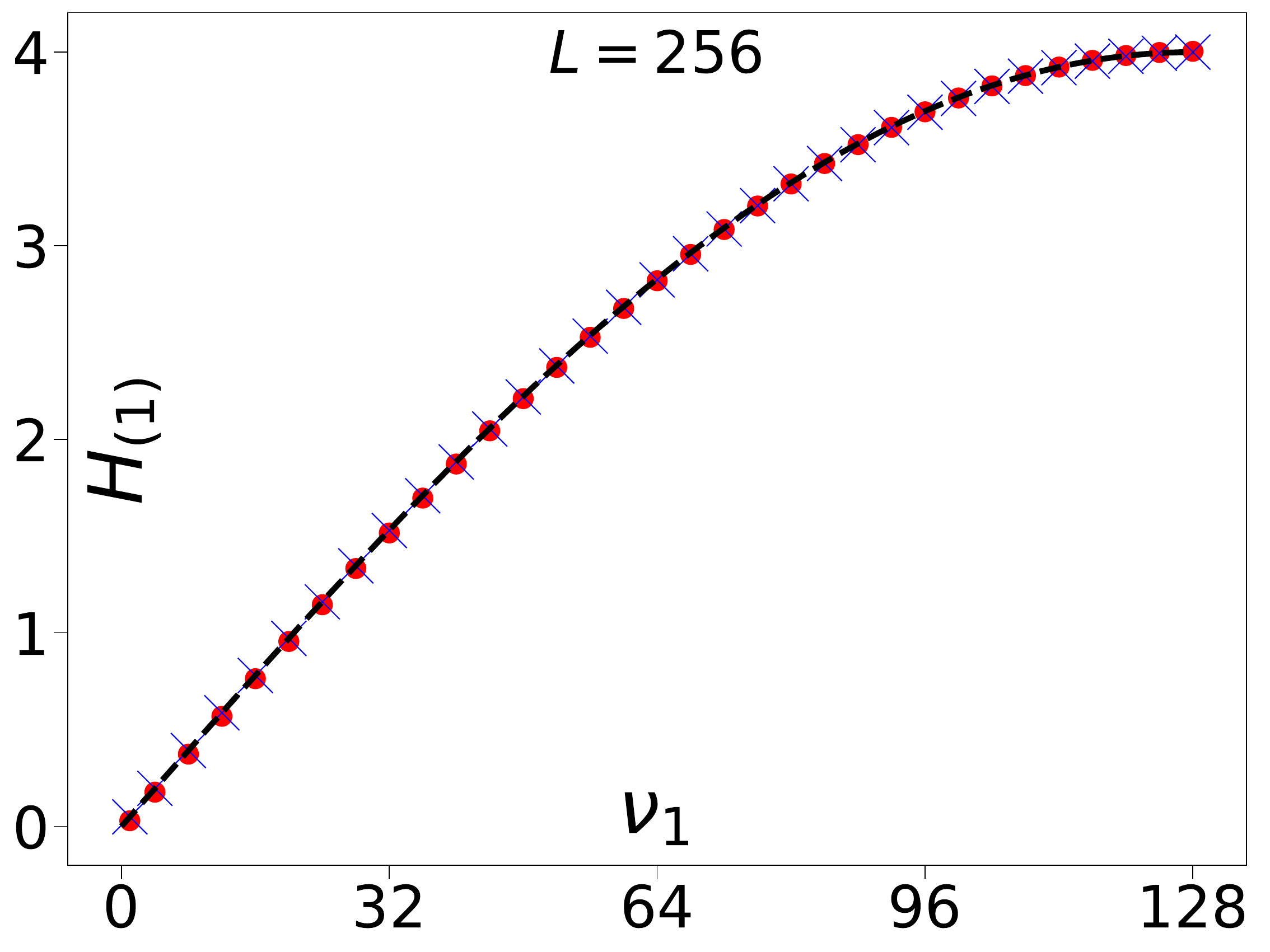}
%
% If no graphics program available, insert a blank space i.e. use
%\picplace{5cm}{2cm} % Give the correct figure height and width in cm
%
\caption{Figure taken from~\cite{Brollo:2023rgp}.
We plot the energy of two-excitations states with mode numbers $\nu_1=-\nu_2>0$ for various values of the volume in the lightcone gauge~$L$. The dashed line is the expectation for a free theory, i.e.\ $H_{(1)}=4\sin\tfrac{\pi\nu_1}{L}$. The cross is given by the solution of the asymptotic Bethe equations and the solid circle is the TBA value.}
\label{fig:twomagnons}       % Give a unique label
\end{figure}

\begin{figure}[b]
\sidecaption
% For example, with the graphicx style use
\includegraphics[width=.5\linewidth]{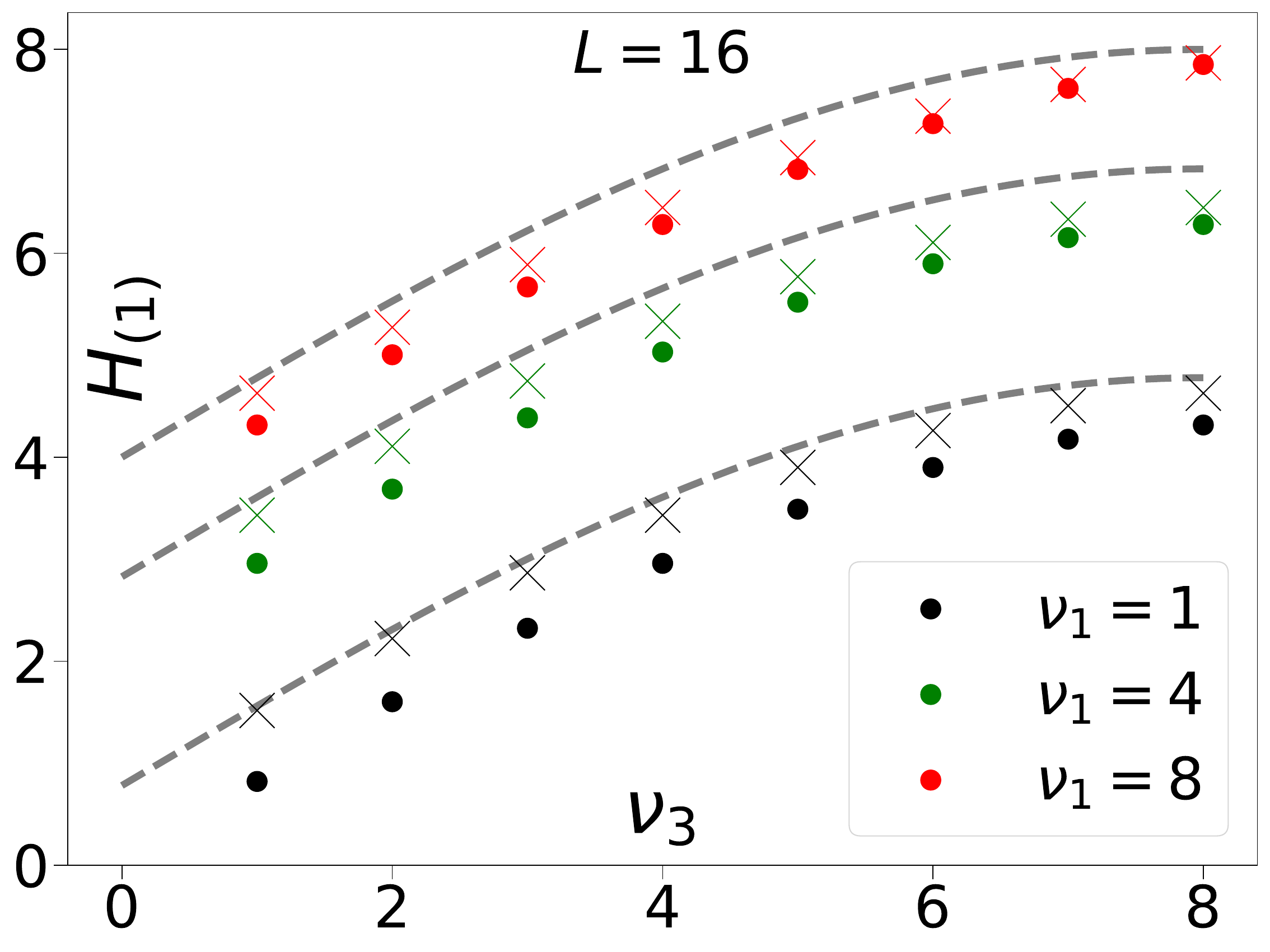}%
\includegraphics[width=.5\linewidth]{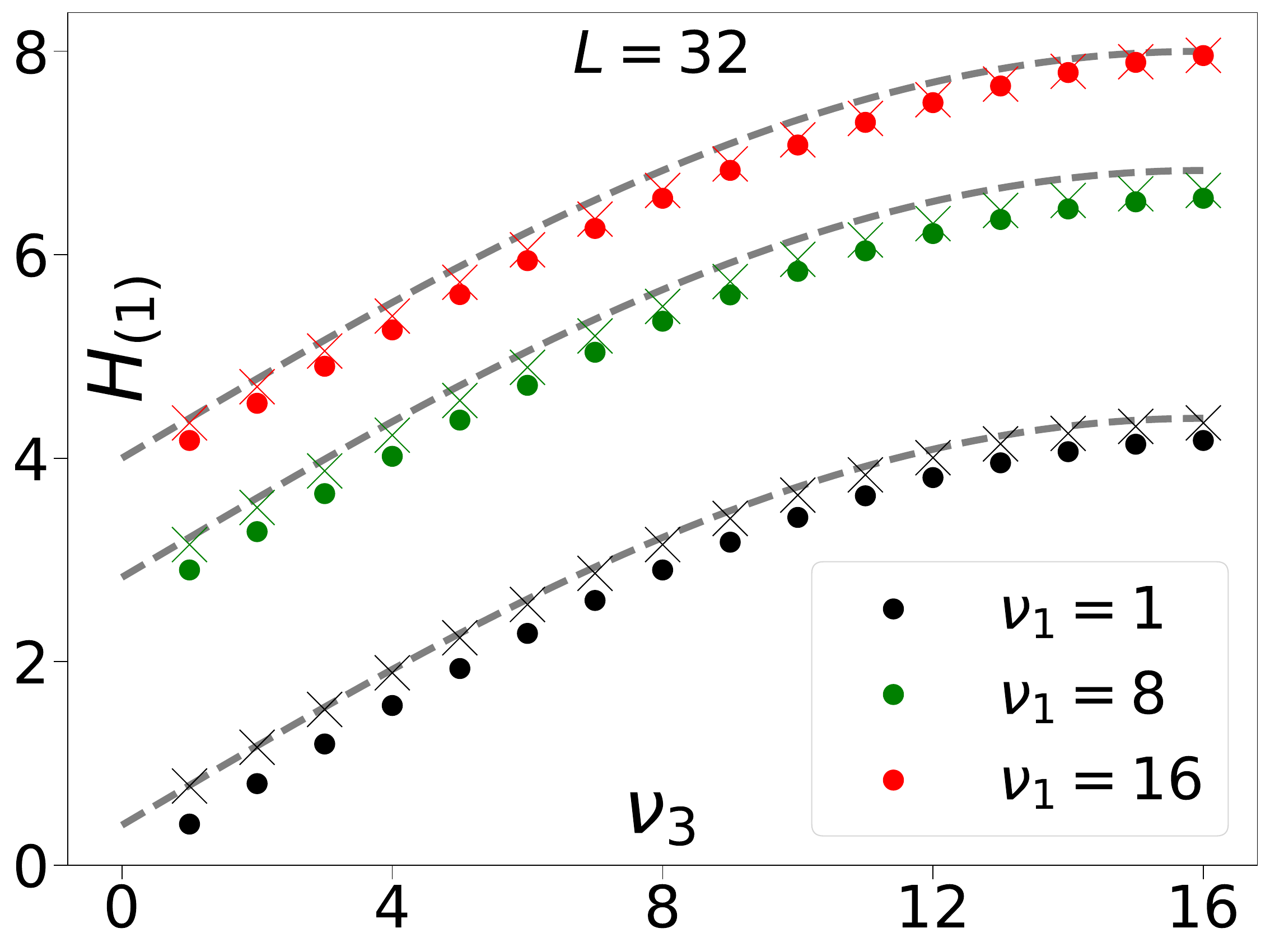}
%
% If no graphics program available, insert a blank space i.e. use
%\picplace{5cm}{2cm} % Give the correct figure height and width in cm
%
\caption{Figure taken from~\cite{Brollo:2023rgp}.
We plot the energy of four-excitations states with mode numbers $\nu_1=-\nu_2>0$ and $\nu_3=-\nu_4>0$ for various values of the volume in the lightcone gauge~$L$. The dashed line is the expectation for a free theory, i.e.\ $H_{(1)}=4\sin\tfrac{\pi\nu_1}{L}+4\sin\tfrac{\pi\nu_3}{L}$. The cross is given by the solution of the asymptotic Bethe equations and the solid circle is the TBA value. Notice how the energy exact is not additive, indicating that the model is interacting.}
\label{fig:fourmagnons}       % Give a unique label
\end{figure}

This is the setup with the smallest amount of tension. Looking at the dispersion relation~\eqref{eq:dispersion} we expect that, at $h\ll1$, the energy of each magnon looks like
\begin{equation}
\label{eq:energyexpanded}
    H(p,\mu)=\sqrt{\mu^2+4h^2\sin^2(p/2)}=
    \begin{cases}
    |\mu|+2h^2\sin^2(p/2)+\mathcal{O}(h^4)&\mu\neq0\,,\\
    2h\,\big|\sin(p/2)\big|&\mu=0\,.
    \end{cases}
\end{equation}
Recall that this formula gives the \textit{asymptotic} energy, i.e.\ it is valid in infinite volume $r\to\infty$. To truly understand what happens to the spectrum of the theory one must consider the mirror TBA equations for a generic state of the model and expand those at $h\ll1$.
This was done in~\cite{Brollo:2023pkl}. Essentially, the careful TBA analysis confirms the qualitative picture from~\eqref{eq:energyexpanded}. At $h=0$, all massive excitations give an integer contribution~$|\mu|$ to the lightcone energy; this is a bit similar to how, in free $\mathcal{N}=4$, excitations contribute only with their ``engineering'' dimension and R-charge. Still at $h=0$, one finds that the $\mu=0$ states (related to the $T^4$ fields and their supersymmetric partners) all carry zero energy, resulting in a very large degeneracy (similar to what one finds in the tensionless limit of flat-space strings).
Things are more interesting at the first nontrivial order in $h\ll1$, that is~$h^1$. One finds that, even when using the full TBA equations, the $\mu\neq0$ modes decouple at this order; one is just left with a set of TBA equation for the massless modes which moreover take a particularly simple form~\cite{Brollo:2023pkl}.%
\footnote{%
Namely, all convolutions are expressed in terms of the standard Cauchy Kernel $K(v_1,v_1)=[2\pi i \cosh(v_1-v_2)]^{-1}$, even though the model remains non-relativistic.}
These can be solved numerically. One chooses the R-charge of a reference BPS state, which is also called the  ``length''~$L$ of the state (in analogy with the $\mathcal{N}=4$ SYM case) and then considers states that contain~$n$ excitations over the vacuum of momentum $p_j$, $j=1,\dots,M$, identified by their mode number~$\nu_j$, see~\eqref{eq:logarithmicBYE}.
In practice this was done for states with $n=2$ or $n=4$. Note that because of the level matching constraint, the sum of the total momenta has to vanish modulo~$2\pi$. Hence, in~\cite{Brollo:2023pkl} the states considered had $\nu_1=-\nu_2$ for $M=2$, while they had $\nu_1=-\nu_2$ and $\nu_3=-\nu_4$ for $M=4$ (this last choice gives just a subset of $M=4$ states). Denoting the full energy of these states and its small-tension ($h\ll1$) expansion as 
\begin{equation}
    \mathbf{H}\, |\nu_1,\dots \nu_M\rangle_{L}
    =H\, |\nu_1,\dots \nu_M\rangle_{L},\qquad
    H= h\ H_{(1)}+\mathcal{O}(h^2)\,,
\end{equation}
the TBA equations can be solved to give, state by state, the numerical value of~$H_{(1)}$. The result is shown in Figure~\ref{fig:twomagnons} for states containing two magnons with $\nu_1=-\nu_2>0$, and in Figure~\ref{fig:fourmagnons} for states containing four magnons with $\nu_1=-\nu_2>0$ and $\nu_3=-\nu_4>0$.

From the figure it is clear that the dual theory is interacting, because the energy of multi-excitation states is not given by the sum of the energies of their constituents.
It would be very interesting to determine what integrable model describes the $\mathcal{O}(h)$ dynamics which gives these spectra. This would be a reduced model made out of four bosons and fermions, without the $AdS_3\times S^3$ modes. Interestingly, because the action of the supersymmetry generators have an action which is of~$\mathcal{O}(1)$ (corresponding to zero-momentum modes of the massive modes), they are not visible at~$\mathcal{O}(h)$. In other words, this model should only contain the highest-weight states of the $\mathfrak{psu}(1,1|2)_{\L}\oplus\mathfrak{psu}(1,1|2)_{\R}$ and their~$\mathcal{A}$ descentants; it would not have superconformal symmetry, as long as we only deal with $\mathcal{O}(h)$. This suggests that this should be a supersymmetric quantum-mechanical model, rather than a two-dimensional CFT, at this order.

A proposal for the dual model, valid order by order in~$h$, was put forward in~\cite{OhlssonSax:2014jtq} based on~\cite{Witten:1997yu}.
One begins by considering open strings stretching between a system of $N_1$ D1 branes and $N_5$ D5 branes. One gets three tyes of supersymmetric multiplets:
\begin{itemize}
    \item[1.] $\mathcal{N}=(8,8)$ vector multiplet for the $U(N_1)$ theory related to strings on D1s;
    \item[2.] $\mathcal{N}=(8,8)$ vector multiplet for the $U(N_5)$ theory related to strings on D5s;
    \item[3.] $\mathcal{N}=(4,4)$ hypermultiplet for the $U(N_1)\times U(N_5)$ strings stretching between D1s and D5s.
\end{itemize}
From the point of view of the two-dimensional theory on the D1s, the $U(N_5)$ dynamics (2.) decouple and $U(N_5)$ becomes a global symmetry; the $U(1)$ in $U(N_1)$ also decouple, leaving us with
\begin{itemize}
    \item[1a.] $\mathcal{N}=(4,4)$ vector multiplet for the $SU(N_1)$ theory;
    \item[1b.] $\mathcal{N}=(4,4)$ adjoint hypermultiplet for the $SU(N_1)$ theory;
    \item[3.] $\mathcal{N}=(4,4)$ fundamental / antifundamental hypermultiplet charged under $SU(N_1)\times U(N_5)$.
\end{itemize}
Hence, $N_1=N_c$ plays the role of color number and $N_5=N_f$ plays the role of flavour number. The fields in 1a., 1b.\ are square $N_c\times N_c$ matrices while those in 3.\ are $N_f\times N_c$ or $N_c\times N_f$ rectangular matrices.
In two dimensions, this supersymmetric gauge theory is not conformal. In the IR it flows to a theory where the kinetic term of the vector multiplet (as well as all the terms related to to it by supersymmetry) vanish. In other terms, the vector multiplet fields may be seen as non-dynamical. 
The proposal of~\cite{OhlssonSax:2014jtq}  is instead to integrate out the fields from the adjoint hypermultiplet. The upshot is that one is then left with $N_c\times N_c$ fields only, which allows to write ``spin-chain-like'' local operators as traces of products of adjoint fields. Note that after this integration, the fields of the vector multiplet propagate (through ``bubbles'' of fields from the fundamental hypermultiples) and interact. The coupling constant is $1/\sqrt{N_5}=1/\sqrt{N_f}$.

\begin{figure}[b]
\sidecaption
% For example, with the graphicx style use
\includegraphics[width=.6\linewidth]{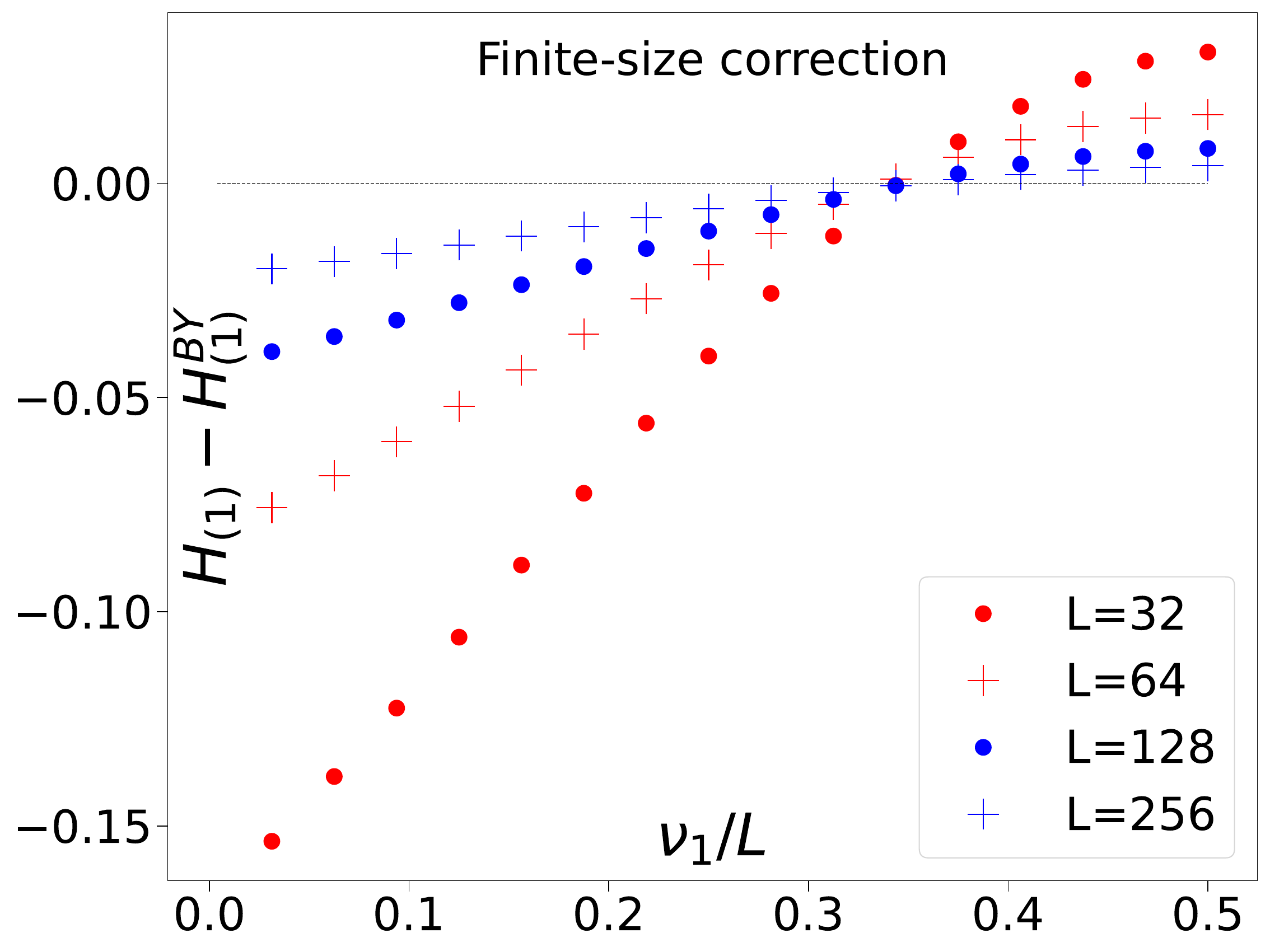}%
%
% If no graphics program available, insert a blank space i.e. use
%\picplace{5cm}{2cm} % Give the correct figure height and width in cm
%
\caption{Figure taken from~\cite{Brollo:2023rgp}. We plot the difference between the asymptotic energy (as predicted from the Asymptotic Bethe Ansatz (Bethe-Yang) equations) and the exact energy (as predicted from the mirror TBA) for a class of two-excitation states with mode numbers~$\nu_1=-\nu_2$ and length (volume)~$L$. The differece, which corresponds to the size of the ``wrapping'' effects, decreses as $1/L$ and $L\to\infty$.}
\label{fig:finitesize}       % Give a unique label
\end{figure}

Let us take a look at the field content of this theory. The vector multiplet contains scalar fields of the type $\Phi^{\alpha\dot{\alpha}}$, in the bifundamental representation of the  $\mathfrak{su}(2)_{\L}\oplus\mathfrak{su}(2)_{\R}$ R-symmetry, with dimension $(1/2, 1/2)$ under $(L^{0}_{\L},L^{0}_{\R})$ (as well as their superpartners, which we will not discuss here, see~\cite{OhlssonSax:2014jtq}). Notice that this dimension is unusual for scalars in a two-dimensional (S)CFT; it is a result of having integrated out the fundamental hypermultiplet fields, which yields a non-standard (non-local) kinetic term to the $\Phi$s. 
One of these fields, say $\Phi^{+\dot{+}}$ can be identified with the half-BPS vacuum of the light-cone gauge; the sphere bosons $y$, $\bar{y}$ correspond to R-symmetry currents relating $\Phi^{+\dot{+}}$ to $\Phi^{-\dot{+}}$ and  $\Phi^{+\dot{+}}$ to $\Phi^{+\dot{-}}$, respectively. This is quite similar to what happens in $\mathcal{N}=4$ SYM.
Conversely, the adjoint hyper contains scalars of the type $T^{a\dot{a}}$, charged under $\mathfrak{su}(2)_\bullet\oplus \mathfrak{su}(2)_\circ$, which have the usual scalar field end hence have dimension $(0, 0)$ under $(L^{0}_{\L},L^{0}_{\R})$ (like a ``regular'' scalar in two dimensions). They correspond to excitations from the $T^4$ part of the string geometry.
This model seems to have nice features, including integrability~\cite{OhlssonSax:2014jtq}. However, starting from a reference BPS state such as
\begin{equation}
|0\rangle_L=\text{Tr}\big[(\Phi^{+\dot{+}})^L\big]\,,
\end{equation}
and considering excitations on top of it, there is the possibility of a huge amount of mixing due to the presence of the dimensionless $T^{a\dot{a}}$ fields and of their superpartners.
This fact is reflected in the presence of strong wrapping corrections (not suppressed in~$h\ll1$) in the TBA. In fact, it can be checked that even at small-$h$ the wrapping effects due to TBA scale as $1/L$, i.e.\ that they are not suppressed at all,%
\footnote{The modifying the S~matrix appearing in the Asymptotic Bethe Ansatz itself would give corrections of order $1/L$.}
see Figure~\ref{fig:finitesize}.

\subsection{The case of \texorpdfstring{$k=1$}{k=1} (symmetric-product orbifold CFT)}

Another notable small-string-tension setup is that of $k=1$ and $h\ll1$ (or $h=0$). This is sometimes called the ``tensionless'' limit (though of course it has ``more tension'' than the $k=0$, $h\ll1$ limit discussed above).
The detailed study of this model was initiated in~\cite{Giribet:2018ada,Gaberdiel:2018rqv,Eberhardt:2018ouy}. As long as we are only interested in the planar string theory spectrum, the correspondence can be simply explained in the language of integrability. The map was recently elucidated in~\cite{Frolov:2024pkz} and we follow (and condense) that presentation here.

The symmetric-product orbifold Sym${}_NT^4$ is obtained by considering $N$ identical copies of the free $\mathcal{N}=(4,4)$ superconformal theory of four bosons and of their superpartners, and modding out by the action of the symmetric group~$S_N$. States are then labeled by the conjugacy classes of~$S_N$, which are products of cycles.  In the planar limit, we take $N\to\infty$ and we are actually interested in states constructed out of a single cycle of length~$w\in\mathbb{N}$. Without loss of generality,%
\footnote{Strictly speaking, one has to sum over the various images of the cycle under $S_N$, which gives an overall prefactor which is important for the scaling in $1/N$ and in $1/w$.}
we can take this cycle to be in the first $w$ copies of the theory, labeled by $I=1,\dots, w$ so that a generic boson $X^{a\dot{a}}_{I}$ from the $I$-th copy has boundary conditions
\begin{equation}
    X^{a\dot{a}}_{I}(\sigma+2\pi,\tau)= X^{a\dot{a}}_{(I\,\text{mod}w)+1}(\sigma,\tau)\,,
\end{equation}
and similarly for the other fields. In practice it is more convenient to work with chiral fields on the plane, e.g.
\begin{equation}
    \partial X^{a\dot{a}}_{I}(z e^{2\pi i})= \partial X^{a\dot{a}}_{(I\,\text{mod}w)+1}(z)\,,\qquad
    \bar\partial X^{a\dot{a}}_{I}(\bar{z} e^{-2\pi i})= \bar\partial X^{a\dot{a}}_{(I\,\text{mod}w)+1}(\bar{z})\,,
\end{equation}
with fermions $\Psi^{\alpha\dot{a}}(z)_I$, $\bar{\Psi}^{\dot\alpha\dot{a}}(\bar{z})_I$ in the NS sector which ensures supersymmetry of the vacuum.

Let us mention that the chiral and antichiral supercurrents are 
\begin{equation}
G^{\alpha a}(z)=\sum_{I=1}^w\frac{i\epsilon_{\dot{a}\dot{b}}}{\sqrt{2}}\Psi^{\alpha\dot{a}}_I(z)\,\partial X^{a\dot{b}}_I(z)\,,\quad
\bar{G}^{\dot{\alpha} a}(\bar{z})=\sum_{I=1}^w\frac{i\epsilon_{\dot{a}\dot{b}}}{\sqrt{2}}\bar{\Psi}^{\dot{\alpha}\dot{a}}_I(\bar{z})\,\bar{\partial} X^{a\dot{b}}_I(\bar{z})\,.
\end{equation}
The supercharges of $Q^{\pm\alpha a}_{\L}$ introduced above are precisely the modes $G^{\alpha a}_{\pm1/2}$ whiles $Q^{\pm\dot{\alpha} a}_{\R}$ are $\bar{G}^{\dot{\alpha} a}_{\pm 1/2}$. From the supercharges, the rest of the $\mathcal{N}=(4,4)$ algebra can be worked out, as well as its globally-defined part which is $\mathfrak{psu}(1,1|2)_{\L}\oplus \mathfrak{psu}(1,1|2)_{\R}$. In particular, one finds that the zero mode of the chiral and antichiral stress tensor $T(z)$, $\bar{T}(\bar{z})$ correspond to the previously introduced Cartan generators $L^{0}_{\L}$ and $L^{0}_{\R}$.
This is just like in any $\mathcal{N}=(4,4)$ theory. Here note however that, as a result of the twisted boundary conditions, the non-zero modes of the fields in this sector are fractionary, \textit{i.e.}~expressed in units of $1/w$.
For instance, we have the mode expansions for the chiral fields
\begin{align}
\nonumber
    \alpha_{-\tfrac{n}{w}}^{a\dot{a}} \equiv\,&  \frac{i}{\sqrt{w}}\oint \frac{\text{d}z}{2\pi i} \sum_{I=1}^w\partial X_I^{a\dot{a}}(z)\,e^{-\frac {2\pi i\, n}{w}(I-1)} z^{-\frac {n}{w}},\\
    \Psi_{-\tfrac{n}{w}-\frac{1}{2}}^{\alpha\dot{a}} \equiv\,&  \frac{1}{\sqrt{w}}\oint \frac{\text{d}z}{2\pi i} \sum_{I=1}^w\Psi^{\alpha\dot{a}}_I(z)\,e^{-\frac {2\pi i\, n}{w}(I-1)} z^{-\frac {n}{w}-1},
\label{eq:modeexpansion}
\end{align}
and similarly for the anti-chiral ones.
Moreover, in the $w$-cycle sector, the vacuum is not the trivial NS vacuum of the CFT but it contains a twist field~$\sigma_w$ which carries scaling dimension%
\footnote{Notice that this vanishes for a trivial cycle ($w=1$) as it should.}
\begin{equation}
    (L^0_{\L},\,L^{0}_{\R})\,|\sigma_{w}\rangle
    = \begin{cases}
        \big(\frac{w}{4},\frac{w}{4}\big)\,|\sigma_{w}\rangle& w\text{ even},\\
        \big(\frac{w^2-1}{4w},\frac{w^2-1}{4w}\big)\,|\sigma_{w}\rangle& w\text{ odd},
    \end{cases}
\end{equation}
as it can be seen \textit{e.g.}\ by expanding the stress energy tensor in terms of the fractionary modes of the fields~\eqref{eq:modeexpansion}; the difference between even and odd sectors is due to Fermions. In fact, in the NS sectors the chiral Fermions have modes indexed by
$(-\tfrac{1}{2}-\tfrac{n}{w})$, cf.~\eqref{eq:modeexpansion}; similarly, the anti-chiral modes are indexed by $(-\tfrac{1}{2}-\tfrac{\tilde{n}}{w})$. Hence, the modes with $n>-w/2$ and $\tilde{n}>-w/2$ are always creation modes, while those with $n<-w/2$ and $\tilde{n}<-w/2$ annihilate $|\sigma_{w}\rangle$; when $w$ is even, the modes with $n=w/2$ and $\tilde{n}=w/2$ give rise to a Clifford module.

Clearly, $|\sigma_{w}\rangle$ is not BPS. However, both in the even-$w$ and odd-$w$ it is possible to dress $|\sigma_{w}\rangle$ to create several 1/2-BPS states~\cite{Lunin:2001pw}. It is simplest to write this for $w$ odd where we define the state
\begin{equation}
    |0\rangle_w=\epsilon_{\dot{a}\dot{b}}\Psi^{+\dot{a}}_{-\tfrac{1}{2}}\bar{\Psi}^{\dot{+}\dot{b}}_{-\tfrac{1}{2}}
    \prod_{\ell=1}^{(w-1)/2}\Big(
    \Psi^{+\dot{2}}_{-\tfrac{1}{2}+\tfrac{\ell}{w}}\bar{\Psi}^{\dot{+}\dot{2}}_{-\tfrac{1}{2}+\frac{\ell}{w}}\Psi^{+\dot{1}}_{-\tfrac{1}{2}+\tfrac{\ell}{w}}\bar{\Psi}^{\dot{+}\dot{1}}_{-\tfrac{1}{2}+\frac{\ell}{w}}\Big)\,|\sigma_w\rangle\,.
\end{equation}
This state has
\begin{equation}
    \big(L^0_{\L},\;J^{12}_{\L},\;\,L^0_{\R},\;J^{\dot{1}\dot{2}}_{\R})\,|0\rangle_w = \Big(\frac{w}{2},\frac{w}{2},\,\frac{w}{2},\frac{w}{2}\Big)\,|0\rangle_w,
\end{equation}
and it is a highest weight state under the global $\mathfrak{psu}(1,1|2)_{\L}\oplus \mathfrak{psu}(1,1|2)_{\R}$ algebra; in fact, it is the only such state in this sector that is a singlet under the $\dot{a},\dot{b}$ indices (i.e., under $\mathfrak{su}(2)_\circ$).
A similar state can be constructed for $w$~even.
The excitations above the BPS vacuum~$|0\rangle_w$ are
\begin{equation}
    \Psi^{-\dot{a}}_{+\tfrac{1}{2}-\tfrac{n}{w}},
    \quad
    \alpha^{a\dot{a}}_{-\tfrac{n}{w}},
    \quad
    \Psi^{+\dot{a}}_{-\tfrac{1}{2}-\tfrac{n}{w}},
    \qquad
    \bar\Psi^{\dot{-}\dot{a}}_{+\tfrac{1}{2}-\tfrac{\tilde{n}}{w}},
    \quad
    \bar\alpha^{a\dot{a}}_{-\tfrac{\tilde{n}}{w}},
    \quad
    \bar\Psi^{\dot{+}\dot{a}}_{-\tfrac{1}{2}-\tfrac{\tilde{n}}{w}},
\end{equation}
with $n>0$ and $\tilde{n}>0$.
The $\mathfrak{sl}(2,\mathbb{R})$ charges can be read off from the above formulae, and it is clear that we can match these excitations with those in Table~\ref{tab:mixedcharges} by setting $k=1$ and~$h=0$, $r=w$ and matching
\begin{equation}
\begin{aligned}
    \Psi^{-\dot{a}}_{+\frac{1}{2}-\frac{n}{w}}|0\rangle_{w}&=\chi^{\dot{a}}(p)|0\rangle_r\,,&\qquad&&\bar\Psi^{-\dot{a}}_{+\frac{1}{2}-\frac{\tilde{n}}{w}}|0\rangle_{w}&=\chi^{\dot{a}}(\bar{p})|0\rangle_r\,,\\
    \alpha^{a\dot{a}}_{-\frac{n}{w}}|0\rangle_w &= T^{a\dot{a}}(p)|0\rangle_r\,,&\qquad&&\bar\alpha^{a\dot{a}}_{-\frac{\tilde{n}}{w}}|0\rangle_w &= T^{a\dot{a}}(\bar{p})|0\rangle_r\,,\\
    \Psi^{+\dot{a}}_{-\frac{1}{2}-\frac{n}{w}}|0\rangle_{w}&=\tilde{\chi}^{\dot{a}}(p)|0\rangle_r\,,&\qquad&&\bar\Psi^{+\dot{a}}_{-\frac{1}{2}-\frac{\tilde{n}}{w}}|0\rangle_{w}&=\tilde{\chi}^{\dot{a}}(\bar{p})|0\rangle_r\,,
\end{aligned}
\end{equation}
with%
\footnote{The $p=0$, or $\bar{p}=0$, modes of the fermions require a little more care~\cite{Frolov:2023pjw}.}
\begin{equation}
\label{eq:pnidentification}
    p=\frac{2\pi n}{w}>0\,,\qquad
    \bar{p}=-\frac{2\pi \tilde{n}}{w}<0\,.
\end{equation}
Clearly, the quantisation conditions on the momentum $p_j$ of the $j$-th excitation on a generic multi-magnon state ($j=1,\dots,M$) can be obtained from a free model,
\begin{equation}
    e^{i p_j r}=e^{i p_j w}=1\,.
\end{equation}
It is worth noting that both in the orbifold description and in the integrability one the co-product is trivial because when~$h=0$ the central extension of the algebra~$\mathcal{A}$ vanishes identically.
Finally, there is an orbifold invariance condition
\begin{equation}
    \sum_{j} n_j-\sum_{j} \tilde{n}_j= 0 \ \text{mod}\,w\,,
\end{equation}
which under the identification~\eqref{eq:pnidentification} maps to the level matching condition~\eqref{eq:levelmatching}.

The above discussion holds for $h=0$, i.e.\ at the free orbifold point. We expect this to be deformed as we turn on the RR flux, $h>0$ (see~\cite{OhlssonSax:2018hgc} for a discussion of the moduli in the string model). In the orbifold theory, there are several exactly marginal operators; those associated with turning on  RR flux come from the $w=2$ sector of the orbifold. Precisely one of these operators is in the singlet representation of~$\mathfrak{su}(2)_{\circ}$~\cite{David:1999ec,Gomis:2002qi,Avery:2010er}.
We denote this state by $|\psi_\mathcal{D}\rangle_2$, and it has dimension
\begin{equation}
    \big(L^0_{\L},\ L^0_{\R}\big)\,|\psi_\mathcal{D}\rangle_2 = (1,\;1)\,|\psi_\mathcal{D}\rangle_2\,,
\end{equation}
while being a singlet under all other generators. This allows one to compute e.g.\ correlation functions order by order in a marginal coupling (deformation parameter)~$g$ in conformal perturbation theory. E.g., for a two-point correlation function $\langle \mathcal{V}_{2}\mathcal{V}_{1}\rangle$ at second order in~$g$ we have the integral
\begin{equation}
    g^2\, \int \de^2\zeta' \int \de^2\zeta\ 
    \big\langle \mathcal{V}_{2}(z_2,\bar{z}_2)\,
     \mathcal{D}(\zeta',\bar{\zeta}')
    \mathcal{D}(\zeta,\bar{\zeta})
      \mathcal{V}_{1}(z_1,\bar{z}_1)\big\rangle\,,
\end{equation}
which needs to be suitably regularised.
The computation of this kind of integrals in symmetric-orbifold CFTs is technically involved but well understood~\cite{Arutyunov:1997gt,Arutyunov:1997gi,Lunin:2000yv,Lunin:2001pw}.
The general structure of the orbifold perturbation theory suggests that in this case the leading corrections to the scaling dimension should be of order~$g^2$. Let us consider, for instance, a state constructed from acting on the half-BPS vacuum by chiral oscillators only
\begin{equation}
    \big|\psi_{\L}\big\rangle_w\equiv\alpha^{a_1\dot{a}_1}_{-\frac{n_1}{w}}\cdots
    \alpha^{a_M\dot{a}_M}_{-\frac{n_M}{w}}\,|0\rangle_w\,.
\end{equation}
In the free orbifold theory
\begin{equation}
     \big(L^0_{\L}-J_{\L}^{12},\ L^0_{\R}-J_{\R}^{12}\big)\,\big|\psi_{\L}\big\rangle_w = \Big(\sum_{j=1}^M\frac{n_j}{w},\;0\Big)\,\big|\psi_{\L}\big\rangle_w\,,\qquad
     \sum_{j=1}^M\frac{n_j}{w}=E_0 \in\mathbb{N}\,,
\end{equation}
and this state is degenerate with many others --- of the order of the number of integer partitions of~$(wE_0)$, if we disregard the $\mathfrak{su}(2)_\bullet$ and  $\mathfrak{su}(2)_\circ$ indices. When turning on the deformation parameter $g$, $|\psi_{\L}\rangle_w$ gets corrected to some
\begin{equation}
    \Big|\psi_{\L}^{(g)}\Big\rangle_w = 
    \big|\psi_{\L}\big\rangle_w+\mathcal{O}(g)\,,
\end{equation}
so that
\begin{equation}
    \big(L^0_{\L}-J_{\L}^{12},\ L^0_{\R}-J_{\R}^{12}\big)\Big|\psi_{\L}^{(g)}\Big\rangle_w=
    \Big(E_0+\frac{1}{2}\delta H,\ \frac{1}{2}\delta H\Big)\Big|\psi_{\L}^{(g)}\Big\rangle_w\,,
\end{equation}
where $\delta H=\mathcal{O}(g^2)$. Because
\begin{equation}
    \big\{\mathbf{Q}_{\R}^a,\,\mathbf{S}_{\R}^b\big\} = \varepsilon^{ab}\Big(L^0_{\R}-J_{\R}^{12}\Big)\,,
\end{equation}
it must therefore be that in the deformed theory
\begin{equation}
\label{eq:perturbedcommutator}
    \Big[\mathbf{Q}_{\R}^a,\;\alpha^{b\dot{a}}_{-\frac{n}{w}}\Big]\neq0\neq\Big[\mathbf{S}_{\R}^a,\;\alpha^{b\dot{a}}_{-\frac{n}{w}}\Big]\qquad\text{at}\quad \mathcal{O}(g)\,.
\end{equation}
This is of course expected from the form of the centrally extended algebra which we discussed above --- which mandates $\{\mathbf{Q}_{\L},\mathbf{Q}_{\R}\}\neq0$ for $h\neq0$ ---  and in particular from Table~\eqref{tab:mixedcharges}. In fact, comparing the above algebraic description at $k=1$ and small~$h$%
\footnote{It is worth keeping in mind that the analysis is drastically different at $k=0$ and small $h$, as discussed in the previous subsetion.}
with this conformal perturbation theory set-up, we find that (as expected) the RR parameter~$h$ and the conformal perturbation theory parameter~$g$ should be related as
\begin{equation}
\label{eq:couplingmatch}
    h(g) = \xi\,g+\mathcal{O}(g^2)\,,
\end{equation}
for some numerical coefficient~$\xi>0$.
The computation of the action~\eqref{eq:perturbedcommutator}, at leading order, boils down to a conformal perturbation theory integral of the form%
\footnote{For simplicity, we are omitting the sums over orbifold representatives and various indices.}
\begin{equation}
    g\,\oint\limits\de \bar{z} \int \de^2\zeta\big\langle\mathcal{V}_{2}^{[w+1]}(\infty)\; \mathcal{D}(\zeta,\bar{\zeta})\;\bar{G}(\bar{z})\;\mathcal{V}_1^{[w]}(0)\big\rangle\,.
\end{equation}
Notice that the twist sector of the two operators change from~$w$ to~$(w+1)$; this is both required by the orbifold construction and by the length-changing nature of the algebra, see eq.~\eqref{eq:lenghtchanging}.
This computation was first discussed in~\cite{Gava:2002xb} (well before the relation between these centrally-extended algebras and integrability came to the fore~\cite{Beisert:2005tm}) and recently revisited in~\cite{Gaberdiel:2023lco} where it was argued that it gives a non-zero result.%
\footnote{Note that it was argued in~\cite{Fabri:2025rok} that there is an inconsistency in the derivation of~\cite{Gaberdiel:2023lco}, specifically in how the large-$w$ limit of the orbifold results is taken. According to~\cite{Fabri:2025rok}, the central extensions $\mathbf{C}$, $\mathbf{C}^\dagger$ would actually vanish in the algebra~\eqref{eq:centralext} when computed from the orbifold, suggesting that a different analysis of that CFT would be needed to construct the central extensions.}
In particular, such a computation would allow to fix the proportionality constant in~\eqref{eq:couplingmatch} to $\xi=1$. Under this assumption, the precise identification of the excitations from the orbifold with the integrability ones was performed in~\cite{Frolov:2023pjw}, where it was also shown that the algebra reproduces the one discussed of~\cite{Lloyd:2014bsa} and therefore yields their S~matrix for the modes with~$\mu=0$. This also fits with the general argument discussed above that, for $k=1$, we only expect excitations with $\mu=0$: recall that the dispersion relation is invariant under shifts of $\mu\to \mu- kn$ and $p\to p+2\pi n$, with~$n\in\mathbb{Z}$, cf.~\eqref{eq:periodicity}. In fact, this is not just an algebraic property of the representations, but it holds for the whole S~matrix (including the dressing factor~\cite{Frolov:2024pkz,Frolov:2025uwz}). Because of this, if we allow the real part of~$p$ to take any value, we can identify the species of particles by $\mu~\text{mod}\,k$, rather than by~$\mu$.

\section{Deformations}
\label{sec:deformations}

In this section we are interested in deformations of the $AdS_3 \times S^3 \times T^4$ superstring that preserve the classical integrability of the theory. We give a brief overview of these deformations, and then focus on two specific examples (the bi-Yang-Baxter+WZ deformation in section \ref{sec:biYB-WZ} and the elliptic deformation in section \ref{sec:elliptic}) to highlight how the string background, the symmetries and the worldsheet S-matrix are affected by the deformation. 

\subsection{Overview of integrable deformations}

Both $AdS_3$ and $S^3$ are symmetric spaces, but also group manifolds
\begin{equation}
    AdS_3 \cong \frac{SO(2,2)}{SO(1,2)} \cong SU(1,1) \cong SL(2;\mathbb{R})\,, \qquad S^3 \cong \frac{SO(4)}{SO(3)} \cong SU(2)\,.
\end{equation}
The bosonic $AdS_3 \times S^3$ string can then equivalently be described by a principal chiral model (PCM) on $G=SL(2;\mathbb{R}) \times SU(2)$, or a symmetric space sigma model on $G/H$ with 
\begin{equation}
   G=SL(2;\mathbb{R}) \times SL(2;\mathbb{R}) \times SU(2) \times SU(2)\,, \qquad H= SO(1,2) \times SO(3)\,. 
\end{equation}
There exist various integrable deformations of the PCM and of the symmetric space sigma model, for a review see~\cite{Hoare:2021dix}. In some cases it is known how to generalise these deformations to the semi-symmetric space sigma model, and how to include the torus directions, leading to integrable deformations of, in particular, the $AdS_3\times S^3 \times T^4$ superstring. Here we shall give an overview of some of the possible integrable deformations, see also~\cite{Seibold:2020ouf}.

\textbf{TsT transformations.} Perhaps the simplest integrable deformations are T-duality-shift-T-duality (TsT) transformations, see~\cite{Lunin:2005jy,Frolov:2005dj,Alday:2005ww}.  These transformations require the presence of two abelian isometries in the background. For concreteness we will assume these to be realised as shifts in the two coordinates $x_1$ and $x_2$. The deformed background is obtained through a T-duality $x_1 \rightarrow \tilde{x}_1$, where $\tilde{x}_1$ denotes the T-dual coordinate, followed by a shift $x_2 \rightarrow x_2 + s \tilde{x}_1$, and finally a T-duality back $\tilde{x}_1 \rightarrow \tilde{\tilde{x}}_1$. Here $s \in \mathbb{R}$ denotes the continuous and real deformation parameter.
What makes TsT deformations particularly simple is that they can be re-absorbed in the boundary conditions of the fields~\cite{Frolov:2005dj,Alday:2005ww}. In the particular case where the TsT is taken (partially or completely) along one of the directions used for the lightcone gauge fixing, the interpretation of the twist of the boundary conditions is more subtle, and can be related to a ``$T \bar{T}$'' deformation of the gauge fixed model, see~\cite{Sfondrini:2019smd,Apolo:2019zai, Idiab:2024bwr}.

\textbf{Yang-Baxter (YB) deformations.} These were initially proposed as deformations of the PCM on an arbitrary Lie group $G$. This model has $G_{\L}\times G_{\R}$ symmetry, corresponding to left and right multiplication by the group. The YB deformation which affects the~$G_{\R}$ symmetry takes the form~\cite{Klimcik:2002zj,Kawaguchi:2014qwa,vanTongeren:2015soa}%
\footnote{%
By construction, this action is invariant under left multiplications $G_{\L}$; some of the $G_{\R}$ symmetries may be preserved too, depending on the explicit form of the deforming operator $\mathcal R$. 
}
\begin{equation}
\label{eq:YBdef}
    S =  -\frac{1}{2} \int \Tr[J \frac{1}{1-\kappa_{\R} \mathcal R} J]\,, \qquad J=g^{-1} \extder g\,, \qquad g \in G.
\end{equation}
Here $\kappa_{\R} \in \mathbb{R}$ is the deformation parameter, with $\kappa_{\R}=0$ corresponding to the (undeformed) PCM. The constant (independent of $g$) deforming operator $\mathcal R: \alg{g} \rightarrow \alg{g}$ is required to satisfy
\begin{align}
    &\Tr[X \mathcal R(Y)] = - \Tr[\mathcal R(X) Y]\,,  \\
    &[\mathcal R(X), \mathcal R(Y)] - \mathcal R([\mathcal R(X),Y]+[X,\mathcal R(Y)])= -c^2 [X,Y]\,,
    \label{eq:cYBE}
\end{align}
for all elements $X,Y \in \alg{g}$%
and with $c \in \mathbb{C}$ %v2
. These two conditions ensure the classical integrability of the model~\cite{Klimcik:2008eq}.
The first condition is the antisymmetry of the operator with respect to the ad-invariant bilinear form $\Tr$, while the second is the classical Yang-Baxter equation. This statement can be made more transparent by defining $r \in \alg{g} \wedge \alg{g}$ and the Casimir $t$ through
\begin{equation}
    \mathcal R(X) = \Tr_2[r (1 \otimes X)]\,, \qquad X=\Tr_2[t (1\otimes X)]\,,
\end{equation}
where the trace acts on the second factor of the tensor product, so that \eqref{eq:cYBE} becomes
\begin{equation}
    [r_{12}^\pm,r_{13}^\pm]+[r_{12}^\pm,r_{23}^\pm]+[r_{13}^\pm,r_{23}^\pm]=0\,, \qquad r^\pm = r \pm ct\,.
\end{equation}
This is the classical limit of the quantum Yang-Baxter equation.%
\footnote{The classical Yang-Baxter equation can be obtained by taking $\mathbf{S}(p_1,p_2)=\Pi\,\mathbf{R}(p_1,p_2)$, where $\Pi$ is the permutation operator, and expanding the quantum R~matrix as $\mathbf{R}(p_1,p_2)=\text{Id}+\varepsilon\,r(p_1,p_2)+\mathcal{O}(\varepsilon^2)$, where $p_1, p_2$ may be replaced by rapidities or spectral parameters. Note that above $r_{ij}$ is a constant matrix (independent from any spectral parameter) acting on the $i$-th and $j$-th copy of the vector space~$\mathfrak{g}$. 
}
The subscript indicates on which factors of the tensor product space $\alg{g} \otimes \alg{g} \otimes \alg{g}$ the $r$-matrix acts.
For $c=0$ this is the homogeneous classical Yang-Baxter equation, leading to \textit{homogeneous} Yang-Baxter deformations, while for $c \neq 0$ this is the modified classical Yang-Baxter equation, giving rise to \textit{inhomogeneous} Yang-Baxter deformations.%
\footnote{Note that up to rescaling of the operator $\mathcal R$, the only relevant cases are $c=0$, $c=1$ and $c=i$.}
We will discuss below at length the difference between these two cases, which is substantial.

Note that one could have defined an equivalent model that is invariant under right multiplications instead,
\begin{equation}
\label{eq:YBdef2}
    S =  -\frac{1}{2} \int \Tr[J \frac{1}{1-\kappa_{\L} \mathcal R_g} J]\,, 
\end{equation}
where the operator $\mathcal R_g = \Ad_g^{-1} \mathcal R \Ad_g$ acts on an element $X\in \alg{g}$ as $\mathcal R_g(X) = g^{-1} \mathcal R(g X g^{-1}) g$.

The Yang-Baxter deformation was extended to symmetric space sigma models in~\cite{Delduc:2013fga}, and to the $AdS_5 \times S^5$ superstring in~\cite{Delduc:2013qra}. The Yang-Baxter deformation of generic semi-symmetric space sigma models on $G/H$ reads
\begin{equation}
    S = \frac{1}{2} \int \STr[ J P \frac{1}{1- \kappa\mathcal R_g P} J]\,, 
\end{equation}
with
\begin{equation}
     P = P^{(2)} + \frac{1+ c^2\eta^2}{2} (P^{(1)} - P^{(3)})\,, \qquad \kappa = \frac{2 \eta}{1+c^2 \eta}\,.
\end{equation}
We recall that the undeformed semi-symmetric space sigma model~\eqref{eq:MT}, obtained by setting $\kappa=0$, is invariant under global left-acting $G$ symmetries and local right-acting $H$ symmetries. 
The deformation parameter $\kappa$ and the deforming operator $\mathcal R$ break (part of) the $G$ symmetries. The deformed model is however still invariant under the right-acting $H$ symmetries, and it is also invariant under a right-acting local fermionic kappa-symmetry.

\textbf{The bi-Yang-Baxter+WZ deformation.} While the above YB deformations can be implemented for any semi-symmetric space $G/H$, there exist other deformations that are specific to $AdS_3$ strings. As mentioned in section \ref{sec:lcgauge}, the curved part of the $AdS_3 \times S^3 \times T^4$ superstring action is described by a semi-symmetric space sigma model with $G = PSU(1,1|2) \times PSU(1,1|2)$. This product structure leads to a very rich space of possible integrable deformations. We already mentioned that one can add a WZ term to the action. On top, one can also deform the first and second copies of $PSU(1,1|2)$  in different ways.  
%First of all, the bosonic truncation can be described by a PCM on $G=SL(2;\mathbb R) \times SU(2)$. 
For the PCM on $G= SL(2;\mathbb{R}) \times SU(2)$ %v2
(which we recall describes the bosonic truncation of the $AdS_3$ superstring) one can have two-parameter YB deformations~\cite{Klimcik:2008eq}, deforming the $G_{\L}$ and $G_{\R}$ symmetries in different ways while preserving the classical integrability~\cite{Klimcik:2014bta}. The bi-YB deformation of the PCM reads
\begin{equation}
\label{eq:biYB}
    S = -\frac{1}{2} \int \Tr\left[J \frac{1}{1-\kappa_{\R} \mathcal R - \kappa_{\L} \mathcal R_{g}} J \right]\,.
\end{equation}
When the deformation parameter $\kappa_{\L}=0$ this is the YB deformation with $G_{\L}$ symmetry \eqref{eq:YBdef} while for $\kappa_{\R}=0$ this is the YB deformation preserving the $G_{\R}$ symmetry \eqref{eq:YBdef2}. This bi-YB deformation can be extended to the model with fermions: the bi-YB deformation of the sigma model on the semi-symmetric space $G_{\L} \times G_{\R}/H$ reads
\begin{equation}
    S = \frac{1}{2} \int \STr \left[ J P \frac{1}{1- \diag(\kappa_{\L}, \kappa_{\R}) \mathcal R_g P} J \right]\,, 
\end{equation}
with
\begin{equation}
    P = P^{(2)} + \frac{\bar{\eta}}{2} (P^{(1)}-P^{(3)})\,, \quad \kappa_{\L,\R} = \frac{2 \eta_{\L,\R}}{\bar{\eta}}\,, \quad \bar{\eta} = \sqrt{(1-\eta_{\L}^2)(1-\eta_{\R}^2)}\,.
\end{equation}
We assumed a parametrisation where $g=(g_{\L},g_{\R})$.
Then, one can also add a WZ term. The expression for the action is a bit more involved. For the PCM it can be written in a compact form as~\cite{Klimcik:2019kkf}
\begin{equation} %v2
    S = -\frac{q}{2} \int_{\Sigma = \partial B} \Tr\left[ J \frac{e^{\chi} + e^{\rho_{\L} \mathcal R_{g}}e^{\rho_{\R} \mathcal R}}{e^{\chi} - e^{\rho_{\L} \mathcal R_{g}}e^{\rho_{\R} \mathcal R}} J\right] - q \int_B \Tr \left[ J \wedge J \wedge J \right]\,.
\end{equation}
As in the previous section, $2 \pi q T = k\in \mathbb{N}$ is the WZ level. The other deformation parameters are $\rho_{\L},\rho_{\R}, \chi \in \mathbb{R}$, with a special combination giving the string tension $T$. %v2
The classical integrability of the bi-YB+WZ model was shown in~\cite{Klimcik:2020fhs}. The bi-YB deformation without WZ term is obtained by taking the limit
\begin{equation} %v2
    \chi= 2 q\,, \qquad \rho_L = 2 q \kappa_L \,, \qquad \rho_R = 2 q \kappa_R \,, \qquad q \rightarrow 0\,.
\end{equation}
The bi-YB+WZ deformation of the PCM can also be recast as a deformation of the symmetric space sigma model, and can also be extended to the semi-symmetric space sigma model~\cite{Delduc:2018xug}. The bi-YB+WZ deformation of the $G_{\L} \times G_{\R}/H$ semi-symmetric space sigma model takes a rather involved form and we refer the reader to the original literature~\cite{Delduc:2018xug} for an explicit action.

\textbf{The elliptic deformation.} Let us finish with the elliptic deformation, focusing on the elliptic deformation of $SU(2)$~\cite{Cherednik:1981df}. A deformation of $SL(2;\mathbb{R})$ can be constructed in a similar fashion, through analytic continuation. We call 
$\sigma_a$ %v2
with $a=1,2,3$ the three generators of $SU(2)$. The elliptic deformation is then defined as 
\begin{equation}
    S = -\frac{1}{2} \int \Tr[J \mathcal O J]\,, \qquad \mathcal O(\sigma_a) = \alpha_a \sigma_a\,, \qquad a=1,2,3\,.
\end{equation}
By construction, the deformed action remains invariant under global left multiplications $g \rightarrow g_{\L} g$ with 
$g_{\L} \in SU(2)$. %v2  
When all deformation parameters are equal this is the undeformed PCM. When two deformation parameters are equal, for instance $\alpha_1=\alpha_2$, this is the trigonometric deformation of the PCM, related to the (unilateral) Yang-Baxter deformation (up to a closed B-field) with Drinfel'd-Jimbo deforming operator
\begin{equation}
    \mathcal R(\sigma_1) = \sigma_2\,, \qquad \mathcal R(\sigma_2) = -\sigma_1\,, \qquad \mathcal R(\sigma_3)=0\,, \qquad \Tr[\sigma_a \sigma_b] = 2 \delta_{ab}\,.
\end{equation}
The equations of motion of the elliptic model also take the form of a flatness equation for a Lax connection and the model is classically integrable. This type of elliptic deformation was also extended to models on $SU(N)$ for arbitrary $N$~\cite{Lacroix:2023qlz}. Contrary to the other deformations, it is not yet known how to generalise it to semi-symmetric spaces. In other words, while it is known how to construct an integrable elliptic deformation of the bosonic $AdS_3 \times S^3 \times T^4$ string, it is not yet clear how to construct an integrable elliptic deformation of the full superstring theory. We present recent advances on the subject in section \ref{sec:elliptic}. Similarly, it is not known how to construct elliptic deformations of e.g.~$AdS_5 \times S^5$, because these spaces can only be realised as cosets, not group manifolds. %v2

\subsection{Homogeneous and inhomogeneous Yang-Baxter deformations}
Let us now explore the difference between homogeneous and inhomogeneous Yang-Baxter deformations.

\textbf{Homogeneous YB deformations.}
Broadly speaking, homogeneous YB deformations modify the structure of a model in a less substantial way than their inhomogeneous counterparts.
For instance, the simplest non-trivial solutions of the classical YB equation with $c=0$ are abelian $r$-matrices, of the form
\begin{equation}
\label{eq:r-abelian}
    r = r^{jk} H_j \wedge H_k\,, \qquad [H_j,H_k]=0\,.
\end{equation}
These implement sequences of TsT transformations~\cite{Osten:2016dvf} discussed above. Another, less straightforward,  class of solutions is given by so-called ``Jordanian'' $r$-matrices~\cite{Kawaguchi:2014qwa,vanTongeren:2015soa}, built out of a Cartan element and a positive (or negative) root,
\begin{equation}
\label{eq:r-jordanian}
    r = H \wedge E^+\,, \qquad [H,E^+] \sim E^+\,.
\end{equation}
All Jordanian $r$-matrices on $\alg{g} = \alg{psu}(2,2|4)$, relevant for strings on $AdS_5 \times S^5$, were constructed and classified in~\cite{Borsato:2022ubq}. Jordanian $r$-matrices on $\alg{g} = \alg{psu}(1,1|2) \oplus \alg{psu}(1,1|2)$, relevant for strings on $AdS_3 \times S^3 \times T^4$, can be obtained by truncation. 
Homogeneous YB deformations can be interpreted as a twist of the boundary conditions on the fields in the model~\cite{Borsato:2021fuy}, just like for TsT transformations~\cite{Frolov:2005dj,Alday:2005ww}. In other words, the deformed model (in which the fields have periodic boundary conditions) can equivalently be formulated as an undeformed model (no deforming operator) with fields having twisted boundary conditions. 
From such a formulation it follows that in the deformed theory one has a twisted version of the symmetry algebra~\cite{vanTongeren:2015uha}. The commutation relations are left unchanged, but the coproduct gets dressed by a Drinfel'd twist,
\begin{equation}
    \Delta_\mathcal F(X) = \mathcal F^{-1} \Delta(X) \mathcal F\,, \qquad \forall X \in \alg{g}\,,
\end{equation}
with the cocycle condition
\begin{equation}
    (\mathcal F \otimes 1) (\Delta \otimes 1) \mathcal F = (1\otimes \mathcal F)(1 \otimes \Delta) \mathcal F\,.
\end{equation}
Further assuming that we have a deformation
\begin{equation}
    \mathcal F = 1 \otimes 1 + \alpha \mathcal F^{(1)} + \mathcal O(\alpha^2)\,,
\end{equation}
for any Drinfel'd twist one can construct a solution to the classical YB deformation through
\begin{equation}
    r_{12} = \frac{1}{2} (\mathcal F_{12}^{(1)} - \mathcal F_{21}^{(1)})\,.
\end{equation}
Conversely, for any solution to the classical YB deformation there exists an associated Drinfel'd twist, but there is no general recipe for its construction. For the abelian $r$-matrix \eqref{eq:r-abelian} the associated Drinfel'd twist is
\begin{equation}
    \mathcal F = e^{-i r}\,.
\end{equation} 
This recovers the simple twist of boundary conditions which we expect from the case of TsT~\cite{Frolov:2005dj,Alday:2005ww}.

Assuming that the smaller symmetry algebra $\mathcal A$ of the light-cone gauge fixed theory inherits this twist, which is the case so long as the twist does not act on lightcone directions~\cite{vanTongeren:2021jhh} we can derive the deformed algebra and S~matrix. If we call $(\Delta_{\mathcal F}, S_{\mathcal F})$ and $(\Delta, S)$ the coproduct and the S-matrix of the YB and undeformed models respectively, then
\begin{equation}
    \Delta_{\mathcal F}(X) = \mathcal F^{-1} \Delta(X) \mathcal F, \qquad S_\mathcal F = \mathcal F_{21}^{-1} S_{12} \mathcal F_{12}\,,
\end{equation}
where $X$ is any element of $\mathcal A$.
%The story is however more involved when the $r$-matrix involves the light-cone directions. 
%For instance, we already mentioned that TsT transformations involving light-cone directions leads to current-current deformations in the light-cone gauge fixed theory.
Finally, let us mention that homogenous YB deformations of the  NSNS theory (the WZW model) correspond~\cite{Borsato:2018spz} to marginal current-current deformations~\cite{Forste:2003km}.

\textbf{Inhomogeneous YB deformations.} These are deformations constructed using a deforming operator satisfying the modified classical YB equation, with $c \neq 0$.
Inhomogeneous YB deformations are also called $\eta$-deformations in the literature. A special solution to the modified YB equation \eqref{eq:cYBE} with $c \neq 0$ is the Drinfel'd-Jimbo $r$-matrix. Its definition relies on a choice of Cartan-Weyl basis for the complexified Lie algebra $\alg{g}^{\mathbb{C}} = \{H_j, E^+_k,E^-_k\}$, with $H_j$ the Cartan generators, $E^+_k$ the positive roots and $E^-_k$ the negative roots. The action of the deforming operator $\mathcal R$ is then
\begin{equation}
    \mathcal R(H_j)=0\,, \qquad \mathcal R(E^+_k)=-c E^+_k\,, \qquad \mathcal R(E^-_k)=+c E^-_k\,.
\end{equation}
For the theories we are interested in, one usually does not have a complex Lie algebra, but impose reality conditions. It is important to check that the action of the operator $\mathcal R$ preserves the chosen real form. For compact real forms, for instance $\alg{g}=\alg{su}(2)$, only solutions with 
$c \in i\mathbb{R}$ %v2
are possible. For non-compact real forms, for instance $\alg{g}=\alg{su}(1,1)$, one can have several different solutions for both
$c\in i \mathbb{R}$ and $c\in \mathbb{R}$. %v2 
Because the superstring theories we are interested in gather these two different Lie algebras into superalgebras, for instance $\alg{psu}(1,1|2)$, only Drinfel'd-Jimbo solutions with 
$c \in i \mathbb{R}$ %v2
will lead to real backgrounds. 
Without loss of generality one can restrict to $c=i$. %v2

As we have seen in section \ref{sec:symmetries-smatrix}, it is important to analyse the fate of the symmetries under the deformation to be able to bootstrap an exact S-matrix. The YB deformation with Drinfel'd-Jimbo operator $\mathcal R$ only manifestly preserves the Cartan $U(1)$ symmetries. There is however a hidden symmetry algebra arising from the integrability of the model. This can be extracted from the Lax connection and expanding the associated monodromy matrix around special values of the spectral parameter~\cite{Delduc:2014kha,Kawaguchi:2011pf}. 
Drinfel'd-Jimbo deformations promote the symmetry algebra of the original theory to a quantum group $\mathcal U_{q}(\alg{g})$, defined as 
\begin{equation}
    [H_j, E^\pm_k] = \pm A_{jk} E^\pm_k\,, \qquad [E^+_j,E^-_k] = \delta_{jk} \frac{q^{H_k} - q^{-H_k}}{q-q^{-1}}\,.
\end{equation}
Carefully taking the $q \rightarrow 1$ limit gives back the undeformed algebra $\alg{g}$ with Cartan matrix $A$. Note that the quantum group is a deformation of the universal enveloping algebra, where objects such as $q^{H_k}$ are well-defined. There are various coassociative coproducts under which the above algebra becomes a Hopf algebra, among which 
\begin{equation}
\begin{aligned}
    \Delta(H_j) &= H_j \otimes 1 + 1 \otimes H_j\,, \\
    \Delta(E^+_j) &= E^+_j \otimes 1 + q^{H_j} \otimes E^+_j\,, \\
    \Delta(E^-_j) &= E^-_j \otimes q^{-H_j} + 1 \otimes E^-_j\,.
\end{aligned}
\end{equation}
At least semi-classically, the deformation parameter $q \in \mathbb{R}$ is related to the deformation parameter $\kappa$ and the tension $T$ in the action through the relation
\begin{equation}
    q= e^{-\kappa/T}\,.
\end{equation}
For the YB deformation \eqref{eq:YBdef} we thus have a $\alg{g}_{\L}\oplus \mathcal U_{q_{\R}}(\alg{g}_{\R})$ symmetry, while for the YB deformation \eqref{eq:YBdef2} we have a $\mathcal U_{q_{\L}}(\alg{g}_{\L}) \oplus \alg{g}_{\R}$ symmetry. More generally, for the bi-YB deformation~\eqref{eq:biYB}, the symmetry algebra is promoted to the quantum group $\mathcal U_{q_{\L}}(\alg{g}_{\L}) \oplus \mathcal U_{q_{\R}}(\alg{g}_{\R})$, with relation
\begin{equation}
    q_{\L}= e^{-\kappa_{\L}/T}\,, \qquad q_{\R}=e^{-\kappa_{\R}/T}\,.
\end{equation} 
For the unilateral YB deformations, it was shown that, at least for the bosonic model, this $q$-deformed algebra structure is still present when adding the WZ term~\cite{Kawaguchi:2013gma} (though the identification of $q$ with the deformation parameters is more involved). Because the deforming operator annihilates the Cartan directions used in the light-cone gauge fixing, also the light-cone algebra $\mathcal A$ inherits the $q$-deformation. This $q$-deformed algebra can then be used to bootstrap an exact $q$-deformed S-matrix.

While for a compact bosonic Lie algebra all Cartan-Weyl bases are essentially equivalent and there is a unique Drinfel'd-Jimbo solution with $c=i$, this is not the case for non-compact algebras, for which one may construct different Drinfel'd-Jimbo $r$-matrices, both for $c=i$ or $c=1$, see for instance the discussion in~\cite{Hoare:2016ibq} for the non-compact $\alg{su}(2,2)$ algebra. For superalgebras, there is another layer of complications, with different Cartan-Weyl bases characterised by different Dynkin diagrams. %v2
In that case, different choices of Cartan-Weyl bases lead to coproducts and S-matrices that are related through Drinfel'd twists,
\begin{equation}
    S' = \mathcal F_{21}^{-1} S \mathcal F_{12}\,,
\end{equation}
with
\begin{equation}
    \mathcal F = \exp \left[ -(q-q^{-1}) E^- \otimes E^+\right]\,,
\end{equation}
where $E^\pm$ are the simple roots involved in the (generalised) Weyl reflection allowing one to go from one Cartan-Weyl basis to the other.

\textbf{Bi-$\lambda$ deformations.} The conserved charges of the inhomogeneous YB deformation obey a $q$-deformed algebra with real parameter $q$. There exists another family of quantum group deformations, called $\lambda$-deformations, for which $q$ is instead a root of unity. These were initially constructed for the PCM~\cite{Sfetsos:2013wia}, then extended to symmetric~\cite{Hollowood:2014rla,Sfetsos:2014cea} and semi-symmetric spaces~\cite{Hollowood:2014qma}. For the PCM
on $G$, %v2
the $\lambda$-deformation interpolates between a WZW model on $G$ (at $\lambda=0$) and the non-abelian T-dual of the PCM on $G$ (at $\lambda \rightarrow 1$; this limit also involves a scaling of the coordinates). Similarly, for cosets $G/H$, the $\lambda$-deformation interpolates between the gauged WZW model and the non-abelian T-dual of the coset model. 
When the symmetry algebra $\alg{g}=\alg{g}_{\L} \oplus \alg{g}_{\R}$, also an integrable two-parameter $\lambda$ deformation can be constructed~\cite{Sfetsos:2017sep,Hoare:2022vnw}. The $\eta$ and $\lambda$ deformations (including their two-parameter generalisations)  are related by Poisson-Lie duality, supplemented by an analytic continuation in the deformation parameters and in the fields~\cite{Sfetsos:2015nya,Hoare:2014pna,Hoare:2015gda,Vicedo:2015pna}. 
Poisson-Lie duality is a generalisation of non-abelian T-duality of string theory to backgrounds that do not necessarily have manifest isometries, but have an underlying Poisson-Lie symmetry. In fact, one can play with this duality to construct many different hybrid $\eta-\lambda$ deformations, by taking the Poisson-Lie dual with respect to a subalgebra rather than the full symmetry algebra~\cite{Hoare:2017ukq}. For superstrings a particularly interesting case is when the Poisson-Lie duality is performed with respect to the bosonic subalgebra~\cite{Hoare:2018ebg}. The model then features the same metric and B-field as the $\lambda$ deformation (obtained through Poisson-Lie duality with respect to the full superalgebra), but different dilaton and RR fluxes. For the purpose of this review we shall call this model 
the $\lambda_B$ deformation. %v2 

\begin{figure}
\begin{tikzpicture}
    \node[rectangle, draw=black, minimum width=3 cm] (A) at (0,0) {$AdS_3$ strings};
    \node[rectangle, draw=black, minimum width=3 cm] (eta1) at (0,-1)  {unilateral YB def.};
    \node[rectangle, draw=black, minimum width=3 cm] (eta2) at (0,-2) {bi-YB def.};
    \node[rectangle, draw=black, minimum width=3 cm] (etaWZ) at (0,-3) {bi-YB+WZ def.};
    \node[rectangle, draw=black, minimum width=3 cm] (lam1) at (4,-1) {unilateral $\lambda$ def.};
    \node[rectangle, draw=black, minimum width=3 cm] (lam2) at (4,-2) {bi-$\lambda$ def.};
    \node[rectangle, draw=black, minimum width=3 cm] (ell) at (-4,-1) {elliptic def.};
    \draw[->] (A)--(eta1);
    \draw[->] (eta1)--(eta2);
    \draw[->] (eta2)--(etaWZ);
    \draw[->] (eta1)--(ell);
    \draw[->] (lam1)--(lam2);
    \draw[<->, dotted] (eta1)--(lam1);
    \draw[<->, dotted] (eta2)--(lam2);
\end{tikzpicture}
\caption{The landscape of integrable deformations of $AdS_3 \times S^3 \times T^4$ strings. The (undeformed) theory preserves 16 SUSYs, the unilateral deformation preserves 8 SUSYs and the bi-YB deformation breaks all the supersymmetries.  Dotted arrows represent Poisson-Lie dualities. Several hybrid YB-$\lambda$ models can be constructed. }
\label{fig:landscape}
\end{figure}
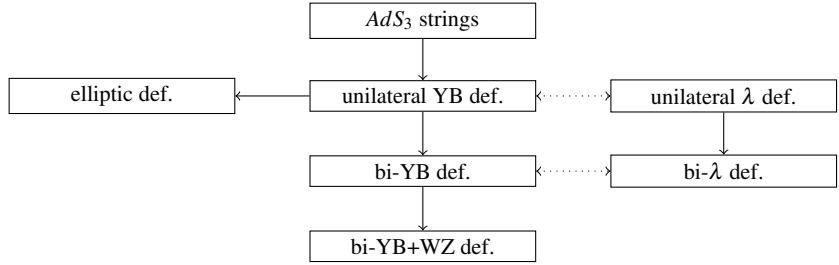

\subsection{Deformed backgrounds}
The deformations modify the background in which the strings propagate. The metric, dilaton, NSNS and RR fluxes depend explicitly on the deformation parameters.  For the YB and $\lambda$ deformations of semi-symmetric space sigma models (without WZ term), a formula expressing the NSNS and RR fluxes in terms of the deforming operators can be found in~\cite{Borsato:2016ose}. For the bi-YB deformation a similar formula was derived~\cite{Seibold:2019dvf}. The deformations were initially constructed in such a way as to preserve the classical integrability of the model. There is no guarantee that the deformed backgrounds satisfy the supergravity equations of motion and that the deformed models correspond to Weyl-invariant string theories. 
In~\cite{Arutyunov:2013ega,Arutyunov:2015qva} the background of a Drinfel'd-Jimbo deformation of the $AdS_5\times S^5$ superstring was extracted. The deformed metric is supported by a RR one-form, three-form and five-form flux, which do not satisfy the type IIB supergravity equations of motion.  Instead, they satisfy a set of ``generalised'' supergravity equations~\cite{Arutyunov:2015mqj}, arising from the one-loop scale invariance and the fermionic kappa-symmetry of the deformed theories~\cite{Arutyunov:2015mqj,Wulff:2016tju}.  For Yang-Baxter models, a sufficient condition (though not necessary, see~\cite{Wulff:2018aku} for examples) for the deformed background to solve the supergravity equations of motion 
%(after appropriately embedding the six-dimensional model into the ten dimensions of the superstring) 
is that the deforming operator $\mathcal R$ satisfies the unimodularity condition~\cite{Borsato:2016ose}
\begin{equation}
\label{eq:unimodularity}
\begin{aligned}
    &\hat{K}^{ab} \STr[[T_a,\mathcal R(T_b)] X]=0\,, \forall X \in \alg{g}\,, \\
    &K_{ab} = \STr[T_a T_b]\,, \qquad \hat{K}^{ab} K_{bc} = \delta^a_c\,.
    \end{aligned}
\end{equation}
Abelian $r$-matrices automatically satisfy this constraint and indeed TsT transformations map supergravity backgrounds to supergravity backgrounds.
Drinfel'd-Jimbo $r$-matrices satisfying this condition were found for the first time in~\cite{Hoare:2018ngg}, and the supergravity backgrounds for $\eta$-deformed $AdS_2 \times S^2 \times T^6$ and $AdS_5 \times S^5$ superstrings were constructed there. In both cases the deformed metric is supported by a RR three-form and a RR five-form flux, together with a non-trivial dilaton. The deformation breaks all supersymmetries. 
Let us present in more detail the $\eta$-deformation of the $AdS_2 \times S^2 \times T^6$ superstring. The undeformed theory has a $\alg{psu}(1,1|2)$ isometry superalgebra, whose complexified superalgebra $\alg{sl}(2|2)$ admits three different Dynkin diagrams, 
\begin{equation}
    D_1 = \Circle - \otimes - \Circle\,, \qquad D_2 = \otimes - \Circle - \otimes\,, \qquad D_3 = \otimes - \otimes -\otimes\,.
\end{equation}
Here we adopt the standard notation where $\Circle$ represents a bosonic simple root, while $\otimes$ represents a fermionic simple root. The ``distinguished'' Dynkin diagram $D_1$ has the bosonic subalgebra $\alg{su}(1,1) \oplus \alg{su}(2)$ as a sub-Dynkin diagram. The three Dynkin diagrams are related through generalised Weyl reflections.
One can then construct different $r$-matrices, based on these three different choices of Dynkin diagrams. While all such $r$-matrices solve the modified classical YB equation, only the ones associated with the fully fermionic Dynkin diagram $D_3$ satisfy the unimodularity condition~\eqref{eq:unimodularity} and give rise to a background solving the supergravity equations of motion~\cite{Hoare:2018ngg}. Even after specialising to a specific Dynkin diagram one can in general still construct different $r$-matrices, but the associated backgrounds will be related by (possibily complex) field redefinitions, and the associated S-matrices will be related by a one-particle change of basis~\cite{Hoare:2016ibq}. Moreover, all $r$-matrices give rise to the same metric and B-field (again, up to possible complex field redefinitions). 
The unimodular $r$-matrix chosen in~\cite{Hoare:2018ngg} leads to an $\eta$-deformed $AdS_2 \times S^2 \times T^6$ with deformed metric
\begin{equation}
\begin{aligned}
ds^2 &= ds_a^2 + ds_b^2 + \sum_{j=1}^4 dx_j^2\,, \\
 ds_a^2 &= \frac{1}{1-\kappa^2 \rho^2} \left( -(1+\rho^2) \de t^2 + \frac{\de \rho^2}{1+\rho^2} \right)\,, \\
  ds_b^2 &= \frac{1}{1+\kappa^2 r^2} \left( (1-r^2) \de \varphi^2 + \frac{\de r^2}{1-r^2} \right)\,,
    \end{aligned}
\end{equation}
the $B$-field is a closed two-form, the dilaton is given by
\begin{equation}
    \Phi = -\frac{1}{2} \log \frac{(1-\kappa^2 \rho^2)(1+\kappa^2 r^2)}{1-\kappa^2 (\rho^2 - r^2 - \rho^2 r^2)}\,.
\end{equation}
The Ricci scalar and the dilaton diverge as $\rho \rightarrow \kappa^{-1}$. This singularity prevents a string situated at $\rho=0$ to reach the $AdS_2$ conformal boundary situated at $\rho = \infty$.  A similar singularity appears for the $\eta$-deformed $AdS_5 \times S^5$ superstring. This makes it difficult to understand these deformations in the context of holography.
This is where the deformations of $AdS_3 \times S^3 \times T^4$ stand out.
In~\cite{Seibold:2019dvf}, two supergravity backgrounds (based on two different unimodular Drinfel'd-Jimbo $r$-matrices) were constructed for the bi-Yang-Baxter deformation of $AdS_3 \times S^3 \times T^4$. For generic deformation parameters $\kappa_{\L}$ and $\kappa_{\R}$ these have non-trivial dilatons and feature both RR 3-forms and 5-form fluxes. The cases with $\kappa_{\L}=0$ or $\kappa_{\R}=0$ are however special, as reviewed below. In particular, they have a trivial dilaton and smooth geometry, which makes them more promising candidates to understand an example of holography between models with $q$-deformed symmetries.  In all these cases, the only Drinfel'd-Jimbo $r$-matrices satisfying the unimodularity conditions are the ones constructed using the fully fermionic Dynkin diagram (with all simple roots being fermionic generators) of the symmetry algebras. Also unimodular Jordanian deformations can be constructed~\cite{vanTongeren:2019dlq}.

Note that the generalised supergravity equations require the presence of isometries in the background. The $\lambda$-deformation does not present such isometries, and the $\lambda$-deformed backgrounds solve the standard supergravity equations. For instance, the supergravity background of the $\lambda$-deformed $AdS_2 \times S^2 \times T^6$ superstring was obtained in~\cite{Borsato:2016zcf}. It features a non-trivial dilaton and a RR 5-form flux. In~\cite{Sfetsos:2014cea} another embedding into supergravity of the $\lambda$-deformed $AdS_2 \times S^2 \times T^6$ metric, with again a non-trivial dilaton and a RR 5-form flux, was proposed. It was later shown in~\cite{Hoare:2018ebg} that this background is the one that arises when taking the Poisson-Lie dual of the $\eta$-deformation with respect to the bosonic subalgebra (what we here call the $\lambda_B$ model). This is an example where a given metric can be embedded into supergravity with two different sets of dilatons and RR fluxes. 
The metric of the $\lambda$-deformed $AdS_3 \times S^3 \times T^4$ string was obtained in~\cite{Sfetsos:2014cea}, where an embedding into supergravity was also provided. These proposed fluxes are also expected to correspond to the $\lambda_B$ model. The RR fluxes of the full Poisson-Lie dual model (what is traditionally called $\lambda$ deformation) is however not yet known. Interestingly, a supergravity embedding of the unilateral $\lambda$ deformation, preserving 8 supersymmetries, was found in~\cite{Itsios:2023kma} for  $AdS_3 \times S^3 \times T^4$ and in~\cite{Itsios:2023uae} for $AdS_3 \times S^3 \times S^3 \times S^1$. These backgrounds were however derived independently from a semi-symmetric space sigma model realisation, and therefore the fate of integrability in these models is an open problem. To answer the question of integrability it would be useful to understand their connection to the supersymmetric unilateral $\lambda$-deformations of~\cite{Hoare:2022vnw} (or their $\lambda_B$ counterparts), which are known to be integrable and preserve 8 supersymmetries -- but whose explicit backgrounds remain unknown.

\section{The bi-Yang-Baxter+WZ deformation}
\label{sec:biYB-WZ}
To illustrate the effect of the deformation on the background and on the worldsheet S-matrix, let us focus on the bi-YB+WZ deformation of the $AdS_3 \times S^3 \times T^4$ string. We introduce  $\{L_1, L_2, L_3\}$ the three generators of $SL(2;\mathbb{R})$ and $\{J_1,J_2,J_3\}$ the three generators of $SU(2)$, with matrix realisation
\begin{equation}
\begin{aligned}
    L_1 &= \sigma_1 \oplus 0\,, &\qquad L_2 &= i \sigma_2 \oplus 0\,, &\qquad L_3 &= \sigma_3 \oplus 0\,, \\
    J_1 &= 0 \oplus i \sigma_1\,, &\qquad J_2 &= 0 \oplus i \sigma_2\,, &\qquad J_3 &= 0 \oplus i\sigma_3\,.
    \end{aligned}
\end{equation} 
The deformation is constructed using the Drinfel'd-Jimbo deforming operator with the following action on these generators,
\begin{equation}
\begin{aligned}
\mathcal R(L_1) &= i L_2\,, &\qquad \mathcal R(L_2) &=i L_1\,, &\qquad \mathcal R(L_3) &=0\,, \\
    \mathcal R(J_1) &= J_2\,, &\qquad \mathcal R(J_2)&=-J_1\,, &\qquad \mathcal R(J_3)&=0\,.
    \end{aligned}
\end{equation}
It is a straightforward exercise to check that this deforming operator satisfies the modified classical YB equation. For the extension to the full $\alg{psu}(1,1|2)_{\L} \oplus \alg{psu}(1,1|2)_{\R}$ algebra we refer the reader to~\cite{Seibold:2019dvf}. 

\subsection{Deformed background} 
To obtain the background fields we use the parametrisation
\begin{equation}
    g = e^{\frac{i}{2} (t+\psi)L_3} e^{\text{arcsinh}(\rho) L_1} e^{\frac{i}{2} (t-\psi) L_3} e^{\frac{1}{2} (\varphi+\phi)J_3} e^{\text{arcsin}(r) J_1} e^{\frac{1}{2} (\varphi-\phi) J_3}\,,
\end{equation}
and add the four torus directions by hand.
The deformed metric reads
\begin{align}
\label{eq:biWZmetric}
\de s^2&= \de s_a^2 + \de s_b^2 + \sum_{j=1}^4 \extder x_j^2, \\
\nonumber
\de s_a^2 &= \frac{1}{F_\rho} \left[- (1+\rho^2) \extder t^2 + \frac{\extder \rho^2}{1+\rho^2} + \rho^2 \extder \psi^2 - \big( \kappa_- (1+\rho^2) \extder t - \kappa_+ \rho^2 \extder \psi + \tilde{q} \rho \extder \rho\big)^2 \right], \\
\de s_b^2 &= \frac{1}{F_r} \left[(1-r^2) \extder \varphi^2 + \frac{\extder r^2}{1-r^2} + r^2 \extder \phi^2 + \big( \kappa_- (1-r^2) \extder \varphi + \kappa_+ r^2 \extder \phi - \tilde{q} r \extder r  \big)^2 \right],
\nonumber
\end{align}
and the deformed B-field is
\begin{align}
B &= -\frac{q}{2F_\rho} \rho^2 \left( 2 + (1+\rho^2)\tilde{q}^2 + (1+\rho^2)\kappa_-^2+(1-\rho^2)\kappa_+^2 \right) \extder t \wedge \extder \psi \nln
& -\frac{q}{2F_r} r^2 \left( 2 + (1-r^2)\tilde{q}^2 + (1-r^2)\kappa_-^2+(1+r^2)\kappa_+^2 \right) \extder \varphi \wedge \extder \phi\,,
\end{align}
where
\begin{equation}
    \begin{aligned}
        F_\rho&= 1+\kappa_-^2(1+\rho^2)-\kappa_+^2\rho^2-\tilde{q}^2 \rho^2(1+\rho^2)\,, \\
        F_r &= 1+\kappa_-^2(1-r^2)+\kappa_+^2r^2+\tilde{q}^2 r^2 (1-r^2)\,,
    \end{aligned}
\end{equation}
and one has the relations
\begin{equation}
     q\sqrt{(\tilde{q}^2 + \kappa_-^2+\kappa_+^2)^2 + 4 (\tilde{q}^2 - \kappa_+^2 \kappa_-^2)} -  2\tilde{q}=0\,, \qquad    \kappa_\pm = \kappa_{\L} \pm \kappa_{\R}\,.
\end{equation}
For convenience, we subtracted a closed contribution to the B-field. 
The background features four $U(1)$ isometries, realised as shifts in $t,\psi,\varphi,\phi$. As in the undeformed case, one may then pick $t,\varphi$ to fix uniform light-cone gauge.   
The perturbative S-matrix was computed to tree-level in~\cite{Bocconcello:2020qkt}. At leading order in the string tension we get the quadratic light-cone Hamiltonian density
\begin{equation}
\begin{aligned}
    \mathcal H_2 &= P_z P_{\bar{z}} + |\acute{z}|^2 + m^2 |z|^2 + i \kappa_+ \kappa_- (z P_{\bar{z}} - \bar{z} P_z) - i \lambda (\bar{z} \acute{z}-z \acute{\bar{z}} ) \\
    &+ P_y P_{\bar{y}} + |\acute{y}|^2 + m^2 |y|^2 + i \kappa_+ \kappa_- (y P_{\bar{y}} - \bar{y} P_y) - i \lambda (\bar{y} \acute{y} - y \acute{\bar{y}}) \\
    &+ P_u P_{\bar{u}} + P_{v} P_{\bar{v}} + |\acute{u}|^2 + |\acute{v}|^2\,,
    \end{aligned}
\end{equation}
with
\begin{equation}
    m^2 = \tilde{q}^2 + (1+\kappa_+^2)(1+\kappa_-^2)\,, \qquad \lambda = \frac{q}{2} (2  + \tilde{q}^2 +\kappa_-^2+\kappa_+^2)\,.
\end{equation}
The corresponding free Hamiltonian reads
\begin{equation}
    \mathbf{H} = \int_{-\infty}^{+\infty} \extder p \sum_{j=1}^8 \omega_j(p) a_j^\dagger(p) a_j(p) \,,
\end{equation}
with dispersion relation
\begin{equation}
    \omega_j(p)= \mu_j \kappa_+ \kappa_- + \sqrt{p^2 + 2 \mu_j \lambda p + (\mu_j m)^2}\,,
\end{equation}
and quantum numbers
\begin{equation}
\begin{aligned}
   \mu_j=(+1,-1,+1,-1,0,0,0,0)\,.
   \end{aligned}
\end{equation}
Let us discuss some special cases for the deformation parameters.
\begin{itemize}
    \item When $\kappa_+=\kappa_-=0$, $\tilde{q}=0$ but $q$ finite, this reproduces the mixed flux background discussed in section \ref{sec:lcgauge}.
    In particular, one has $m^2 =1$ and $\lambda=q$. The dispersion relation then becomes the one in \eqref{eq:nearbmnH}.
    \item When $\kappa_+=\kappa_-=\kappa$, $\tilde{q}=0$  but $q$ finite, this reproduces the unilateral mixed-flux deformation discussed in~\cite{Hoare:2022asa}.  Note that $F_\rho=F_r=1+\kappa^2$, and the background does not present singularities. The metric is the one of warped $AdS_3$ and squashed sphere. The (integrable) embedding into supergravity is known, with in particular a trivial dilaton,
    \begin{equation}
    \begin{aligned}
         \Phi &=0\,, \qquad H_3 =  \frac{\sqrt{1-\hat{q}^2}}{\sqrt{1+\kappa^2}} d \hat{B}\,, \qquad \\
        F_3 &= - \frac{1}{\sqrt{1+\kappa^2}\sqrt{\hat{q}^2+\kappa^2}} \left[\hat{q}^2 d \hat{B}+\kappa^2 d \check{B}\right]\,, \\
        F_5 &= - \frac{1}{\sqrt{1+\kappa^2}\sqrt{\hat{q}^2+\kappa^2}}  \hat{q} \kappa \left[ d \hat{B} - d \check{B} \right]\wedge J_2\,,
        \end{aligned}
    \end{equation}
    where
    \begin{equation}
    \begin{aligned}
        \hat{B} &= \rho^2 \de t \wedge \de \psi + r^2 \de \varphi \wedge d \phi\,, \\
        \check{B} &= \left[(1+\rho^2) \de t - \rho^2  \de \psi\right] \wedge \left[ (1-r^2) \de \varphi + r^2 \de \phi\right]\,, \\
        J_2 &= \de x_1 \wedge \de x_2 - \de x_3 \wedge \de x_4\,,
        \end{aligned}
    \end{equation}
    and
    \begin{equation}
        \frac{k}{2 \pi} = T q\,, \qquad \hat{q}= \sqrt{1-q^2 (1+\kappa^2)}\,.
    \end{equation}
    In fact, the deformed supergravity background can also be obtained from the undeformed one through a sequence of T-duality and S-duality rotations.  The model is invariant under one copy of the $\alg{psu}(1,1|2)$ symmetry algebra and the  background preserves 8 supersymmetries. One can also define a related model through analytic continuation $\tilde{\kappa} = i \kappa$, which interpolates between $AdS_3 \times S^3 \times T^4$ (for $\tilde{\kappa}=0$) and $AdS_2 \times S^2 \times T^6$ (for $\tilde{\kappa}=1$), with real fluxes. 
    \item When $\tilde{q}=q=0$ this reproduces the bi-YB deformation. There are two possible embeddings into supergravity, derived from the supercoset construction in~\cite{Seibold:2019dvf}. The first  supergravity background takes the form
   \begin{equation}
       \begin{aligned}
\label{eq:backbi1}
&e^{2 \Phi} =  \frac{P^2}{F_\rho F_r}\,, \qquad 
P = 1 +\kappa_+^2 -(1+\rho^2)(1-r^2)(\kappa_+^2-\kappa_-^2)\,,  \\
&H_3 =0\,, \\
&F_3 = \de C_2\,, \qquad C_2 = - \sqrt{\frac{1+\kappa_+^2}{1+\kappa_-^2}} \frac{1}{P} \big[ \hat{B}  + \kappa_-^2 \check{B}\big]\,,\\
&F_5 = \de C_4\,, \qquad C_4 = - \sqrt{\frac{1+\kappa_+^2}{1+\kappa_-^2}} \frac{1}{P} \kappa_- \big[ \hat{B} - \check{B}\big]  \wedge J_2\,, 
\end{aligned}
\end{equation}
with
\begin{equation}
\begin{aligned}
 \hat{B} &= \rho^2 \de t \wedge \de  \psi + r^2 \de  \varphi \wedge \de  \phi\,, \\
    \check{B} &= \frac{1}{\kappa_+ \kappa_-} \left[\kappa_-(1+\rho^2) \de t - \kappa_+ \rho^2  \de \psi\right] \wedge \left[ \kappa_-(1-r^2) \de \varphi + \kappa_+ r^2 \de \phi\right]\,, \\
    J_2 &= \de x_1 \wedge \de x_2 - \de x_3 \wedge \de x_4\,.
    \end{aligned}
\end{equation}
The second supergravity background can be obtained from the above through
\begin{equation}
    \rho \rightarrow i \sqrt{1-\rho^2}\,, \qquad r \rightarrow \sqrt{1-r^2}\,, \quad t \leftrightarrow \psi\,, \quad \varphi \leftrightarrow \phi\,, \quad \kappa_+ \leftrightarrow \kappa_-\,,
\end{equation}
which leaves the metric and $\check{B}$ invariant and keeps the dilaton and fluxes real. 
    The deformation breaks all supersymmetries. There is however a hidden quantum group symmetry. The exact S-matrix (without dressing phases) is known~\cite{Hoare:2014oua} and reviewed below.
    \item For generic deformation parameters, both the embedding into supergravity and the exact S-matrix are not yet known.  
\end{itemize}

\subsection{Deformed symmetries and S-matrix} 
The bi-YB+WZ deformation is a classically integrable model. As for the undeformed setup, one can make the assumption that integrability carries over to the quantum theory. An important step is then to use the symmetries of the light-cone gauge fixed theory to bootstrap an exact 2-to-2 worldsheet S-matrix. 
The conjecture of~\cite{Hoare:2014oua} is that under the bi-YB deformation the light-cone symmetries are promoted to quantum groups, obeying the relations
\begin{equation}
    \{\mathbf{Q}_{\L}^a,\mathbf{S}_{\L}^b\} = \epsilon^{ab}[\mathbf{H}_{\L}]_{q_{\L}}\,, \qquad \{\mathbf{Q}_{\R}^a,\mathbf{S}_{\R}^b\} = \epsilon^{ba}[\mathbf{H}_{\R}]_{q_{\R}}\,, 
\end{equation}
with
\begin{equation}
    \mathbf{H}_{\L} =\frac{1}{2}(\mathbf{H}+\mathbf{M})\,, \qquad \mathbf{H}_{\R} =\frac{1}{2}(\mathbf{H}-\mathbf{M})\,, \qquad [x]_q = \frac{q^x-q^{-x}}{q-q^{-1}}\,.
\end{equation}
This quantum group deformation also admits a central extension,
\begin{equation}
\begin{aligned}
    &\{ \mathbf{Q}_{\L}^a, \mathbf{Q}_{\R}^b\} = \epsilon^{ab}\mathbf{P}\,, &\qquad &\{ \mathbf{S}_{\L}^a, \mathbf{S}_{\R}^b\} = \epsilon^{ba}\mathbf{K}\,.
    \end{aligned}
\end{equation}
Let us mention that the superalgebra $\alg{psu}(1|1)$ only admits one Dynkin diagram $\otimes$ (with one single fermionic simple root), and hence there is essentially only one choice of quantum group deformation. The deformed coproduct, including the braiding $\mathbf{U}$, reads (for $\star=\{\L,\R\}$ and dropping the $a$ indices),
\begin{align}
&\Delta[\mathbf{H}_\star] = \mathbf{H}_\star \otimes 1 + 1 \otimes \mathbf{H}_\star\,, \qquad &&\Delta[\mathbf{U}] = \mathbf{U} \otimes \mathbf{U}\,, \\
    &\Delta[\mathbf{Q}_{\star}] = \mathbf{Q}_\star \otimes 1 + q_\star^{\mathbf{H}_\star} \mathbf{U}^{+1} \otimes \mathbf{Q}_\star\,, \qquad &&\Delta[\mathbf{S}_{\star}] = \mathbf{S}_\star \otimes q_\star^{-\mathbf{H}_\star} + \mathbf{U}^{-1} \otimes \mathbf{S}_\star\,.
\nonumber
\end{align}
Requiring co-commutativity of the central elements,
\begin{equation}
    \Delta[\mathbf{P}] = \tau \circ \Delta[\mathbf{P}]\,, \qquad \Delta[\mathbf{K}] = \tau \circ \Delta[\mathbf{K}]\,, \qquad \tau(x \otimes y) = y \otimes x\,,
\end{equation}
then fixes
\begin{equation}
        \mathbf{P}  = \frac{i\nu}{2} (q_{\L}^{\mathbf{H}_{\L}}q_{\R}^{\mathbf{H}_{\R}} \mathbf{U}^2-1)\,, \qquad
    \mathbf{K} = \frac{i\nu}{2} (q_{\L}^{\mathbf{H}_{\L}}q_{\R}^{\mathbf{H}_{\R}}- \mathbf{U}^{-2})\,.
\end{equation}
Note that $\mathbf{P}^\dagger \neq \mathbf{K}$, but this is consistent with the fact that in the $q$-deformed algebra the reality conditions also get dressed by factors of $q^{\mathbf{H}_{\L,\R}}$.~\footnote{One could have chosen an alternative coproduct compatible with $\mathbf{Q}^\dagger = \mathbf{S}$ and $\mathbf{P}^\dagger = \mathbf{K}$ instead.}
The representation theory of quantum groups with $q \in \mathbb{R}$ is similar to the one of the undeformed Lie algebra. Short representations are also four dimensional, and obey the $q$-deformed constraint
\begin{equation}
    [\mathbf{H}_{\L}]_{q_{\L}} [\mathbf{H}_{\R}]_{q_{\R}} = \mathbf{P}\, \mathbf{K}\,.
\end{equation}
To compare with the perturbative results, in the case of the bi-YB deformation (without WZ term) one assumes  the following action on single-particle states,
\begin{equation}
\label{eq:identification1}
    \mathbf{M} \left|p,\mu\right> = \mu \left|p,\mu\right>\,,  \qquad \mathbf{U} \left|p,\mu\right> = e^{ip/2} \left|p,\mu\right>\,,
\end{equation}
as well as the following semi-classical relation between the parameters
\begin{equation}
\label{eq:identification2}
    \nu = \frac{h}{\sqrt{1+\kappa_+^2}}\,, \qquad q_\star = e^{-\kappa_\star/h}\,.
\end{equation}
The two-particle S-matrix is required to satisfy
\begin{equation}
    \tau \circ\Delta[\mathbf{X}] S = S \Delta[\mathbf{X}]\,,
\end{equation}
for all elements $\mathbf{X}$ in the light-cone symmetry algebra.
The hidden quantum-group symmetry is still powerful enough to bootstrap the exact S-matrix~\cite{Hoare:2014oua}. This exact S-matrix, together with the identification \eqref{eq:identification1} and \eqref{eq:identification2} is in agreement with the perturbative results for the bi-YB deformation without WZ term~\cite{Seibold:2021lju}. The S-matrices associated with the two different supergravity backgrounds are related through a one-particle change of basis. 
 For the model with WZ term, the identification of parameters that reproduces the perturbative dispersion relation and perturbative S-matrix in the large tension limit is not yet known, and expected to be more complicated than the naive replacement $\mu \rightarrow \mu + \frac{k}{2 \pi} p$ which works for the undeformed mixed flux theory~\cite{Bocconcello:2020qkt}.

\section{Elliptic deformation}
\label{sec:elliptic}
In this section we consider the elliptic deformation on $SL(2;\mathbb{R}) \times SU(2)$. 
To obtain the explicit background one may choose the following parametrisation:
\begin{equation}
    g=e^{T L_2} e^{U L_3} e^{V L_1} e^{\Phi J_2} e^{X J_3} e^{Y J_1}\,.
\end{equation}
The metric then reads
\begin{equation}
\label{eq:metric_elliptic}
\begin{aligned}
    \de s^2 &= \de s_a^2 + \de s_b^2 +\sum_{j=1}^4 \de x_j^2\,, \\
    \de s_a^2 &= \alpha_1 (\sinh 2U \de T - \de V)^2 - \alpha_2 (\cosh 2 U \cosh 2 V \de T + \sinh 2 V \de U)^2 \\
    &\qquad + \alpha_3 (\cosh 2 U \sinh 2 V \de T + \cosh 2 V \de U)^2\,, \\
    \de s_b^2 &= \alpha_1 (\sin 2X \de \Phi - \de Y)^2 + \alpha_2 (\cos 2 X \cos 2 Y \de \Phi - \sin 2 Y \de X)^2 \\
    &\qquad + \alpha_3 (\cos 2 X \sin 2 Y \de \Phi + \cos 2 Y \de X)^2\,,   
\end{aligned}
\end{equation}
where $\alpha_1, \alpha_2,\alpha_3 >0$ are the deformation parameters and the torus directions have been added by hand. The B-field vanishes. The metric presents two $U(1)$ isometries, realised as shifts in the coordinates $T, \Phi$. These can then be used to implement the uniform light-cone gauge fixing. Note that in the rational (undeformed) limit, $T$ does not represent global $AdS_3$ time. This corresponds to an alternative light-cone gauge fixing~\cite{Borsato:2023oru}.  At leading order we obtain the light-cone Hamiltonian~\cite{Hoare:2023zti}
\begin{equation}
\begin{aligned}
    \mathcal H_2 &= P_z P_{\bar{z}} + i \gamma_3(z P_z - \bar{z} P_{\bar{z}})+ |\acute{z}|^2 +(\gamma_1^2-\gamma_2^2) |z|^2 - \gamma_2 \gamma_3 (z^2 + \bar{z}^2)  \\
    &+P_y P_{\bar{y}} + i \gamma_3(y P_y - \bar{y} P_{\bar{y}})+ |\acute{y}|^2 +(\gamma_1^2-\gamma_2^2) |y|^2 - \gamma_2 \gamma_3 (y^2 + \bar{y}^2) \\
    &+ P_u P_{\bar{u}} + P_v P_{\bar{v}} + |\acute{u}|^2 + |\acute{v}|^2\,,
    \end{aligned}
\end{equation}
with
\begin{equation}
    \gamma_1 = \frac{\alpha_2}{\sqrt{\alpha_1 \alpha_2 \alpha_3}}\,, \qquad \gamma_2 = \frac{\alpha_1-\alpha_3}{\sqrt{\alpha_1 \alpha_2 \alpha_3}}\,, \qquad \gamma_3 = \frac{\alpha_1-\alpha_2+\alpha_3}{\sqrt{\alpha_1 \alpha_2 \alpha_3}}\,.
\end{equation}
Note that when $\gamma_2 =0$ (trigonometric deformation) then $\mathcal H_2$ has two $U(1)$ symmetries realised as $Y \rightarrow e^{i \beta} Y$ and $Z\rightarrow e^{i \beta} Z$. These are however lost for the elliptic deformation, and the equations of motion couple $Z$ with its conjugate $\bar{Z}$, as well as $Y$ with its conjugate $\bar{Y}$.
To diagonalise the quadratic Hamiltonian we then make the following Ansatz,
\begin{equation}
\begin{aligned}
    Z = \frac{1}{\sqrt{2}} \int \extder p \sum_{k=1}^2 \left( A_{zk}(p) e^{-i \omega_k(p) \tau + i p \sigma} a_k(p) + B_{zk}(p) e^{i \omega_k(p) \tau - i p \sigma} a_k^\dagger(p) \right)\,, \\
    Y = \frac{1}{\sqrt{2}} \int \extder p \sum_{k=3}^4 \left( A_{yk}(p) e^{-i \omega_k(p) \tau + i p \sigma} a_k(p) + B_{yk}(p) e^{i \omega_k(p) \tau - i p \sigma} a_k^\dagger(p) \right)\,.
    \end{aligned}
\end{equation}
The unusual form of this Ansatz is due to the fact that the $\mathfrak{u}(1)$ symmetry of the complex fields is broken by the term $z^2+\bar{z}{}^2$ (and similarly for~$y$).
The creation and annihilation operators satisfy the canonical commutation relations
\begin{equation}
\begin{aligned}
    &[a_j(p_1),a_k^\dagger(p_2)]=\delta_{jk}\delta(p_1-p_2)\,, \\
    &[a_j(p_1),a_k(p_2)]=[a_j^\dagger(p_1),a_k^\dagger(p_2)]=0\,,
    \end{aligned}
\end{equation}
and the functions $A_{zk}(p), B_{zk}(p), A_{yk}(p), B_{yk}(p)$ as well as $\omega_k(p)$ are fixed by requiring that the equations of motion are satisfied and that the fields and conjugate momenta obey canonical commutation relations.
Then,
\begin{equation}
    \mathbf{H} = \int_{-\infty}^{+\infty} \extder p \sum_{j=1}^8 \omega_j(p) a_j^\dagger(p) a_j(p)\,,
\end{equation}
with dispersion relations~$\omega_j(p)$ such that
\begin{equation}
    \sqrt{\omega_j(p)^2 + (\mu_j\gamma_2)^2} = \mu_j \gamma_3  + \sqrt{p^2+(\mu_j\gamma_1)^2}\,,
\end{equation}
and quantum numbers
\begin{equation}
    \mu_j = (+1,-1,+1,-1,0,0,0,0)\,.
\end{equation}
The tree-level S-matrix describing the scattering between the above excitations was obtained in~\cite{Hoare:2023zti} and it satisfies the classical YB equation. 
While it is not yet known how to generalise the elliptic deformation to the semi-symmetric space sigma model, the metric \eqref{eq:metric_elliptic} can be embedded into supergravity with RR 3-form and 5-form fluxes, as well as NSNS 3-form fluxes (in case one adds a WZ term) and a trivial dilaton~\cite{Hoare:2023zti}. These fluxes are most easily written in terms of the vielbein
\begin{equation}
\begin{aligned}
    e^0 &= \sqrt{\alpha_2} (-\cosh 2 U \cosh 2 V dT -\sinh 2 V dU)\,, \\
    e^1 &= \sqrt{\alpha_1} (-\sinh 2 U dT + d V)\,, \\
    e^2 &= \sqrt{\alpha_3} (\cosh 2 U \sinh 2 V dT + \cosh 2 V dU)\,, \\
    e^3 &= \sqrt{\alpha_2} (-\cos 2 X \cos 2 Y d\Phi + \sin 2 Y dX)\,, \\
    e^4 &= \sqrt{\alpha_1} (\sin 2 X d\Phi - dY)\,, \\
    e^5 &= \sqrt{\alpha_3} (-\cos 2 X \sin 2 Y d\Phi - \cos 2 Y dX)\,, \\
    e^6 &= \de x_1\,, \qquad e^7 = \de x_2\,, \qquad e^8 = \de x_3\,, \qquad e^9 = \de x_4\,.
    \end{aligned}
\end{equation}
Let us  make the following Ansatz for the NSNS and RR fluxes (while $B=0$ and hence $H_3=0$ for the elliptic deformation, here we add an H-flux for completeness)
\begin{equation}\begin{gathered}
\Phi = 0 \,, \qquad H_3 = F_3^{(5)}\,, \qquad 
F_1 = 0 \,, \qquad
F_3 = F_3^{(4)} \,, \qquad
F_5 = \sum_{i=1}^3 F_3^{(i)} \wedge J_2^{(i)} \,.
\end{gathered}\end{equation}
The 2-forms
\begin{equation}
\begin{aligned}
J_2^{(1)} &= \de x_1 \wedge \de x_2 - \de x_3 \wedge \de x_4 \,, \\
J_2^{(2)} &= \de x_1 \wedge \de x_3 + \de x_2 \wedge \de x_4 \,, \\
J_2^{(3)} &= \de x_1 \wedge \de x_4 - \de x_2 \wedge \de x_3 \,,
\end{aligned}
\end{equation}
are three orthogonal self-dual 2-forms on the 4-torus.
The three-form fluxes are given by
\begin{equation}\label{eq:fluxes}
F_3^{(i)} =  \mathbf{x}^{(1)}_i f_3^{(1)}
+ \mathbf{x}^{(2)}_i f_3^{(2)}
+  \mathbf{x}^{(3)}_i f_3^{(3)}
+ \mathbf{x}^{(4)}_i f_3^{(4)} \,,
\end{equation}
with auxiliary quantities
\begin{equation}
\begin{aligned}
    f_3^{(1)} &= 2  (
-e^0 \wedge e^2 \wedge e^4
+  e^1 \wedge e^3 \wedge e^5) \,,
\\
f_3^{(2)} &= 2 (
- e^1 \wedge e^2 \wedge e^3
- e^0 \wedge e^4 \wedge e^5) \,,
\\
f_3^{(3)} &=  2 (
- e^0 \wedge e^1 \wedge e^5
+ e^2 \wedge e^3 \wedge e^4) \,,
\\
f_3^{(4)} &= 2(e^0 \wedge e^1 \wedge e^2 + e^3 \wedge e^4 \wedge e^5) \,.
\end{aligned}
\end{equation}
For this background to solve the supergravity equations of motion the coefficients of the fluxes must satisfy 
\begin{equation}\begin{aligned}\label{eq:solution}
||\mathbf{x}^{(1)}||^2 & = \frac{(\alpha_2 + \alpha_3 - \alpha_1)}{\alpha_2 \alpha_3} - ||\mathbf{x}^{(4)}||^2 \,,
& \qquad
\mathbf{x}^{(1)} \cdot \mathbf{x}^{(4)} = -\mathbf{x}^{(2)} \cdot \mathbf{x}^{(3)} \,,
\\
||\mathbf{x}^{(2)}||^2 & = \frac{(\alpha_2 - \alpha_1 - \alpha_3)}{\alpha_1 \alpha_3} + ||\mathbf{x}^{(4)}||^2 \,,
& \qquad
\mathbf{x}^{(2)} \cdot \mathbf{x}^{(4)} = +\mathbf{x}^{(1)} \cdot \mathbf{x}^{(3)} \,,
\\
||\mathbf{x}^{(3)}||^2 & = \frac{(\alpha_2 + \alpha_1 - \alpha_3)}{\alpha_1 \alpha_2} - ||\mathbf{x}^{(4)}||^2 \,,
& \qquad
\mathbf{x}^{(3)} \cdot \mathbf{x}^{(4)} = -\mathbf{x}^{(1)} \cdot \mathbf{x}^{(2)} \,.
\end{aligned}
\end{equation}
The classical and quantum integrability of the proposed elliptic deformation of the $AdS_3 \times S^3 \times T^4$ superstring is an open question. Insight can be obtained through computing the perturbative S-matrix including the contributions coming from the scattering with fermions~\cite{Hoare:2025rtl}.
The fate of the broken right-acting symmetries under the elliptic deformation is not yet known, and the exact S-matrix remains to be found.

\section{Outlook}
\label{sec:outlook}
From this discussion, several outstanding questions emerge clearly.
In the case of the undeformed~$AdS_3\times S^3\times T^4$ model, a pressing question is to complete the integrability construction of the mirror TBA (or QSC) for mixed-flux backgrounds. The recent results of~\cite{Frolov:2024pkz} on the dressing factors are very promising in this regard. Having such a system of equations would allow to study various interesting limits, in particular the perturbations around the WZW points (fixed~$k$, $h\ll1$) which could hopefully then be compared with the results of other approaches --- e.g.\ based on the RNS approach or on the ``hybrid'' formalism of~\cite{Berkovits:1999im}. It is worth emphasising that, despite some recent work~\cite{Cho:2018nfn, Eberhardt:2018exh}, the computation of the spectrum in the worldsheet-CFT remains challenging even for fixed~$k$ and $h\ll1$.

Coming back to integrability, even in the pure-RR case, it would be appropriate to better understand the relation between the existing proposals for the spectrum~\cite{Ekhammar:2021pys, Cavaglia:2021eqr, Frolov:2021bwp} and to compare them with semiclassical computations.
Moreover, again in the case of the pure-RR background it would be interesting to study in more detail the tensionless limit ($k=0$, $h\ll1$). It would be important to work out the dual CFT either starting from the original description of the dual~\cite{Maldacena:1997re} as the sigma model on the moduli space of~$N_1$ instantons for a~$U(N_5)$ gauge theory on~$T^4$,%
\footnote{%
See~\cite{Aharony:2024fid} (and references therein) for a recent discussion of this idea; this was also recently discussed for the case of the $AdS_3\times S^3\times S^3\times S^1$ background in~\cite{Witten:2024yod}.}
or from the Higgs branch CFT~\cite{Witten:1997yu,OhlssonSax:2014jtq}. A less ambitious  (but still very interesting) related question is to work out the integrable model that reproduce the TBA spectrum at leading order in the weak-tension limit --- that is, $k=0$ and $T=\mathcal{O}(h)$ with $h\ll1$. From the TBA spectrum, it looks like an interacting supersymmetric quantum mechanical model of four bosons and fermions --- very different from the symmetric-product orbifold appearing at~$k=1$.

In a similar vein, virtually all of the questions which we have discussed above, such as the investigation of the spectrum (in the S~matrix / TBA / QSC approach) and of the duals could be asked for the $AdS_3\times S^3\times S^3\times S^1$ background. It is natural to expect that this would be substantially more complicated technically, due to the smaller amount of residual supersymmetry after lightcone gauge fixing --- which means more irreducible representations, more distinct dressing factors, etc. --- and it seems reasonable to first try and get a stronger grip on the  $AdS_3\times S^3\times T^4$ background before moving forward on the $AdS_3\times S^3\times S^3\times S^1$ background.
On the other hand, some interesting results for the deformations of $AdS_3\times S^3\times S^3\times S^1$ have recently been found~\cite{papero}.

Concerning the deformed backgrounds, it would be interesting to learn more about their spectrum. Various integrability techniques have been adapted to describe $q$-deformed $AdS_5 \times S^5$ superstrings, including the asymptotic Bethe Ansatz~\cite{Beisert:2008tw,Seibold:2021rml}, the TBA~\cite{Arutyunov:2012zt,Arutyunov:2012ai} and the QSC~\cite{Klabbers:2017vtw}. It would be interesting to extend these to $q$-deformed $AdS_3$ strings. The algebraic Bethe Ansatz for $AdS_3 \times S^3 \times T^4$ strings was worked out in~\cite{Seibold:2022mgg}, also with an abelian twist. Perhaps this could help identify their dual, maybe starting from the case in which the undeformed model is dual to the symmetric-orbifold CFT.

Among the deformations of $AdS_3$ backgrounds, the ones preserving some amount of supersymmetry are especially interesting; they share many interesting properties with the undeformed theory, with in particular a vanishing scalar curvature, a trivial dilaton, and relatively simple (mixed) fluxes~\cite{Hoare:2022asa}. This then gives us hope to be able to find a brane construction that gives rise to these backgrounds in some near-horizon limit, and opens up the possibility to study these deformations in the context of holography. They also allow to interpolate between well-studied integrable backgrounds, going for instance from $AdS_3$ strings to $AdS_2$ strings, or contracting certain spheres from $S^3$ to $S^2$.

Finally, it is worth noting that the very large majority of the literature of deformations focuses on the PCM/coset model, and from there tries to complete the study of the model through its S~matrix by analysing the (deformed) symmetries. This is also the route which we followed in this review.
One could in principle adopt another approach, initiated in~\cite{deLeeuw:2021ufg}, where one tries to bootstrap the most general 6-vertex or 8-vertex $S$-matrix satisfying the quantum Yang-Baxter equation. In that case, the main difficulty is to understand the physical interpretation of those S~matrices.

\begin{acknowledgement}
We thank the participants of the
MATRIX program ``New Deformations of Quantum Field and Gravity Theories'' in Creswick, Australia, and of the ``Integrability in low-supersymmetry theories'' school \& workshop in Trani, Italy, for stimulating discussions.
We are grateful to the MATRIX Institute and the Simons Foundation for financial support that allowed us to take part in the aforementioned MATRIX program. We are also grateful to the Kavli Institute for Theoretical Physics in Santa Barbara for the hospitality during the follow-on of the Integrable22 workshop, where part of this review was written.
FS is supported by the European Union Horizon
2020 research and innovation programme under the Marie Sk{\l}odowska-Curie grant agreement number 101027251, and by the Deutsche Forschungsgemeinschaft (DFG) under the Collaborative Research Center 1624 ``Higher structures, moduli spaces and integrability'', project number 506632645. FS would like to thank B.~Hoare and A.~Tseytlin for collaborations on related topics.
AS would like to acknowledge support from the European Union – NextGenerationEU, from the program STARS@UNIPD, under project ``Exact-Holography -- A
new exact approach to holography: harnessing the power of string theory, conformal field theory, and integrable
models'', from the PRIN Project n. 2022ABPBEY, ``Understanding quantum field theory through its deformations'', from the CARIPLO Foundation ``Supporto ai giovani talenti italiani nelle competizioni dell'European Research Council'' grant n.~2022-1886 ``Nuove basi per la teoria delle stringhe", from the CARIPARO Foundation Grant ``Ricerca Scientifica di Eccellenza 2023'' under project n.~68079 ``New Tools for String Theory'', and from the Australian Research Council (ARC) Discovery Project DP240101409.
\end{acknowledgement}

%
% BibTeX users please use
\bibliographystyle{spmpsci}
\bibliography{refs}
\end{document}